\documentclass[aps, prl, reprint, superscriptaddress]{revtex4-1}
\usepackage[utf8]{inputenc}
\usepackage{amsmath,amssymb}
\usepackage{physics}
\usepackage{graphicx}
\usepackage{newunicodechar}
\newunicodechar{。}{.}
\graphicspath{{pics/}}
\usepackage{subfigure}
\usepackage{tabularx,booktabs}
\usepackage{tikz}
\definecolor{airforceblue}{rgb}{0.36, 0.54, 0.66}
\definecolor{corn}{rgb}{0.98, 0.93, 0.36}
\definecolor{antiquewhite}{rgb}{0.98, 0.92, 0.84}
\usetikzlibrary{arrows.meta,shapes.arrows,shapes.geometric}
\usepackage{hyperref}

\begin{document}

\title{Tensor complex renormalization with generalized symmetry and topological bootstrap}
\author{Dong-Yu Bao}
\affiliation{Department of Physics, The Chinese University of Hong Kong, Sha Tin, New Territories, Hong Kong, China}
\author{Gong Cheng}
\affiliation{Department of Physics, Virginia Tech, Blacksburg, VA 24060, USA}
\author{Hong-Hao Song}
\affiliation{International Center for Quantum Materials, School of Physics, Peking University, Beijing 100871, China}
\author{Zheng-Cheng Gu}
\email{zcgu@phy.cuhk.edu.hk}
\affiliation{Department of Physics, The Chinese University of Hong Kong, Sha Tin, New Territories, Hong Kong, China}

\begin{abstract}
Recent progress in generalized symmetry and topological holography has shown that, in conformal field theory (CFT), topological data from one dimensional higher can play a key role in determining local dynamics. Based on this insight, a fixed-point (FP) tensor complex (TC) for CFT has recently been constructed.
In this work, we develop a TC renormalization (TCR) algorithm adapted to this CFT-based structure, forming a renormalization-group (RG) framework with generalized symmetry. We show that the full FP tensor can emerge from the RG flow starting with only the three-point function of the primary fields. Remarkably, even when starting solely from topological data, the RG process can still reconstruct the full FP tensor—a method we call as topological bootstrap. This approach deepens the connection between the topological and dynamical aspects of CFT and suggests pathways toward a fully algebraic description of gapless quantum states, with potential extensions to higher dimensions.
\end{abstract}

\maketitle

\emph{Introduction} — In recent years, generalized global symmetries have emerged as a profound organizing principle in the study of two-dimensional conformal field theories (CFTs). These symmetries reveal that topological data from one higher dimension can significantly constrain—or even fully determine—the local dynamical properties of a CFT. In the case of rational CFTs (RCFTs), this structure is algebraically captured by a fusion category \cite{mooreClassicalQuantumConformal1989,Moore1990,VERLINDE1988360}, which plays a profound role in the analysis of topological defects \cite{fuchs_tft_2002, fuchs_conformal_2002, fuchs_tft_2004,FUCHS2004277, FUCHS2005539, Frohlich:2006ch, PETKOVA2001157}.
This algebraic framework has become central to modern reformulations of two-dimensional statistical models, where it provides a natural bridge between 2D RCFTs and topological field theories (TFTs). 
This relationship is formalized through the symmetric TFT (SymTFT) correspondence, also known as topological holography \cite{PhysRevResearch.2.033417, PhysRevResearch.2.043086, Kong:2020jne, Freed:2022qnc, PhysRevB.107.155136,Kong2019AMT}.
A particularly concrete and powerful realization of this correspondence is provided by the strange correlator construction \cite{youWaveFunctionStrange2014,vanhoveMappingTopologicalConformal2018, Aasen:2020jwb,aasenTopologicalDefectsLattice2016, Vanhove_2022, hung_2d-cft_2025}. 
Building on this perspective, 
the holographic modular bootstrap \cite{PhysRevB.108.075105,RUELLE1998650,PhysRevD.107.125025, PhysRevD.103.125001, PhysRevResearch.1.033054, PhysRevB.102.045139,jzfv-ygmr} exploits modular consistency of twisted partition functions in the bulk TFT to bound operator dimensions and the density of states. Fusion-category symmetry imposes additional selection rules and asymptotic constraints \cite{Lin:2022dhv,Nakayama:2025mrm}.
Together, these developments demonstrate that minimal topological data can tightly constrain the CFT structure.

Over the past two decades, tensor networks, especially tensor network renormalization (TNR), have become powerful tools for studying both quantum and classical many-body systems. In particular, entanglement renormalization has been highly successful in classifying gapped topological phases \cite{PhysRevLett.99.220405,PhysRevB.78.205116,PhysRevB.79.144108,PhysRevA.79.040301,guTensorentanglementfilteringRenormalizationApproach2009,PhysRevLett.102.180406}
and in extracting conformal data at critical points for low-dimensional systems \cite{levinTensorRenormalizationGroup2007,xieSecondRenormalizationTensorNetwork2009,guTensorentanglementfilteringRenormalizationApproach2009,xieCoarsegrainingRenormalizationHigherorder2012,evenblyTensorNetworkRenormalization2015,yangLoopOptimizationTensor2017,evenblyAlgorithmsTensorNetwork2017a,PhysRevB.105.L060402,PhysRevResearch.6.043102}.
However, most of these methods do not directly incorporate the generalized symmetry and algebraic structure of CFTs. A recent advance introduced an exact fixed-point (FP) tensor construction for RCFTs~\cite{chengPrecisionReconstructionRational2025} , in which the FP tensor is represented as a tensor complex (TC) built from primary and descendant fields, with components determined by the fusion rules, structure constants, and conformal blocks of CFTs.
This construction naturally incorporates generalized symmetries and points toward a new type of generalized symmetry-preserving renormalization scheme that is fully capture the algebirac structure of the CFT.

In this work, we adapt the Loop-TNR algorithm \cite{yangLoopOptimizationTensor2017} to the TC structure and obtain a TC renormalization (TCR) framework. We first show that, starting from only the primary fields, the RG flow reconstructs the full FP tensor of the RCFT, even though the exact FP tensor formally requires an infinite bond dimension. Remarkably, the resulting flows are as stable as those produced by conventional TNR initialized from lattice models at criticality.
Furthermore, we show that it is unnecessary to supply the full operator product expansion (OPE) data
at the beginning. Using only the topological fusion data and partial scaling dimensions of the primary fields, the RG flow converges to the same FP tensor with comparable stability.
We validate this approach across multiple minimal models and discuss how this framework potentially points toward a  topological bootstrap for CFTs.

\emph{TCR with generalized symmetry} —
Here we briefly outline how the Loop-TNR scheme with generalized symmetry can be applied to the TC. 
Since the FP tensor for a CFT provides a way to encode generalized symmetry, we begin with its explicit structure.
An FP tensor is represented by a rank-3 TC $\mathcal{T}^{abc}_{(i,I)(j,J)(k,K)}$ with nine indices, where $a,b,c$ denote conformal boundary conditions (CBCs); $i,j,k$ label the boundary-changing operators (BCOs) associated with the primary fields of the CFT; and $I,J,K$ label the corresponding descendant fields: 
\begin{equation}\label{eq:FPtensor}
    \mathcal{T}^{abc}_{(i,I)(j,J)(l,K)}=\alpha^{ijk}_{IJK}C^{abc}_{ijk},
\end{equation}
where $\alpha^{ijk}_{IJK}$ is the three-point conformal block determined entirely by conformal symmetry, and $C^{abc}_{ijk}$ is the structure constant appearing in the OPE of BCOs. Additional details on the construction of the TC are provided in the Supplementary Material. 

\begin{figure}
    \centering
    \subfigure{
    \includegraphics[width=.7\linewidth]{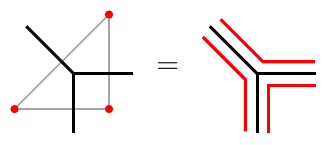}
    }
    \subfigure{
    \includegraphics[width=.7\linewidth]{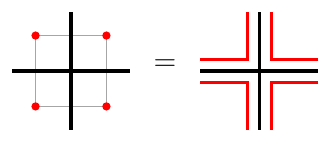}
    }
    \caption{Representing three-point and four-point correlation functions in the form of the TC as triple-line tensors. Red dots/lines denote CBCs, while black lines denote BCOs labeled by primary and descendant fields.}
    \label{fig:tripleline}
\end{figure}

The key observation is that the TC naturally exhibits a block structure (superselection sectors): each block is specified by the set of CBCs at the corners of a triangle, and the states on the edges are those compatible with the fusion rules. The TC therefore decomposes as a direct sum over all such blocks, with different blocks not mixing. This block structure allows us to equivalently represent the TC as a ``triple-line tensor", as illustrated in Fig.~\ref{fig:tripleline}, where each red leg encodes the CBC and each black leg carries the data associated with the primary fields and their corresponding descendants. In our TCR algorithm, however, we do not work with general triple-line tensors directly, as the computational cost is high and the generalized symmetry is not manifest. Instead, we perform all linear algebra operations in a blockwise manner, keeping each superselection sector separate and preserving the TC structure throughout the algorithm.


\begin{figure}
    \centering
    \includegraphics[width=\linewidth]{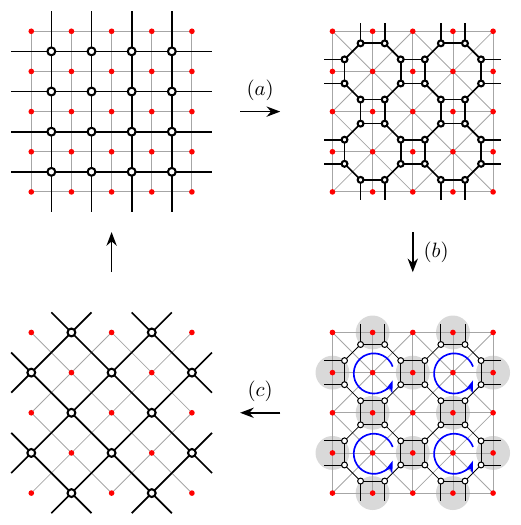}
    \caption{One RG step of the Loop-TNR algorithm for the TC. Red dots denote CBCs. (a) Each rank-4 tensor is decomposed into two rank-3 tensors by performing SVD in two different directions. (b) Loop optimization is carried out on the octagons indicated by the blue arrows. (c) Each shaded region containg four rank-3 tensors is contracted, with the internal CBC summed over, to produce a new coarse-grained rank-4 tensor.}
\label{fig:rgstep}
\end{figure}

The basic ingredients of one RG step of Loop-TNR for the TC are illustrated in Fig.~\ref{fig:rgstep}. In step $(a)$, we perform singular value decompositions (SVDs) on rank-4 tensors to obtain rank-3 tensors. In step $(b)$, we carry out loop optimization on these newly obtained rank-3 tensors inside the octagons indicated by blue arrows, which removes short-range entanglement. Both the SVD and loop optimization are performed in a blockwise manner. In step $(c)$, all small shaded squares are contracted to form new coarse-grained rank-4 tensors, with the internal CBCs summed over. Throughout, we assume that the tensor network is uniform so that we only need to work with a $2\times 2$ unit cell.

\begin{figure}
    \centering
    \includegraphics[width=\linewidth]{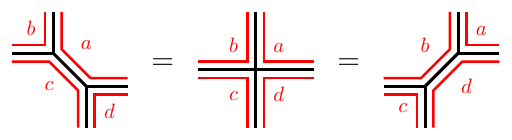}
    \caption{Two types of SVDs are performed on each rank-4 tensor block to obtain two rank-3 tensors. For the SVD shown on the right, we fix the boundary conditions $b,d$ and group together all tensor blocks with different $a,c$ into a single large matrix, on which the SVD is performed for each pair $(b,d)$. Similarly, for the SVD shown on the left, we fix $a,c$ and group over $b,d$.}
    \label{fig:svd}
\end{figure}

We now demonstrate in more detail how to perform blockwise SVD with truncation. The triple-line structure naturally guides this process. As shown in Fig.~\ref{fig:svd}, consider the decomposition on the right: for each pair of CBCs $(b,d)$ on the two sides of the intermediate leg, we group all CBCs $a,c$ when forming the SVD matrix. In other words, for a fixed pair $(b,d)$, the tensor blocks corresponding to different $a,c$ are assembled into a large block matrix, with its row and column indices labeled by $a$ and $c$, respectively. We then perform an SVD on this large block matrix and keep only the largest $\chi$ number of singular values and singular vectors. The complete truncated SVD of the rank-4 TC is obtained by repeating this procedure for all pairs of boundary conditions $(b,d)$. Note that if more than one primary appears in the fusion of CBCs $b$ and $d$, they mix under this SVD scheme through a gauge transformation. However, the states/BCOs associated with different pairs of CBCs—i.e., different superselection sectors—remain strictly separated.

\begin{figure}
    \centering
    \includegraphics[width=\linewidth]{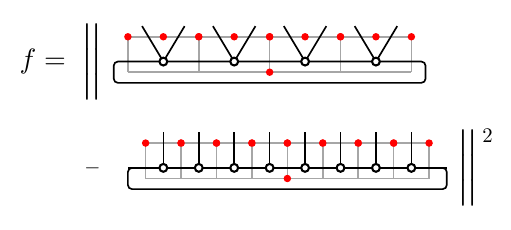}
    \caption{Cost function defined as the distance between two periodic MPS wave functions with TC structure. In both MPS wave functions, the leftmost CBCs are identified with the rightmost CBCs.}
    \label{fig:cost}
\end{figure}

In the loop optimization for the TC, we minimize the cost function shown in Fig.~\ref{fig:cost} within an octagon (indicated by the blue arrow in Fig.~\ref{fig:rgstep}) by optimizing the tensors site by site, following the standard Loop-TNR procedure \cite{yangLoopOptimizationTensor2017}. The key differences from the usual Loop-TNR are twofold. First, tensor contractions are replaced by contractions of TCs, where all internal CBC indices are summed over. Second, when solving the linear equation associated with optimizing the tensor at each site, we perform the optimization in a blockwise manner. Analogous to the SVD procedure, we fix a subset of CBC indices, group the remaining ones into the row or column indices of enlarged block matrices, and then solve the resulting linear systems for all previously fixed CBC choices. We sweep through all sites repeatedly until the total cost function converges to a small value. As in conventional Loop-TNR, we further improve the initial tensors by performing an SVD after a sequence of QR/LQ decompositions on the TC. Additional details are provided in the Supplementary Material. The overall computational cost scales as $O(n^{5}\chi^{6})$ (with an appropriate order of contractions), where $n$ is the number of CBCs.

In the final coarse-graining step, the CBCs in the middle of the shaded squares must be summed and weighted by factors $\omega_i$ proportional to the quantum dimensions, in order to satisfy the coarse-graining condition of the FP tensor \cite{chengPrecisionReconstructionRational2025}. Equivalently, we may absorb this weight into the initial rank-4 tensors by multiplying each block with CBCs $a,b,c,d$ by a factor $\omega_a^{1/4}\omega_b^{1/4}\omega_c^{1/4}\omega_d^{1/4}$ (see the middle panel of Fig.~\ref{fig:svd}). This ensures that, during coarse-graining, each boundary condition $i$ is weighted by exactly $\omega_i$ in the summation. We can then perform standard blockwise tensor contractions without explicitly tracking $\omega_i$ thereafter. 
From the rank-4 tensors produced at each RG step, and after appropriate normalization, we extract the conformal data—such as the central charge, scaling dimensions, and conformal spins—by constructing the corresponding transfer matrices. Further details are provided in the Supplementary Material. 


\begin{figure}
    \centering
    \includegraphics[width=0.9\linewidth]{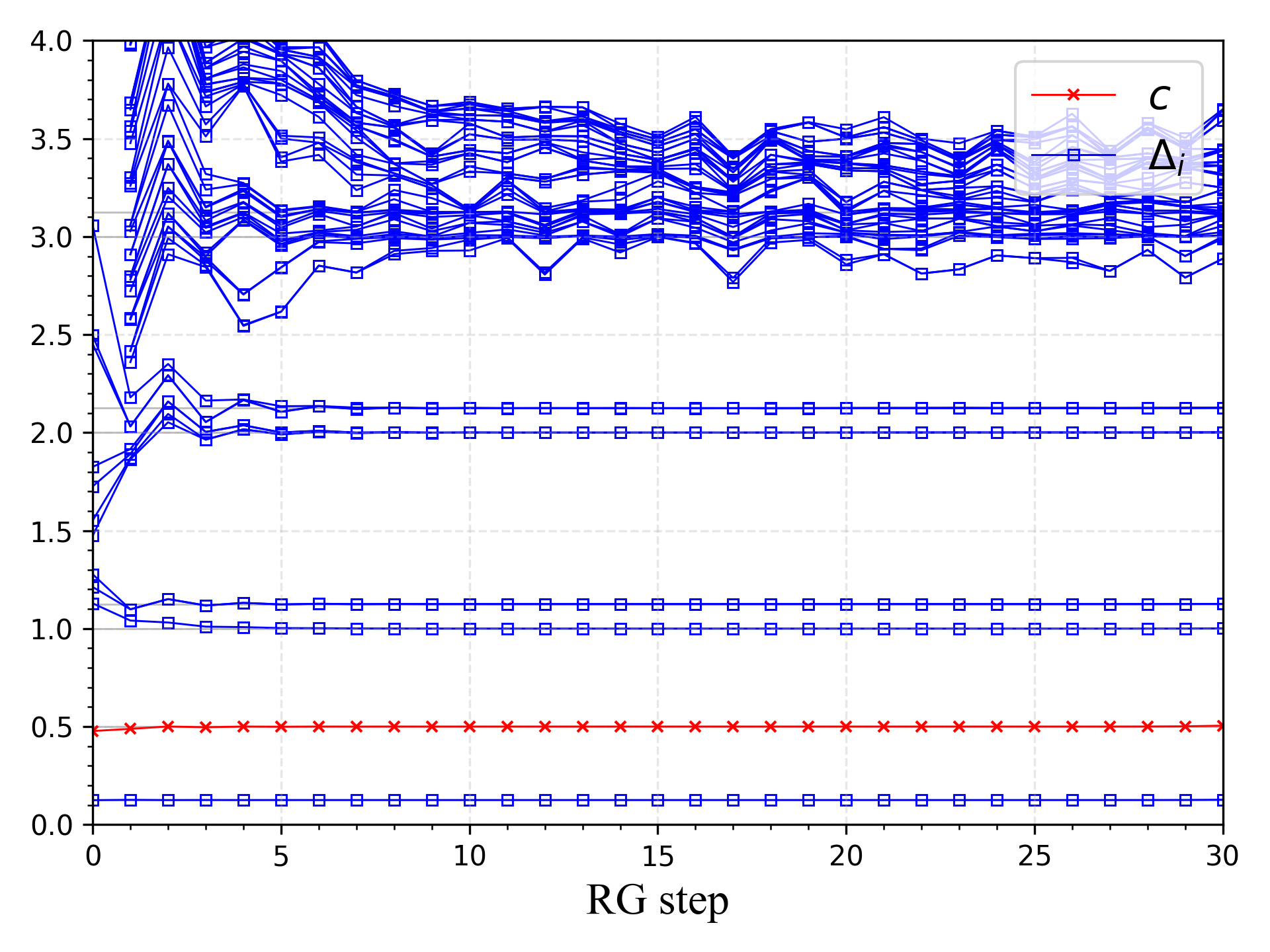}
    \caption{Ising CFT spectrum in the disk gauge, with conformal block initialized using only primary components, $\chi=16$.}
    \label{fig:ising_full_disk}
\end{figure}
\begin{figure}
    \centering
    \includegraphics[width=0.9\linewidth]{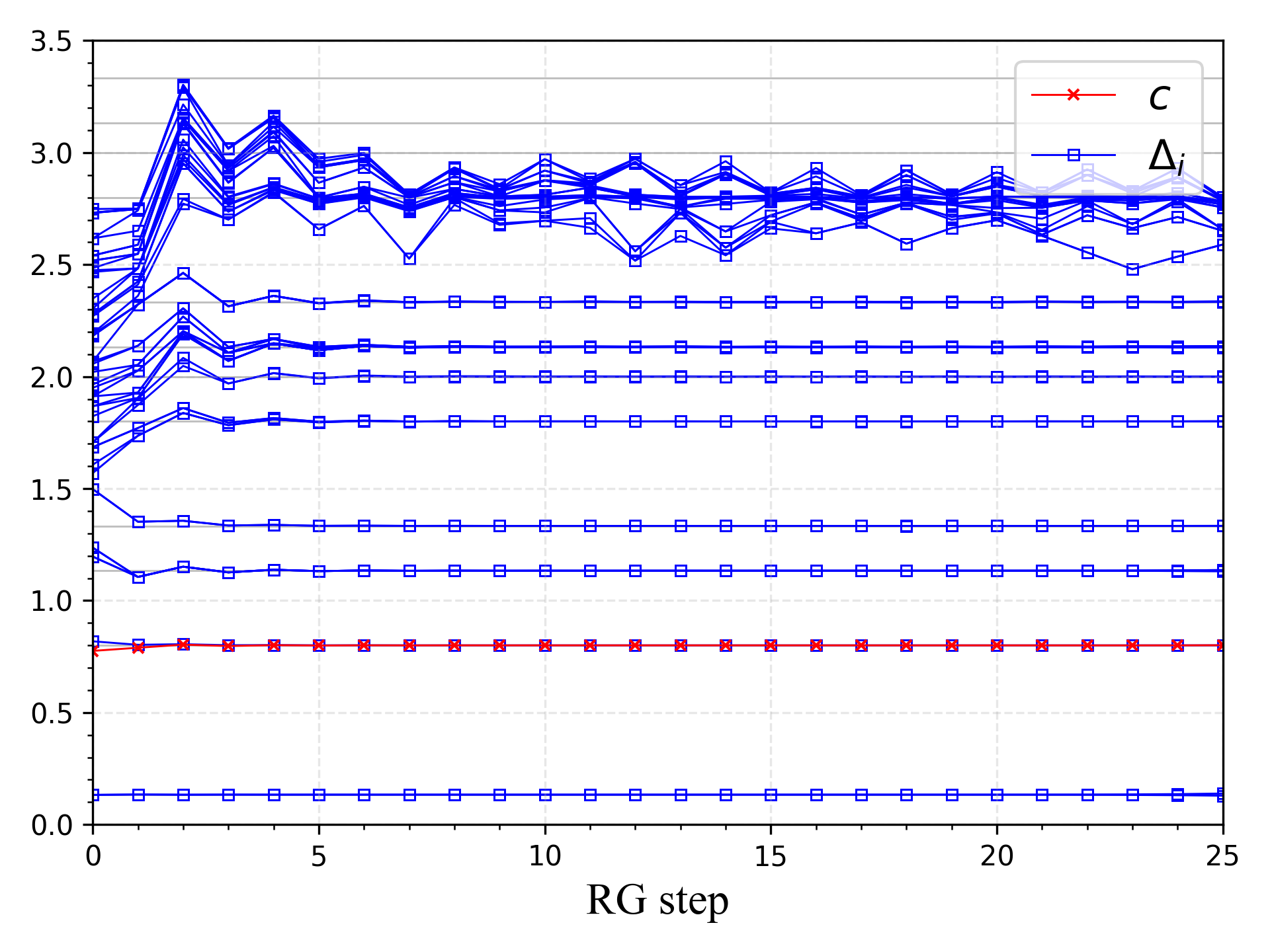}
    \caption{3-state Potts CFT spectrum in the disk gauge, with conformal block initialized using only primary components, $\chi=16$.}
    \label{fig:3potts_full_disk}
\end{figure}

\emph{Simple examples} —
Here we present the results of the TCR algorithm applied to two minimal models: the Ising CFT $\mathcal{M}(4,3)$ in the $A$-series and the 3-state Potts CFT $\mathcal{M}(6,5)$ in the $D$-series. 
The Ising CFT has three primaries $\Bqty{I,\sigma,\varepsilon}$ with holomorphic conformal dimensions $h_I=0,h_\sigma=1/16$, and $h_{\varepsilon}=1/2$. The three CBCs $\Bqty{+,f,-}$ correspond, respectively, to the Cardy states built from these primaries. The structure constants $C^{abc}_{ijk}$ for the Ising CFT are computed in \cite{LEWELLEN1992654}. In the 3-state Potts CFT, we use six primary fields, labeled by their conformal dimensions $\{0, \frac 1{15},\frac 2 5, \frac 2 3, \frac7 5, 3\}$, and six conformal boundary conditions classified in Ref.~\cite{CARDY1989581}. The structure constants can be computed using the general algorithm of Ref.~\cite{Runkel:1999dz}, derived from the fusion matrix of the corresponding A-series model. A complete list of the structure constants of the 3-state Potts CFT can be found in the Supplementary Material.

We first demonstrate that, when initialized with only the primary components of the tensor in Eq.~\eqref{eq:FPtensor}, our TCR algorithm produces the correct FP tensors. Concretely, the input tensor is chosen as
\begin{equation}
\mathcal{T}^{abc}_{ijk} = \alpha^{ijk}_{000} C^{abc}_{ijk},
\end{equation}
where $\alpha^{ijk}_{000}$ denotes the three-point conformal block of the primary fields. In the disk gauge (see Supplementary Material for a review),
its leading behavior is
\begin{equation}\label{eq:alpha_disk}
\alpha^{ijk}_{000} \approx 0.266^{h_i+h_j}0.704^{h_k}.
\end{equation}

The spectra obtained from the resulting FP tensors are shown in Fig.~\ref{fig:ising_full_disk}--\ref{fig:3potts_full_disk}. We find that these low-dimensional input data yield stable RG flows that accurately reproduce both the conformal dimensions and the central charge of the underlying CFTs.
The conformal spins are also correctly extracted using twisted transfer matrices, as shown in the Supplementary Material.
These results demonstrate that the basic three-point functions of the primary BCOs are sufficient to generate the entire CFT fixed-point tensors and thus capture the essential features of the theory. We note that, as shown in Ref.~\cite{chengPrecisionReconstructionRational2025}, the inclusion of descendant fields is necessary to satisfy the FP conditions exactly. Our analysis further suggests that the contributions associated with descendants can in fact be generated dynamically through the RG flow acting on the primary input data.

\emph{Topological bootstrap} —
The information contained in the input tensor can be separated into a topological/categorical part and a conformal/geometric part. The topological data are encoded in the $6j$-symbols, denoted by $\hat{C}_{ijk}^{abc}$ (up to normalization), which are related to the structure constants $C_{ijk}^{abc}$ in the OPE of BCOs
via
\begin{equation}
    C_{ijk}^{abc}=\hat{C}_{ijk}^{abc}/\mathcal{N}_{ijk},
\end{equation}
where the factor $\mathcal{N}_{ijk}$ is determined by the bulk OPE coefficients:
\begin{equation}
    \mathcal N_{ijk}=\frac 1 {\sqrt{C_{ijk}^{\text{bulk}}}}.
\end{equation}
Unlike the $6j$-symbols, the factors $\mathcal N_{ijk}$ are not topological and need to be solved by bootstrap methods \cite{DOTSENKO1985291}. 

To isolate the purely topological input, we set $\mathcal N_{ijk}=1$ and retain only $\hat{C}_{ijk}^{abc}$ and $\alpha^{ijk}_{000}$ in the input tensor:
\begin{equation}
\mathcal{T}^{abc}_{ijk} = \alpha^{ijk}_{000} \hat{C}^{abc}_{ijk}.
\end{equation}
Note that the primary three-point conformal block $\alpha^{ijk}_{000}$ depends only on the conformal dimensions, which can be determined from the Dehn twists up to an additive integer ambiguity \cite{Li:1989hs,Rowell:2007dge, Aasen:2020jwb}.

\begin{figure}
    \centering
    \includegraphics[width=0.9\linewidth]{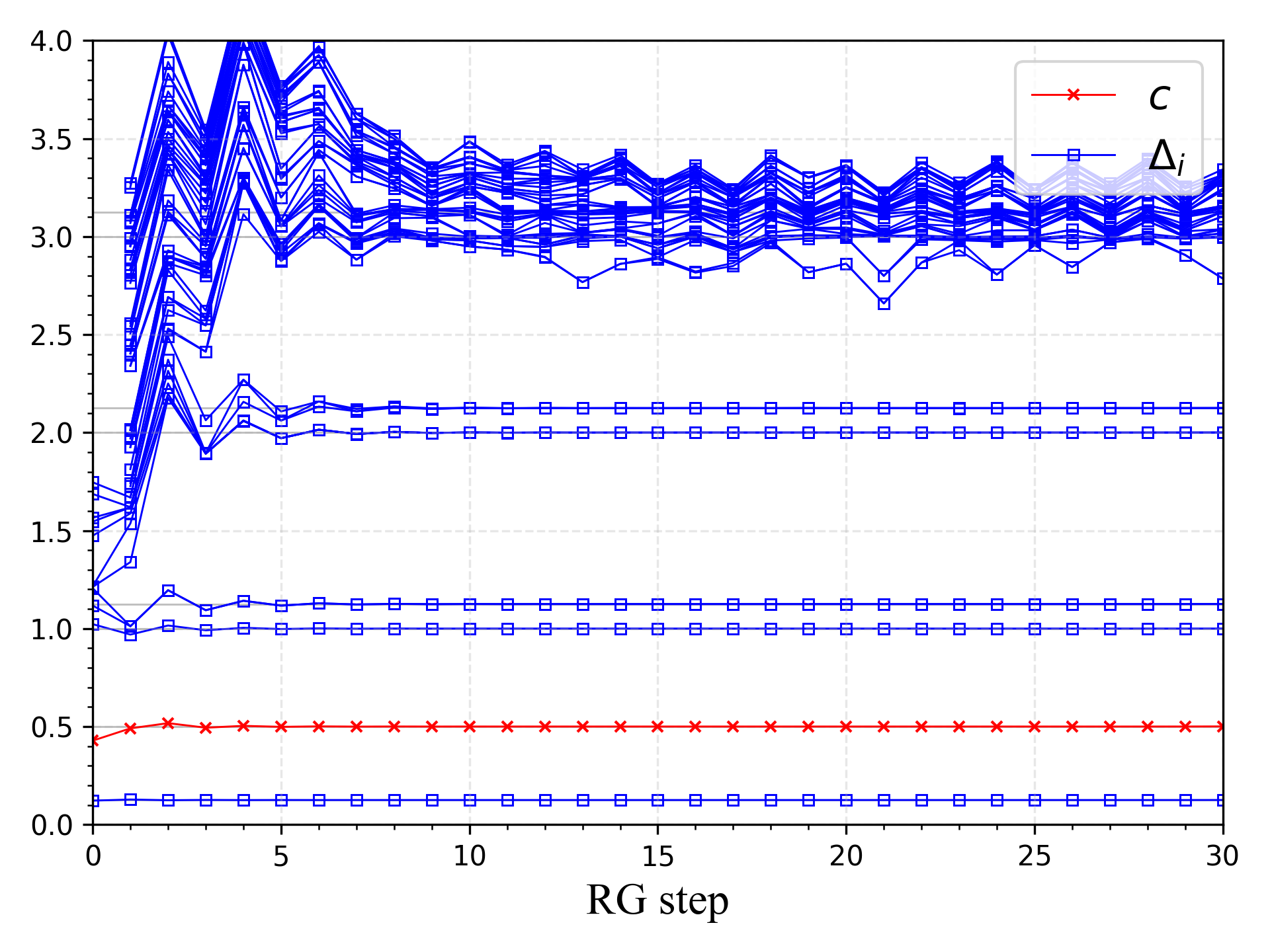}
    \caption{Ising CFT spectrum in the disk gauge, $\mathcal{N}=1$, $\chi=16$.}
    \label{fig:ising_N=1_disk}
\end{figure}
\begin{figure}
    \centering
    \includegraphics[width=0.9\linewidth]{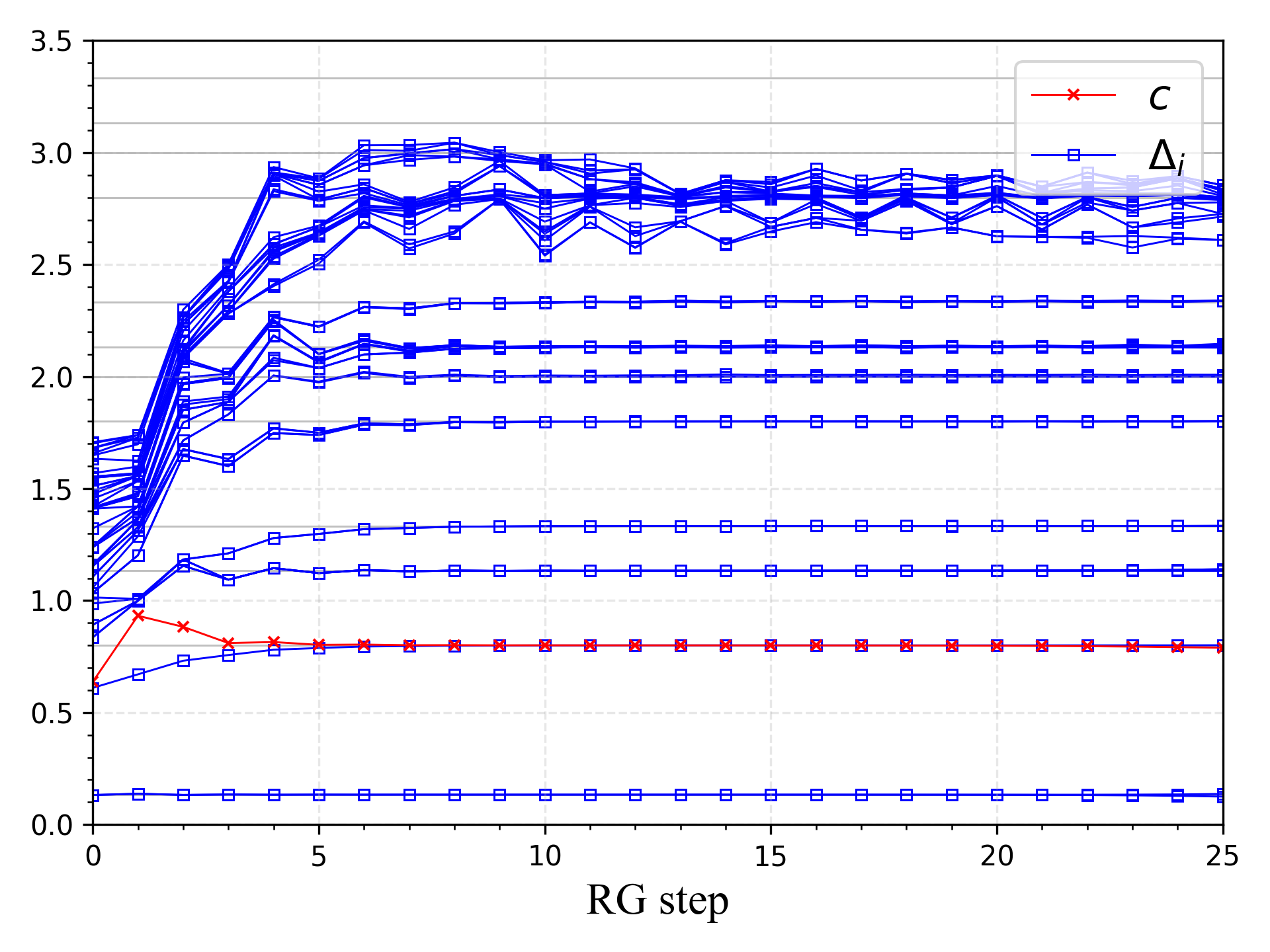}
    \caption{3-state Potts CFT spectrum in the disk gauge, $\mathcal{N}=1$, $\chi=16$.}
    \label{fig:3potts_N=1_disk}
\end{figure}

We tested the RG flows using these reduced input data. As shown in Fig.~\ref{fig:ising_N=1_disk}--\ref{fig:3potts_N=1_disk}, although the spectra are not correct at the beginning, the RG flows gradually approach the correct spectra and become stable after several RG steps. 
We further examined other $A$-series minimal models—including the tricritical Ising CFT $\mathcal{M}(5,4)$, the tetracritical Ising CFT $\mathcal{M}(6,5)$, as well as the simplest non-unitary Yang-Lee CFT $\mathcal{M}(5,2)$—in both the disk gauge and an alternative pants gauge with a different expression of the conformal block $\alpha^{ijk}_{000}$ (we note that in the non-unitary case, for the pants gauge with primary components only, we need to add a factor of $\exp(-Ih_i)$ on each bond, where $I$ is a length parameter and $h_i$ is the conformal dimension of the corresponding BCO, see Supplementary Material for more details).
In all cases, we find that the $\mathcal{N}=1$ topological bootstrap works consistently well. These results demonstrate that the reduced topological input data remain sufficient to accurately reproduce the FP tensors under RGs flows that preserve the generalized symmetry -- a topological bootstrap scheme. 

\emph{Conclusions and discussions} — 
To conclude, we have proposed a TCR method built upon a recent construction of the FP tensors of CFT. We find that accurate CFT FP tensors can be reproduced when the three-point conformal blocks in the FP tensors are initialized with only the primary components. Remarkably, the RG flows remain stable even when all detailed conformal or geometric information is removed: purely topological input, supplemented only by partial information about the conformal dimensions, is already sufficient to recover the full CFT FP tensors. 
We have validated this result across several minimal models, including Ising, tricritical Ising, tetracritical Ising CFT in the $A$-series, the 3-state Potts model in the $D$-series, and the non-unitary Yang-Lee CFT. These results point to a topological bootstrap scheme for CFTs, in which the full dynamical content of the theory can be recovered from its topological data under a generalized-symmetry-preserving RG flow, with potential generalizations to CFTs beyond the minimal models. In this regard, the method also serves as a powerful numerical tool for identifying new CFTs directly from their underlying fusion-category data.

We emphasize that the topological bootstrap mechanism developed in this work differs in several important aspects from previous generalized-symmetry–based bootstrap approaches \cite{PhysRevB.108.075105, RUELLE1998650, PhysRevD.107.125025, PhysRevD.103.125001, PhysRevResearch.1.033054, PhysRevB.102.045139, jzfv-ygmr, Lin:2022dhv, Nakayama:2025mrm}. Traditional bootstrap methods impose consistency relations—such as crossing symmetry, modular constraints, or defect-fusion selection rules—to define the allowed region of conformal data, typically yielding bounds or islands for operator dimensions and OPE coefficients.
In contrast, our approach employs a generalized-symmetry-constrained tensor network renormalization procedure that constructively generates the conformal data through RG flow, rather than inferring it indirectly from inequalities or consistency conditions. Although the present implementation focuses on two-dimensional CFTs, the underlying philosophy naturally extends to higher dimensions. 


\bibliographystyle{apsrev4-1}
\bibliography{refs}

\begin{thebibliography}{61}%
\makeatletter
\providecommand \@ifxundefined [1]{%
 \@ifx{#1\undefined}
}%
\providecommand \@ifnum [1]{%
 \ifnum #1\expandafter \@firstoftwo
 \else \expandafter \@secondoftwo
 \fi
}%
\providecommand \@ifx [1]{%
 \ifx #1\expandafter \@firstoftwo
 \else \expandafter \@secondoftwo
 \fi
}%
\providecommand \natexlab [1]{#1}%
\providecommand \enquote  [1]{``#1''}%
\providecommand \bibnamefont  [1]{#1}%
\providecommand \bibfnamefont [1]{#1}%
\providecommand \citenamefont [1]{#1}%
\providecommand \href@noop [0]{\@secondoftwo}%
\providecommand \href [0]{\begingroup \@sanitize@url \@href}%
\providecommand \@href[1]{\@@startlink{#1}\@@href}%
\providecommand \@@href[1]{\endgroup#1\@@endlink}%
\providecommand \@sanitize@url [0]{\catcode `\\12\catcode `\$12\catcode
  `\&12\catcode `\#12\catcode `\^12\catcode `\_12\catcode `\%12\relax}%
\providecommand \@@startlink[1]{}%
\providecommand \@@endlink[0]{}%
\providecommand \url  [0]{\begingroup\@sanitize@url \@url }%
\providecommand \@url [1]{\endgroup\@href {#1}{\urlprefix }}%
\providecommand \urlprefix  [0]{URL }%
\providecommand \Eprint [0]{\href }%
\providecommand \doibase [0]{http://dx.doi.org/}%
\providecommand \selectlanguage [0]{\@gobble}%
\providecommand \bibinfo  [0]{\@secondoftwo}%
\providecommand \bibfield  [0]{\@secondoftwo}%
\providecommand \translation [1]{[#1]}%
\providecommand \BibitemOpen [0]{}%
\providecommand \bibitemStop [0]{}%
\providecommand \bibitemNoStop [0]{.\EOS\space}%
\providecommand \EOS [0]{\spacefactor3000\relax}%
\providecommand \BibitemShut  [1]{\csname bibitem#1\endcsname}%
\let\auto@bib@innerbib\@empty
\bibitem [{\citenamefont {Moore}\ and\ \citenamefont
  {Seiberg}(1989)}]{mooreClassicalQuantumConformal1989}%
  \BibitemOpen
  \bibfield  {author} {\bibinfo {author} {\bibfnamefont {G.}~\bibnamefont
  {Moore}}\ and\ \bibinfo {author} {\bibfnamefont {N.}~\bibnamefont
  {Seiberg}},\ }\href {\doibase 10.1007/BF01238857} {\bibfield  {journal}
  {\bibinfo  {journal} {Communications in Mathematical Physics}\ }\textbf
  {\bibinfo {volume} {123}},\ \bibinfo {pages} {177} (\bibinfo {year}
  {1989})}\BibitemShut {NoStop}%
\bibitem [{\citenamefont {Moore}\ and\ \citenamefont
  {Seiberg}(1990)}]{Moore1990}%
  \BibitemOpen
  \bibfield  {author} {\bibinfo {author} {\bibfnamefont {G.}~\bibnamefont
  {Moore}}\ and\ \bibinfo {author} {\bibfnamefont {N.}~\bibnamefont
  {Seiberg}},\ }\enquote {\bibinfo {title} {Lectures on rcft},}\ in\ \href
  {\doibase 10.1007/978-1-4615-3802-8_8} {\emph {\bibinfo {booktitle} {Physics,
  Geometry and Topology}}},\ \bibinfo {editor} {edited by\ \bibinfo {editor}
  {\bibfnamefont {H.~C.}\ \bibnamefont {Lee}}}\ (\bibinfo  {publisher}
  {Springer US},\ \bibinfo {address} {Boston, MA},\ \bibinfo {year} {1990})\
  pp.\ \bibinfo {pages} {263--361}\BibitemShut {NoStop}%
\bibitem [{\citenamefont {Verlinde}(1988)}]{VERLINDE1988360}%
  \BibitemOpen
  \bibfield  {author} {\bibinfo {author} {\bibfnamefont {E.}~\bibnamefont
  {Verlinde}},\ }\href {\doibase https://doi.org/10.1016/0550-3213(88)90603-7}
  {\bibfield  {journal} {\bibinfo  {journal} {Nuclear Physics B}\ }\textbf
  {\bibinfo {volume} {300}},\ \bibinfo {pages} {360} (\bibinfo {year}
  {1988})}\BibitemShut {NoStop}%
\bibitem [{\citenamefont {Fuchs}\ \emph
  {et~al.}(2002{\natexlab{a}})\citenamefont {Fuchs}, \citenamefont {Runkel},\
  and\ \citenamefont {Schweigert}}]{fuchs_tft_2002}%
  \BibitemOpen
  \bibfield  {author} {\bibinfo {author} {\bibfnamefont {J.}~\bibnamefont
  {Fuchs}}, \bibinfo {author} {\bibfnamefont {I.}~\bibnamefont {Runkel}}, \
  and\ \bibinfo {author} {\bibfnamefont {C.}~\bibnamefont {Schweigert}},\
  }\href {\doibase 10.1016/S0550-3213(02)00744-7} {\bibfield  {journal}
  {\bibinfo  {journal} {Nuclear Physics B}\ }\textbf {\bibinfo {volume}
  {646}},\ \bibinfo {pages} {353} (\bibinfo {year} {2002}{\natexlab{a}})},\
  \bibinfo {note} {arXiv:hep-th/0204148}\BibitemShut {NoStop}%
\bibitem [{\citenamefont {Fuchs}\ \emph
  {et~al.}(2002{\natexlab{b}})\citenamefont {Fuchs}, \citenamefont {Runkel},\
  and\ \citenamefont {Schweigert}}]{fuchs_conformal_2002}%
  \BibitemOpen
  \bibfield  {author} {\bibinfo {author} {\bibfnamefont {J.}~\bibnamefont
  {Fuchs}}, \bibinfo {author} {\bibfnamefont {I.}~\bibnamefont {Runkel}}, \
  and\ \bibinfo {author} {\bibfnamefont {C.}~\bibnamefont {Schweigert}},\
  }\href {\doibase 10.1016/S0550-3213(01)00638-1} {\bibfield  {journal}
  {\bibinfo  {journal} {Nuclear Physics B}\ }\textbf {\bibinfo {volume}
  {624}},\ \bibinfo {pages} {452} (\bibinfo {year} {2002}{\natexlab{b}})},\
  \bibinfo {note} {arXiv:hep-th/0110133}\BibitemShut {NoStop}%
\bibitem [{\citenamefont {Fuchs}\ \emph
  {et~al.}(2004{\natexlab{a}})\citenamefont {Fuchs}, \citenamefont {Runkel},\
  and\ \citenamefont {Schweigert}}]{fuchs_tft_2004}%
  \BibitemOpen
  \bibfield  {author} {\bibinfo {author} {\bibfnamefont {J.}~\bibnamefont
  {Fuchs}}, \bibinfo {author} {\bibfnamefont {I.}~\bibnamefont {Runkel}}, \
  and\ \bibinfo {author} {\bibfnamefont {C.}~\bibnamefont {Schweigert}},\
  }\href {\doibase 10.1016/j.nuclphysb.2003.11.026} {\bibfield  {journal}
  {\bibinfo  {journal} {Nuclear Physics B}\ }\textbf {\bibinfo {volume}
  {678}},\ \bibinfo {pages} {511} (\bibinfo {year} {2004}{\natexlab{a}})},\
  \bibinfo {note} {arXiv:hep-th/0306164}\BibitemShut {NoStop}%
\bibitem [{\citenamefont {Fuchs}\ \emph
  {et~al.}(2004{\natexlab{b}})\citenamefont {Fuchs}, \citenamefont {Runkel},\
  and\ \citenamefont {Schweigert}}]{FUCHS2004277}%
  \BibitemOpen
  \bibfield  {author} {\bibinfo {author} {\bibfnamefont {J.}~\bibnamefont
  {Fuchs}}, \bibinfo {author} {\bibfnamefont {I.}~\bibnamefont {Runkel}}, \
  and\ \bibinfo {author} {\bibfnamefont {C.}~\bibnamefont {Schweigert}},\
  }\href {\doibase https://doi.org/10.1016/j.nuclphysb.2004.05.014} {\bibfield
  {journal} {\bibinfo  {journal} {Nuclear Physics B}\ }\textbf {\bibinfo
  {volume} {694}},\ \bibinfo {pages} {277} (\bibinfo {year}
  {2004}{\natexlab{b}})}\BibitemShut {NoStop}%
\bibitem [{\citenamefont {Fuchs}\ \emph {et~al.}(2005)\citenamefont {Fuchs},
  \citenamefont {Runkel},\ and\ \citenamefont {Schweigert}}]{FUCHS2005539}%
  \BibitemOpen
  \bibfield  {author} {\bibinfo {author} {\bibfnamefont {J.}~\bibnamefont
  {Fuchs}}, \bibinfo {author} {\bibfnamefont {I.}~\bibnamefont {Runkel}}, \
  and\ \bibinfo {author} {\bibfnamefont {C.}~\bibnamefont {Schweigert}},\
  }\href {\doibase https://doi.org/10.1016/j.nuclphysb.2005.03.018} {\bibfield
  {journal} {\bibinfo  {journal} {Nuclear Physics B}\ }\textbf {\bibinfo
  {volume} {715}},\ \bibinfo {pages} {539} (\bibinfo {year}
  {2005})}\BibitemShut {NoStop}%
\bibitem [{\citenamefont {Frohlich}\ \emph {et~al.}(2007)\citenamefont
  {Frohlich}, \citenamefont {Fuchs}, \citenamefont {Runkel},\ and\
  \citenamefont {Schweigert}}]{Frohlich:2006ch}%
  \BibitemOpen
  \bibfield  {author} {\bibinfo {author} {\bibfnamefont {J.}~\bibnamefont
  {Frohlich}}, \bibinfo {author} {\bibfnamefont {J.}~\bibnamefont {Fuchs}},
  \bibinfo {author} {\bibfnamefont {I.}~\bibnamefont {Runkel}}, \ and\ \bibinfo
  {author} {\bibfnamefont {C.}~\bibnamefont {Schweigert}},\ }\href {\doibase
  10.1016/j.nuclphysb.2006.11.017} {\bibfield  {journal} {\bibinfo  {journal}
  {Nucl. Phys. B}\ }\textbf {\bibinfo {volume} {763}},\ \bibinfo {pages} {354}
  (\bibinfo {year} {2007})},\ \Eprint {http://arxiv.org/abs/hep-th/0607247}
  {arXiv:hep-th/0607247} \BibitemShut {NoStop}%
\bibitem [{\citenamefont {Petkova}\ and\ \citenamefont
  {Zuber}(2001)}]{PETKOVA2001157}%
  \BibitemOpen
  \bibfield  {author} {\bibinfo {author} {\bibfnamefont {V.}~\bibnamefont
  {Petkova}}\ and\ \bibinfo {author} {\bibfnamefont {J.-B.}\ \bibnamefont
  {Zuber}},\ }\href {\doibase https://doi.org/10.1016/S0370-2693(01)00276-3}
  {\bibfield  {journal} {\bibinfo  {journal} {Physics Letters B}\ }\textbf
  {\bibinfo {volume} {504}},\ \bibinfo {pages} {157} (\bibinfo {year}
  {2001})}\BibitemShut {NoStop}%
\bibitem [{\citenamefont {Ji}\ and\ \citenamefont
  {Wen}(2020)}]{PhysRevResearch.2.033417}%
  \BibitemOpen
  \bibfield  {author} {\bibinfo {author} {\bibfnamefont {W.}~\bibnamefont
  {Ji}}\ and\ \bibinfo {author} {\bibfnamefont {X.-G.}\ \bibnamefont {Wen}},\
  }\href {\doibase 10.1103/PhysRevResearch.2.033417} {\bibfield  {journal}
  {\bibinfo  {journal} {Phys. Rev. Res.}\ }\textbf {\bibinfo {volume} {2}},\
  \bibinfo {pages} {033417} (\bibinfo {year} {2020})}\BibitemShut {NoStop}%
\bibitem [{\citenamefont {Kong}\ \emph
  {et~al.}(2020{\natexlab{a}})\citenamefont {Kong}, \citenamefont {Lan},
  \citenamefont {Wen}, \citenamefont {Zhang},\ and\ \citenamefont
  {Zheng}}]{PhysRevResearch.2.043086}%
  \BibitemOpen
  \bibfield  {author} {\bibinfo {author} {\bibfnamefont {L.}~\bibnamefont
  {Kong}}, \bibinfo {author} {\bibfnamefont {T.}~\bibnamefont {Lan}}, \bibinfo
  {author} {\bibfnamefont {X.-G.}\ \bibnamefont {Wen}}, \bibinfo {author}
  {\bibfnamefont {Z.-H.}\ \bibnamefont {Zhang}}, \ and\ \bibinfo {author}
  {\bibfnamefont {H.}~\bibnamefont {Zheng}},\ }\href {\doibase
  10.1103/PhysRevResearch.2.043086} {\bibfield  {journal} {\bibinfo  {journal}
  {Phys. Rev. Res.}\ }\textbf {\bibinfo {volume} {2}},\ \bibinfo {pages}
  {043086} (\bibinfo {year} {2020}{\natexlab{a}})}\BibitemShut {NoStop}%
\bibitem [{\citenamefont {Kong}\ \emph
  {et~al.}(2020{\natexlab{b}})\citenamefont {Kong}, \citenamefont {Lan},
  \citenamefont {Wen}, \citenamefont {Zhang},\ and\ \citenamefont
  {Zheng}}]{Kong:2020jne}%
  \BibitemOpen
  \bibfield  {author} {\bibinfo {author} {\bibfnamefont {L.}~\bibnamefont
  {Kong}}, \bibinfo {author} {\bibfnamefont {T.}~\bibnamefont {Lan}}, \bibinfo
  {author} {\bibfnamefont {X.-G.}\ \bibnamefont {Wen}}, \bibinfo {author}
  {\bibfnamefont {Z.-H.}\ \bibnamefont {Zhang}}, \ and\ \bibinfo {author}
  {\bibfnamefont {H.}~\bibnamefont {Zheng}},\ }\href {\doibase
  10.1007/JHEP09(2020)093} {\bibfield  {journal} {\bibinfo  {journal} {JHEP}\
  }\textbf {\bibinfo {volume} {09}},\ \bibinfo {pages} {093} (\bibinfo {year}
  {2020}{\natexlab{b}})},\ \Eprint {http://arxiv.org/abs/2003.08898}
  {arXiv:2003.08898 [math-ph]} \BibitemShut {NoStop}%
\bibitem [{\citenamefont {Freed}\ \emph {et~al.}(2022)\citenamefont {Freed},
  \citenamefont {Moore},\ and\ \citenamefont {Teleman}}]{Freed:2022qnc}%
  \BibitemOpen
  \bibfield  {author} {\bibinfo {author} {\bibfnamefont {D.~S.}\ \bibnamefont
  {Freed}}, \bibinfo {author} {\bibfnamefont {G.~W.}\ \bibnamefont {Moore}}, \
  and\ \bibinfo {author} {\bibfnamefont {C.}~\bibnamefont {Teleman}},\
  }\href@noop {} {\  (\bibinfo {year} {2022})},\ \Eprint
  {http://arxiv.org/abs/2209.07471} {arXiv:2209.07471 [hep-th]} \BibitemShut
  {NoStop}%
\bibitem [{\citenamefont {Chatterjee}\ and\ \citenamefont
  {Wen}(2023{\natexlab{a}})}]{PhysRevB.107.155136}%
  \BibitemOpen
  \bibfield  {author} {\bibinfo {author} {\bibfnamefont {A.}~\bibnamefont
  {Chatterjee}}\ and\ \bibinfo {author} {\bibfnamefont {X.-G.}\ \bibnamefont
  {Wen}},\ }\href {\doibase 10.1103/PhysRevB.107.155136} {\bibfield  {journal}
  {\bibinfo  {journal} {Phys. Rev. B}\ }\textbf {\bibinfo {volume} {107}},\
  \bibinfo {pages} {155136} (\bibinfo {year} {2023}{\natexlab{a}})}\BibitemShut
  {NoStop}%
\bibitem [{\citenamefont {Kong}\ and\ \citenamefont
  {Zheng}(2019)}]{Kong2019AMT}%
  \BibitemOpen
  \bibfield  {author} {\bibinfo {author} {\bibfnamefont {L.}~\bibnamefont
  {Kong}}\ and\ \bibinfo {author} {\bibfnamefont {H.}~\bibnamefont {Zheng}},\
  }\href {https://api.semanticscholar.org/CorpusID:152282355} {\bibfield
  {journal} {\bibinfo  {journal} {Journal of High Energy Physics}\ }\textbf
  {\bibinfo {volume} {2020}},\ \bibinfo {pages} {1} (\bibinfo {year}
  {2019})}\BibitemShut {NoStop}%
\bibitem [{\citenamefont {You}\ \emph {et~al.}(2014)\citenamefont {You},
  \citenamefont {Bi}, \citenamefont {Rasmussen}, \citenamefont {Slagle},\ and\
  \citenamefont {Xu}}]{youWaveFunctionStrange2014}%
  \BibitemOpen
  \bibfield  {author} {\bibinfo {author} {\bibfnamefont {Y.-Z.}\ \bibnamefont
  {You}}, \bibinfo {author} {\bibfnamefont {Z.}~\bibnamefont {Bi}}, \bibinfo
  {author} {\bibfnamefont {A.}~\bibnamefont {Rasmussen}}, \bibinfo {author}
  {\bibfnamefont {K.}~\bibnamefont {Slagle}}, \ and\ \bibinfo {author}
  {\bibfnamefont {C.}~\bibnamefont {Xu}},\ }\href {\doibase
  10.1103/PhysRevLett.112.247202} {\bibfield  {journal} {\bibinfo  {journal}
  {Physical Review Letters}\ }\textbf {\bibinfo {volume} {112}},\ \bibinfo
  {pages} {247202} (\bibinfo {year} {2014})}\BibitemShut {NoStop}%
\bibitem [{\citenamefont {Vanhove}\ \emph {et~al.}(2018)\citenamefont
  {Vanhove}, \citenamefont {Bal}, \citenamefont {Williamson}, \citenamefont
  {Bultinck}, \citenamefont {Haegeman},\ and\ \citenamefont
  {Verstraete}}]{vanhoveMappingTopologicalConformal2018}%
  \BibitemOpen
  \bibfield  {author} {\bibinfo {author} {\bibfnamefont {R.}~\bibnamefont
  {Vanhove}}, \bibinfo {author} {\bibfnamefont {M.}~\bibnamefont {Bal}},
  \bibinfo {author} {\bibfnamefont {D.~J.}\ \bibnamefont {Williamson}},
  \bibinfo {author} {\bibfnamefont {N.}~\bibnamefont {Bultinck}}, \bibinfo
  {author} {\bibfnamefont {J.}~\bibnamefont {Haegeman}}, \ and\ \bibinfo
  {author} {\bibfnamefont {F.}~\bibnamefont {Verstraete}},\ }\href {\doibase
  10.1103/PhysRevLett.121.177203} {\bibfield  {journal} {\bibinfo  {journal}
  {Physical Review Letters}\ }\textbf {\bibinfo {volume} {121}},\ \bibinfo
  {pages} {177203} (\bibinfo {year} {2018})}\BibitemShut {NoStop}%
\bibitem [{\citenamefont {Aasen}\ \emph {et~al.}(2020)\citenamefont {Aasen},
  \citenamefont {Fendley},\ and\ \citenamefont {Mong}}]{Aasen:2020jwb}%
  \BibitemOpen
  \bibfield  {author} {\bibinfo {author} {\bibfnamefont {D.}~\bibnamefont
  {Aasen}}, \bibinfo {author} {\bibfnamefont {P.}~\bibnamefont {Fendley}}, \
  and\ \bibinfo {author} {\bibfnamefont {R.~S.~K.}\ \bibnamefont {Mong}},\
  }\href@noop {} {\  (\bibinfo {year} {2020})},\ \Eprint
  {http://arxiv.org/abs/2008.08598} {arXiv:2008.08598 [cond-mat.stat-mech]}
  \BibitemShut {NoStop}%
\bibitem [{\citenamefont {Aasen}\ \emph {et~al.}(2016)\citenamefont {Aasen},
  \citenamefont {Mong},\ and\ \citenamefont
  {Fendley}}]{aasenTopologicalDefectsLattice2016}%
  \BibitemOpen
  \bibfield  {author} {\bibinfo {author} {\bibfnamefont {D.}~\bibnamefont
  {Aasen}}, \bibinfo {author} {\bibfnamefont {R.~S.~K.}\ \bibnamefont {Mong}},
  \ and\ \bibinfo {author} {\bibfnamefont {P.}~\bibnamefont {Fendley}},\ }\href
  {\doibase 10.1088/1751-8113/49/35/354001} {\bibfield  {journal} {\bibinfo
  {journal} {Journal of Physics A: Mathematical and Theoretical}\ }\textbf
  {\bibinfo {volume} {49}},\ \bibinfo {pages} {354001} (\bibinfo {year}
  {2016})},\ \Eprint {http://arxiv.org/abs/1601.07185} {arXiv:1601.07185
  [cond-mat]} \BibitemShut {NoStop}%
\bibitem [{\citenamefont {Vanhove}\ \emph {et~al.}(2022)\citenamefont
  {Vanhove}, \citenamefont {Lootens}, \citenamefont {Tu},\ and\ \citenamefont
  {Verstraete}}]{Vanhove_2022}%
  \BibitemOpen
  \bibfield  {author} {\bibinfo {author} {\bibfnamefont {R.}~\bibnamefont
  {Vanhove}}, \bibinfo {author} {\bibfnamefont {L.}~\bibnamefont {Lootens}},
  \bibinfo {author} {\bibfnamefont {H.-H.}\ \bibnamefont {Tu}}, \ and\ \bibinfo
  {author} {\bibfnamefont {F.}~\bibnamefont {Verstraete}},\ }\href {\doibase
  10.1088/1751-8121/ac68b1} {\bibfield  {journal} {\bibinfo  {journal} {Journal
  of Physics A: Mathematical and Theoretical}\ }\textbf {\bibinfo {volume}
  {55}},\ \bibinfo {pages} {235002} (\bibinfo {year} {2022})}\BibitemShut
  {NoStop}%
\bibitem [{\citenamefont {Hung}\ \emph {et~al.}(2025)\citenamefont {Hung},
  \citenamefont {Ji}, \citenamefont {Shen}, \citenamefont {Wan},\ and\
  \citenamefont {Zhao}}]{hung_2d-cft_2025}%
  \BibitemOpen
  \bibfield  {author} {\bibinfo {author} {\bibfnamefont {L.-Y.}\ \bibnamefont
  {Hung}}, \bibinfo {author} {\bibfnamefont {K.}~\bibnamefont {Ji}}, \bibinfo
  {author} {\bibfnamefont {C.}~\bibnamefont {Shen}}, \bibinfo {author}
  {\bibfnamefont {Y.}~\bibnamefont {Wan}}, \ and\ \bibinfo {author}
  {\bibfnamefont {Y.}~\bibnamefont {Zhao}},\ }\href {\doibase
  10.48550/arXiv.2506.05324} {\enquote {\bibinfo {title} {A {2D}-{CFT}
  {Factory}: {Critical} {Lattice} {Models} from {Competing} {Anyon}
  {Condensation} {Processes} in {SymTO}/{SymTFT}},}\ } (\bibinfo {year}
  {2025}),\ \bibinfo {note} {arXiv:2506.05324 [cond-mat]}\BibitemShut {NoStop}%
\bibitem [{\citenamefont {Chatterjee}\ and\ \citenamefont
  {Wen}(2023{\natexlab{b}})}]{PhysRevB.108.075105}%
  \BibitemOpen
  \bibfield  {author} {\bibinfo {author} {\bibfnamefont {A.}~\bibnamefont
  {Chatterjee}}\ and\ \bibinfo {author} {\bibfnamefont {X.-G.}\ \bibnamefont
  {Wen}},\ }\href {\doibase 10.1103/PhysRevB.108.075105} {\bibfield  {journal}
  {\bibinfo  {journal} {Phys. Rev. B}\ }\textbf {\bibinfo {volume} {108}},\
  \bibinfo {pages} {075105} (\bibinfo {year} {2023}{\natexlab{b}})}\BibitemShut
  {NoStop}%
\bibitem [{\citenamefont {Ruelle}\ and\ \citenamefont
  {Verhoeven}(1998)}]{RUELLE1998650}%
  \BibitemOpen
  \bibfield  {author} {\bibinfo {author} {\bibfnamefont {P.}~\bibnamefont
  {Ruelle}}\ and\ \bibinfo {author} {\bibfnamefont {O.}~\bibnamefont
  {Verhoeven}},\ }\href {\doibase
  https://doi.org/10.1016/S0550-3213(98)00639-7} {\bibfield  {journal}
  {\bibinfo  {journal} {Nuclear Physics B}\ }\textbf {\bibinfo {volume}
  {535}},\ \bibinfo {pages} {650} (\bibinfo {year} {1998})}\BibitemShut
  {NoStop}%
\bibitem [{\citenamefont {Lin}\ and\ \citenamefont
  {Shao}(2023)}]{PhysRevD.107.125025}%
  \BibitemOpen
  \bibfield  {author} {\bibinfo {author} {\bibfnamefont {Y.-H.}\ \bibnamefont
  {Lin}}\ and\ \bibinfo {author} {\bibfnamefont {S.-H.}\ \bibnamefont {Shao}},\
  }\href {\doibase 10.1103/PhysRevD.107.125025} {\bibfield  {journal} {\bibinfo
   {journal} {Phys. Rev. D}\ }\textbf {\bibinfo {volume} {107}},\ \bibinfo
  {pages} {125025} (\bibinfo {year} {2023})}\BibitemShut {NoStop}%
\bibitem [{\citenamefont {Lin}\ and\ \citenamefont
  {Shao}(2021)}]{PhysRevD.103.125001}%
  \BibitemOpen
  \bibfield  {author} {\bibinfo {author} {\bibfnamefont {Y.-H.}\ \bibnamefont
  {Lin}}\ and\ \bibinfo {author} {\bibfnamefont {S.-H.}\ \bibnamefont {Shao}},\
  }\href {\doibase 10.1103/PhysRevD.103.125001} {\bibfield  {journal} {\bibinfo
   {journal} {Phys. Rev. D}\ }\textbf {\bibinfo {volume} {103}},\ \bibinfo
  {pages} {125001} (\bibinfo {year} {2021})}\BibitemShut {NoStop}%
\bibitem [{\citenamefont {Ji}\ and\ \citenamefont
  {Wen}(2019)}]{PhysRevResearch.1.033054}%
  \BibitemOpen
  \bibfield  {author} {\bibinfo {author} {\bibfnamefont {W.}~\bibnamefont
  {Ji}}\ and\ \bibinfo {author} {\bibfnamefont {X.-G.}\ \bibnamefont {Wen}},\
  }\href {\doibase 10.1103/PhysRevResearch.1.033054} {\bibfield  {journal}
  {\bibinfo  {journal} {Phys. Rev. Res.}\ }\textbf {\bibinfo {volume} {1}},\
  \bibinfo {pages} {033054} (\bibinfo {year} {2019})}\BibitemShut {NoStop}%
\bibitem [{\citenamefont {Chen}\ \emph {et~al.}(2020)\citenamefont {Chen},
  \citenamefont {Jian}, \citenamefont {Kong}, \citenamefont {You},\ and\
  \citenamefont {Zheng}}]{PhysRevB.102.045139}%
  \BibitemOpen
  \bibfield  {author} {\bibinfo {author} {\bibfnamefont {W.-Q.}\ \bibnamefont
  {Chen}}, \bibinfo {author} {\bibfnamefont {C.-M.}\ \bibnamefont {Jian}},
  \bibinfo {author} {\bibfnamefont {L.}~\bibnamefont {Kong}}, \bibinfo {author}
  {\bibfnamefont {Y.-Z.}\ \bibnamefont {You}}, \ and\ \bibinfo {author}
  {\bibfnamefont {H.}~\bibnamefont {Zheng}},\ }\href {\doibase
  10.1103/PhysRevB.102.045139} {\bibfield  {journal} {\bibinfo  {journal}
  {Phys. Rev. B}\ }\textbf {\bibinfo {volume} {102}},\ \bibinfo {pages}
  {045139} (\bibinfo {year} {2020})}\BibitemShut {NoStop}%
\bibitem [{\citenamefont {Chatterjee}\ \emph {et~al.}(2025)\citenamefont
  {Chatterjee}, \citenamefont {Ji},\ and\ \citenamefont {Wen}}]{jzfv-ygmr}%
  \BibitemOpen
  \bibfield  {author} {\bibinfo {author} {\bibfnamefont {A.}~\bibnamefont
  {Chatterjee}}, \bibinfo {author} {\bibfnamefont {W.}~\bibnamefont {Ji}}, \
  and\ \bibinfo {author} {\bibfnamefont {X.-G.}\ \bibnamefont {Wen}},\ }\href
  {\doibase 10.1103/jzfv-ygmr} {\bibfield  {journal} {\bibinfo  {journal}
  {Phys. Rev. B}\ }\textbf {\bibinfo {volume} {112}},\ \bibinfo {pages}
  {115142} (\bibinfo {year} {2025})}\BibitemShut {NoStop}%
\bibitem [{\citenamefont {Lin}\ \emph {et~al.}(2023)\citenamefont {Lin},
  \citenamefont {Okada}, \citenamefont {Seifnashri},\ and\ \citenamefont
  {Tachikawa}}]{Lin:2022dhv}%
  \BibitemOpen
  \bibfield  {author} {\bibinfo {author} {\bibfnamefont {Y.-H.}\ \bibnamefont
  {Lin}}, \bibinfo {author} {\bibfnamefont {M.}~\bibnamefont {Okada}}, \bibinfo
  {author} {\bibfnamefont {S.}~\bibnamefont {Seifnashri}}, \ and\ \bibinfo
  {author} {\bibfnamefont {Y.}~\bibnamefont {Tachikawa}},\ }\href {\doibase
  10.1007/JHEP03(2023)094} {\bibfield  {journal} {\bibinfo  {journal} {JHEP}\
  }\textbf {\bibinfo {volume} {03}},\ \bibinfo {pages} {094} (\bibinfo {year}
  {2023})},\ \Eprint {http://arxiv.org/abs/2208.05495} {arXiv:2208.05495
  [hep-th]} \BibitemShut {NoStop}%
\bibitem [{\citenamefont {Nakayama}\ and\ \citenamefont
  {Onagi}(2025)}]{Nakayama:2025mrm}%
  \BibitemOpen
  \bibfield  {author} {\bibinfo {author} {\bibfnamefont {Y.}~\bibnamefont
  {Nakayama}}\ and\ \bibinfo {author} {\bibfnamefont {T.}~\bibnamefont
  {Onagi}},\ }\href@noop {} {\  (\bibinfo {year} {2025})},\ \Eprint
  {http://arxiv.org/abs/2511.00386} {arXiv:2511.00386 [hep-th]} \BibitemShut
  {NoStop}%
\bibitem [{\citenamefont {Vidal}(2007)}]{PhysRevLett.99.220405}%
  \BibitemOpen
  \bibfield  {author} {\bibinfo {author} {\bibfnamefont {G.}~\bibnamefont
  {Vidal}},\ }\href {\doibase 10.1103/PhysRevLett.99.220405} {\bibfield
  {journal} {\bibinfo  {journal} {Phys. Rev. Lett.}\ }\textbf {\bibinfo
  {volume} {99}},\ \bibinfo {pages} {220405} (\bibinfo {year}
  {2007})}\BibitemShut {NoStop}%
\bibitem [{\citenamefont {Gu}\ \emph {et~al.}(2008)\citenamefont {Gu},
  \citenamefont {Levin},\ and\ \citenamefont {Wen}}]{PhysRevB.78.205116}%
  \BibitemOpen
  \bibfield  {author} {\bibinfo {author} {\bibfnamefont {Z.-C.}\ \bibnamefont
  {Gu}}, \bibinfo {author} {\bibfnamefont {M.}~\bibnamefont {Levin}}, \ and\
  \bibinfo {author} {\bibfnamefont {X.-G.}\ \bibnamefont {Wen}},\ }\href
  {\doibase 10.1103/PhysRevB.78.205116} {\bibfield  {journal} {\bibinfo
  {journal} {Phys. Rev. B}\ }\textbf {\bibinfo {volume} {78}},\ \bibinfo
  {pages} {205116} (\bibinfo {year} {2008})}\BibitemShut {NoStop}%
\bibitem [{\citenamefont {Evenbly}\ and\ \citenamefont
  {Vidal}(2009{\natexlab{a}})}]{PhysRevB.79.144108}%
  \BibitemOpen
  \bibfield  {author} {\bibinfo {author} {\bibfnamefont {G.}~\bibnamefont
  {Evenbly}}\ and\ \bibinfo {author} {\bibfnamefont {G.}~\bibnamefont
  {Vidal}},\ }\href {\doibase 10.1103/PhysRevB.79.144108} {\bibfield  {journal}
  {\bibinfo  {journal} {Phys. Rev. B}\ }\textbf {\bibinfo {volume} {79}},\
  \bibinfo {pages} {144108} (\bibinfo {year} {2009}{\natexlab{a}})}\BibitemShut
  {NoStop}%
\bibitem [{\citenamefont {Pfeifer}\ \emph {et~al.}(2009)\citenamefont
  {Pfeifer}, \citenamefont {Evenbly},\ and\ \citenamefont
  {Vidal}}]{PhysRevA.79.040301}%
  \BibitemOpen
  \bibfield  {author} {\bibinfo {author} {\bibfnamefont {R.~N.~C.}\
  \bibnamefont {Pfeifer}}, \bibinfo {author} {\bibfnamefont {G.}~\bibnamefont
  {Evenbly}}, \ and\ \bibinfo {author} {\bibfnamefont {G.}~\bibnamefont
  {Vidal}},\ }\href {\doibase 10.1103/PhysRevA.79.040301} {\bibfield  {journal}
  {\bibinfo  {journal} {Phys. Rev. A}\ }\textbf {\bibinfo {volume} {79}},\
  \bibinfo {pages} {040301} (\bibinfo {year} {2009})}\BibitemShut {NoStop}%
\bibitem [{\citenamefont {Gu}\ and\ \citenamefont
  {Wen}(2009)}]{guTensorentanglementfilteringRenormalizationApproach2009}%
  \BibitemOpen
  \bibfield  {author} {\bibinfo {author} {\bibfnamefont {Z.-C.}\ \bibnamefont
  {Gu}}\ and\ \bibinfo {author} {\bibfnamefont {X.-G.}\ \bibnamefont {Wen}},\
  }\href {\doibase 10.1103/PhysRevB.80.155131} {\bibfield  {journal} {\bibinfo
  {journal} {Physical Review B}\ }\textbf {\bibinfo {volume} {80}},\ \bibinfo
  {pages} {155131} (\bibinfo {year} {2009})}\BibitemShut {NoStop}%
\bibitem [{\citenamefont {Evenbly}\ and\ \citenamefont
  {Vidal}(2009{\natexlab{b}})}]{PhysRevLett.102.180406}%
  \BibitemOpen
  \bibfield  {author} {\bibinfo {author} {\bibfnamefont {G.}~\bibnamefont
  {Evenbly}}\ and\ \bibinfo {author} {\bibfnamefont {G.}~\bibnamefont
  {Vidal}},\ }\href {\doibase 10.1103/PhysRevLett.102.180406} {\bibfield
  {journal} {\bibinfo  {journal} {Phys. Rev. Lett.}\ }\textbf {\bibinfo
  {volume} {102}},\ \bibinfo {pages} {180406} (\bibinfo {year}
  {2009}{\natexlab{b}})}\BibitemShut {NoStop}%
\bibitem [{\citenamefont {Levin}\ and\ \citenamefont
  {Nave}(2007)}]{levinTensorRenormalizationGroup2007}%
  \BibitemOpen
  \bibfield  {author} {\bibinfo {author} {\bibfnamefont {M.}~\bibnamefont
  {Levin}}\ and\ \bibinfo {author} {\bibfnamefont {C.~P.}\ \bibnamefont
  {Nave}},\ }\href {\doibase 10.1103/PhysRevLett.99.120601} {\bibfield
  {journal} {\bibinfo  {journal} {Physical Review Letters}\ }\textbf {\bibinfo
  {volume} {99}},\ \bibinfo {pages} {120601} (\bibinfo {year}
  {2007})}\BibitemShut {NoStop}%
\bibitem [{\citenamefont {Xie}\ \emph {et~al.}(2009)\citenamefont {Xie},
  \citenamefont {Jiang}, \citenamefont {Chen}, \citenamefont {Weng},\ and\
  \citenamefont {Xiang}}]{xieSecondRenormalizationTensorNetwork2009}%
  \BibitemOpen
  \bibfield  {author} {\bibinfo {author} {\bibfnamefont {Z.~Y.}\ \bibnamefont
  {Xie}}, \bibinfo {author} {\bibfnamefont {H.~C.}\ \bibnamefont {Jiang}},
  \bibinfo {author} {\bibfnamefont {Q.~N.}\ \bibnamefont {Chen}}, \bibinfo
  {author} {\bibfnamefont {Z.~Y.}\ \bibnamefont {Weng}}, \ and\ \bibinfo
  {author} {\bibfnamefont {T.}~\bibnamefont {Xiang}},\ }\href {\doibase
  10.1103/PhysRevLett.103.160601} {\bibfield  {journal} {\bibinfo  {journal}
  {Physical Review Letters}\ }\textbf {\bibinfo {volume} {103}},\ \bibinfo
  {pages} {160601} (\bibinfo {year} {2009})}\BibitemShut {NoStop}%
\bibitem [{\citenamefont {Xie}\ \emph {et~al.}(2012)\citenamefont {Xie},
  \citenamefont {Chen}, \citenamefont {Qin}, \citenamefont {Zhu}, \citenamefont
  {Yang},\ and\ \citenamefont
  {Xiang}}]{xieCoarsegrainingRenormalizationHigherorder2012}%
  \BibitemOpen
  \bibfield  {author} {\bibinfo {author} {\bibfnamefont {Z.~Y.}\ \bibnamefont
  {Xie}}, \bibinfo {author} {\bibfnamefont {J.}~\bibnamefont {Chen}}, \bibinfo
  {author} {\bibfnamefont {M.~P.}\ \bibnamefont {Qin}}, \bibinfo {author}
  {\bibfnamefont {J.~W.}\ \bibnamefont {Zhu}}, \bibinfo {author} {\bibfnamefont
  {L.~P.}\ \bibnamefont {Yang}}, \ and\ \bibinfo {author} {\bibfnamefont
  {T.}~\bibnamefont {Xiang}},\ }\href {\doibase 10.1103/PhysRevB.86.045139}
  {\bibfield  {journal} {\bibinfo  {journal} {Physical Review B}\ }\textbf
  {\bibinfo {volume} {86}},\ \bibinfo {pages} {045139} (\bibinfo {year}
  {2012})}\BibitemShut {NoStop}%
\bibitem [{\citenamefont {Evenbly}\ and\ \citenamefont
  {Vidal}(2015)}]{evenblyTensorNetworkRenormalization2015}%
  \BibitemOpen
  \bibfield  {author} {\bibinfo {author} {\bibfnamefont {G.}~\bibnamefont
  {Evenbly}}\ and\ \bibinfo {author} {\bibfnamefont {G.}~\bibnamefont
  {Vidal}},\ }\href {\doibase 10.1103/PhysRevLett.115.180405} {\bibfield
  {journal} {\bibinfo  {journal} {Physical Review Letters}\ }\textbf {\bibinfo
  {volume} {115}},\ \bibinfo {pages} {180405} (\bibinfo {year}
  {2015})}\BibitemShut {NoStop}%
\bibitem [{\citenamefont {Yang}\ \emph {et~al.}(2017)\citenamefont {Yang},
  \citenamefont {Gu},\ and\ \citenamefont
  {Wen}}]{yangLoopOptimizationTensor2017}%
  \BibitemOpen
  \bibfield  {author} {\bibinfo {author} {\bibfnamefont {S.}~\bibnamefont
  {Yang}}, \bibinfo {author} {\bibfnamefont {Z.-C.}\ \bibnamefont {Gu}}, \ and\
  \bibinfo {author} {\bibfnamefont {X.-G.}\ \bibnamefont {Wen}},\ }\href
  {\doibase 10.1103/PhysRevLett.118.110504} {\bibfield  {journal} {\bibinfo
  {journal} {Physical Review Letters}\ }\textbf {\bibinfo {volume} {118}},\
  \bibinfo {pages} {110504} (\bibinfo {year} {2017})}\BibitemShut {NoStop}%
\bibitem [{\citenamefont
  {Evenbly}(2017)}]{evenblyAlgorithmsTensorNetwork2017a}%
  \BibitemOpen
  \bibfield  {author} {\bibinfo {author} {\bibfnamefont {G.}~\bibnamefont
  {Evenbly}},\ }\href {\doibase 10.1103/PhysRevB.95.045117} {\bibfield
  {journal} {\bibinfo  {journal} {Physical Review B}\ }\textbf {\bibinfo
  {volume} {95}} (\bibinfo {year} {2017}),\
  10.1103/PhysRevB.95.045117}\BibitemShut {NoStop}%
\bibitem [{\citenamefont {Adachi}\ \emph {et~al.}(2022)\citenamefont {Adachi},
  \citenamefont {Okubo},\ and\ \citenamefont {Todo}}]{PhysRevB.105.L060402}%
  \BibitemOpen
  \bibfield  {author} {\bibinfo {author} {\bibfnamefont {D.}~\bibnamefont
  {Adachi}}, \bibinfo {author} {\bibfnamefont {T.}~\bibnamefont {Okubo}}, \
  and\ \bibinfo {author} {\bibfnamefont {S.}~\bibnamefont {Todo}},\ }\href
  {\doibase 10.1103/PhysRevB.105.L060402} {\bibfield  {journal} {\bibinfo
  {journal} {Phys. Rev. B}\ }\textbf {\bibinfo {volume} {105}},\ \bibinfo
  {pages} {L060402} (\bibinfo {year} {2022})}\BibitemShut {NoStop}%
\bibitem [{\citenamefont {Homma}\ \emph {et~al.}(2024)\citenamefont {Homma},
  \citenamefont {Okubo},\ and\ \citenamefont
  {Kawashima}}]{PhysRevResearch.6.043102}%
  \BibitemOpen
  \bibfield  {author} {\bibinfo {author} {\bibfnamefont {K.}~\bibnamefont
  {Homma}}, \bibinfo {author} {\bibfnamefont {T.}~\bibnamefont {Okubo}}, \ and\
  \bibinfo {author} {\bibfnamefont {N.}~\bibnamefont {Kawashima}},\ }\href
  {\doibase 10.1103/PhysRevResearch.6.043102} {\bibfield  {journal} {\bibinfo
  {journal} {Phys. Rev. Res.}\ }\textbf {\bibinfo {volume} {6}},\ \bibinfo
  {pages} {043102} (\bibinfo {year} {2024})}\BibitemShut {NoStop}%
\bibitem [{\citenamefont {Cheng}\ \emph {et~al.}(2025)\citenamefont {Cheng},
  \citenamefont {Chen}, \citenamefont {Gu},\ and\ \citenamefont
  {Hung}}]{chengPrecisionReconstructionRational2025}%
  \BibitemOpen
  \bibfield  {author} {\bibinfo {author} {\bibfnamefont {G.}~\bibnamefont
  {Cheng}}, \bibinfo {author} {\bibfnamefont {L.}~\bibnamefont {Chen}},
  \bibinfo {author} {\bibfnamefont {Z.-C.}\ \bibnamefont {Gu}}, \ and\ \bibinfo
  {author} {\bibfnamefont {L.-Y.}\ \bibnamefont {Hung}},\ }\href {\doibase
  10.1103/PhysRevX.15.011073} {\bibfield  {journal} {\bibinfo  {journal}
  {Physical Review X}\ }\textbf {\bibinfo {volume} {15}},\ \bibinfo {pages}
  {011073} (\bibinfo {year} {2025})}\BibitemShut {NoStop}%
\bibitem [{\citenamefont {Lewellen}(1992)}]{LEWELLEN1992654}%
  \BibitemOpen
  \bibfield  {author} {\bibinfo {author} {\bibfnamefont {D.~C.}\ \bibnamefont
  {Lewellen}},\ }\href {\doibase https://doi.org/10.1016/0550-3213(92)90370-Q}
  {\bibfield  {journal} {\bibinfo  {journal} {Nuclear Physics B}\ }\textbf
  {\bibinfo {volume} {372}},\ \bibinfo {pages} {654} (\bibinfo {year}
  {1992})}\BibitemShut {NoStop}%
\bibitem [{\citenamefont {Cardy}(1989)}]{CARDY1989581}%
  \BibitemOpen
  \bibfield  {author} {\bibinfo {author} {\bibfnamefont {J.~L.}\ \bibnamefont
  {Cardy}},\ }\href {\doibase https://doi.org/10.1016/0550-3213(89)90521-X}
  {\bibfield  {journal} {\bibinfo  {journal} {Nuclear Physics B}\ }\textbf
  {\bibinfo {volume} {324}},\ \bibinfo {pages} {581} (\bibinfo {year}
  {1989})}\BibitemShut {NoStop}%
\bibitem [{\citenamefont {Runkel}(2000)}]{Runkel:1999dz}%
  \BibitemOpen
  \bibfield  {author} {\bibinfo {author} {\bibfnamefont {I.}~\bibnamefont
  {Runkel}},\ }\href {\doibase 10.1016/S0550-3213(99)00707-5} {\bibfield
  {journal} {\bibinfo  {journal} {Nucl. Phys. B}\ }\textbf {\bibinfo {volume}
  {579}},\ \bibinfo {pages} {561} (\bibinfo {year} {2000})},\ \Eprint
  {http://arxiv.org/abs/hep-th/9908046} {arXiv:hep-th/9908046} \BibitemShut
  {NoStop}%
\bibitem [{\citenamefont {Dotsenko}\ and\ \citenamefont
  {Fateev}(1985)}]{DOTSENKO1985291}%
  \BibitemOpen
  \bibfield  {author} {\bibinfo {author} {\bibfnamefont {V.}~\bibnamefont
  {Dotsenko}}\ and\ \bibinfo {author} {\bibfnamefont {V.}~\bibnamefont
  {Fateev}},\ }\href {\doibase https://doi.org/10.1016/0370-2693(85)90366-1}
  {\bibfield  {journal} {\bibinfo  {journal} {Physics Letters B}\ }\textbf
  {\bibinfo {volume} {154}},\ \bibinfo {pages} {291} (\bibinfo {year}
  {1985})}\BibitemShut {NoStop}%
\bibitem [{\citenamefont {Li}\ and\ \citenamefont {Yu}(1990)}]{Li:1989hs}%
  \BibitemOpen
  \bibfield  {author} {\bibinfo {author} {\bibfnamefont {M.}~\bibnamefont
  {Li}}\ and\ \bibinfo {author} {\bibfnamefont {M.}~\bibnamefont {Yu}},\ }\href
  {\doibase 10.1007/BF02096502} {\bibfield  {journal} {\bibinfo  {journal}
  {Commun. Math. Phys.}\ }\textbf {\bibinfo {volume} {127}},\ \bibinfo {pages}
  {195} (\bibinfo {year} {1990})}\BibitemShut {NoStop}%
\bibitem [{\citenamefont {Rowell}\ \emph {et~al.}(2009)\citenamefont {Rowell},
  \citenamefont {Stong},\ and\ \citenamefont {Wang}}]{Rowell:2007dge}%
  \BibitemOpen
  \bibfield  {author} {\bibinfo {author} {\bibfnamefont {E.}~\bibnamefont
  {Rowell}}, \bibinfo {author} {\bibfnamefont {R.}~\bibnamefont {Stong}}, \
  and\ \bibinfo {author} {\bibfnamefont {Z.}~\bibnamefont {Wang}},\ }\href
  {\doibase 10.1007/s00220-009-0908-z} {\bibfield  {journal} {\bibinfo
  {journal} {Commun. Math. Phys.}\ }\textbf {\bibinfo {volume} {292}},\
  \bibinfo {pages} {343} (\bibinfo {year} {2009})},\ \Eprint
  {http://arxiv.org/abs/0712.1377} {arXiv:0712.1377 [math.QA]} \BibitemShut
  {NoStop}%
\bibitem [{\citenamefont {Wang}\ and\ \citenamefont
  {Verstraete}(2011)}]{wangClusterUpdateTensor2011}%
  \BibitemOpen
  \bibfield  {author} {\bibinfo {author} {\bibfnamefont {L.}~\bibnamefont
  {Wang}}\ and\ \bibinfo {author} {\bibfnamefont {F.}~\bibnamefont
  {Verstraete}},\ }\href {\doibase 10.48550/arXiv.1110.4362} {\enquote
  {\bibinfo {title} {Cluster update for tensor network states},}\ } (\bibinfo
  {year} {2011}),\ \Eprint {http://arxiv.org/abs/1110.4362} {arXiv:1110.4362
  [cond-mat]} \BibitemShut {NoStop}%
\bibitem [{\citenamefont
  {Cardy}(1986)}]{cardyOperatorContentTwodimensional1986a}%
  \BibitemOpen
  \bibfield  {author} {\bibinfo {author} {\bibfnamefont {J.~L.}\ \bibnamefont
  {Cardy}},\ }\href {\doibase 10.1016/0550-3213(86)90552-3} {\bibfield
  {journal} {\bibinfo  {journal} {Nuclear Physics B}\ }\textbf {\bibinfo
  {volume} {270}},\ \bibinfo {pages} {186} (\bibinfo {year}
  {1986})}\BibitemShut {NoStop}%
\bibitem [{\citenamefont {{Chenfeng
  Bao}}(2019)}]{chenfengbaoLoopOptimizationTensor2019}%
  \BibitemOpen
  \bibfield  {author} {\bibinfo {author} {\bibnamefont {{Chenfeng Bao}}},\
  }\emph {\bibinfo {title} {Loop {{Optimization}} of {{Tensor Network
  Renormalization}}: {{Algorithms}} and {{Applications}}}},\ \href@noop {}
  {Ph.D. thesis},\ \bibinfo  {school} {The University of Waterloo} (\bibinfo
  {year} {2019})\BibitemShut {NoStop}%
\bibitem [{\citenamefont {Di~Francesco}\ \emph {et~al.}(1997)\citenamefont
  {Di~Francesco}, \citenamefont {Mathieu},\ and\ \citenamefont
  {S{\'e}n{\'e}chal}}]{difrancescoConformalFieldTheory1997}%
  \BibitemOpen
  \bibfield  {author} {\bibinfo {author} {\bibfnamefont {P.}~\bibnamefont
  {Di~Francesco}}, \bibinfo {author} {\bibfnamefont {P.}~\bibnamefont
  {Mathieu}}, \ and\ \bibinfo {author} {\bibfnamefont {D.}~\bibnamefont
  {S{\'e}n{\'e}chal}},\ }\href@noop {} {\emph {\bibinfo {title} {Conformal
  Field Theory}}},\ Graduate Texts in Contemporary Physics\ (\bibinfo
  {publisher} {Springer},\ \bibinfo {address} {New York},\ \bibinfo {year}
  {1997})\BibitemShut {NoStop}%
\bibitem [{\citenamefont {Hauru}\ \emph {et~al.}(2016)\citenamefont {Hauru},
  \citenamefont {Evenbly}, \citenamefont {Ho}, \citenamefont {Gaiotto},\ and\
  \citenamefont {Vidal}}]{hauruTopologicalConformalDefects2016}%
  \BibitemOpen
  \bibfield  {author} {\bibinfo {author} {\bibfnamefont {M.}~\bibnamefont
  {Hauru}}, \bibinfo {author} {\bibfnamefont {G.}~\bibnamefont {Evenbly}},
  \bibinfo {author} {\bibfnamefont {W.~W.}\ \bibnamefont {Ho}}, \bibinfo
  {author} {\bibfnamefont {D.}~\bibnamefont {Gaiotto}}, \ and\ \bibinfo
  {author} {\bibfnamefont {G.}~\bibnamefont {Vidal}},\ }\href {\doibase
  10.1103/PhysRevB.94.115125} {\bibfield  {journal} {\bibinfo  {journal}
  {Physical Review B}\ }\textbf {\bibinfo {volume} {94}},\ \bibinfo {pages}
  {115125} (\bibinfo {year} {2016})}\BibitemShut {NoStop}%
\bibitem [{\citenamefont {Cardy}(1985)}]{cardyConformalInvarianceYangLee1985}%
  \BibitemOpen
  \bibfield  {author} {\bibinfo {author} {\bibfnamefont {J.~L.}\ \bibnamefont
  {Cardy}},\ }\href {\doibase 10.1103/PhysRevLett.54.1354} {\bibfield
  {journal} {\bibinfo  {journal} {Physical Review Letters}\ }\textbf {\bibinfo
  {volume} {54}},\ \bibinfo {pages} {1354} (\bibinfo {year}
  {1985})}\BibitemShut {NoStop}%
\bibitem [{\citenamefont {Furlan}\ \emph {et~al.}(1990)\citenamefont {Furlan},
  \citenamefont {Ganchev},\ and\ \citenamefont {Petkova}}]{Furlan:1989ra}%
  \BibitemOpen
  \bibfield  {author} {\bibinfo {author} {\bibfnamefont {P.}~\bibnamefont
  {Furlan}}, \bibinfo {author} {\bibfnamefont {A.~C.}\ \bibnamefont {Ganchev}},
  \ and\ \bibinfo {author} {\bibfnamefont {V.~B.}\ \bibnamefont {Petkova}},\
  }\href {\doibase 10.1142/S0217751X90001252} {\bibfield  {journal} {\bibinfo
  {journal} {Int. J. Mod. Phys. A}\ }\textbf {\bibinfo {volume} {5}},\ \bibinfo
  {pages} {2721} (\bibinfo {year} {1990})},\ \bibinfo {note} {[Erratum:
  Int.J.Mod.Phys.A 5, 3641 (1990)]}\BibitemShut {NoStop}%
\bibitem [{\citenamefont {Kirillov}\ and\ \citenamefont
  {Reshetikhin}(1989)}]{kirillovRepresentationsAlgebraUqsl21989}%
  \BibitemOpen
  \bibfield  {author} {\bibinfo {author} {\bibfnamefont {A.}~\bibnamefont
  {Kirillov}}\ and\ \bibinfo {author} {\bibfnamefont {N.}~\bibnamefont
  {Reshetikhin}},\ }\href@noop {} {\bibfield  {journal} {\bibinfo  {journal}
  {Infinite-dimensional Lie algebras and groups}\ ,\ \bibinfo {pages} {285}}
  (\bibinfo {year} {1989})}\BibitemShut {NoStop}%
\bibitem [{\citenamefont {Behrend}\ \emph {et~al.}(1998)\citenamefont
  {Behrend}, \citenamefont {Pearce},\ and\ \citenamefont
  {Zuber}}]{Behrend_1998}%
  \BibitemOpen
  \bibfield  {author} {\bibinfo {author} {\bibfnamefont {R.~E.}\ \bibnamefont
  {Behrend}}, \bibinfo {author} {\bibfnamefont {P.~A.}\ \bibnamefont {Pearce}},
  \ and\ \bibinfo {author} {\bibfnamefont {J.-B.}\ \bibnamefont {Zuber}},\
  }\href {\doibase 10.1088/0305-4470/31/50/001} {\bibfield  {journal} {\bibinfo
   {journal} {Journal of Physics A: Mathematical and General}\ }\textbf
  {\bibinfo {volume} {31}},\ \bibinfo {pages} {L763} (\bibinfo {year}
  {1998})}\BibitemShut {NoStop}%
\end{thebibliography}%

\clearpage
\appendix
\onecolumngrid
\section{\large Supplementary Material}
\vspace{2em}
\setcounter{equation}{0}
\setcounter{figure}{0}
\setcounter{table}{0}
\renewcommand{\theequation}{S\arabic{equation}}
\renewcommand{\thefigure}{S\arabic{figure}}
\renewcommand{\thetable}{S\arabic{table}}
\makeatletter
\renewcommand{\p@subfigure}{\thefigure}
\makeatother

\newcounter{supsec}      
\newcounter{supsubsec}   
\renewcommand{\thesupsec}{\Roman{supsec}}
\renewcommand{\thesupsubsec}{\Alph{supsubsec}}

\newcommand{\supsection}[1]{
  \refstepcounter{supsec}
  \setcounter{supsubsec}{0}
  \section*{\thesupsec.\quad #1}
}

\newcommand{\supsubsection}[1]{
  \refstepcounter{supsubsec}
  \subsection*{\thesupsubsec.\quad #1}
}

\supsection{Review of the fixed-point tensor construction}
\label{sup:fptensor}
Consider a triangulation of a 2D manifold. The CFT path integral is performed on each triangle and the resulting contributions are then glued together via tensor contractions. To regulate each triangle, we smooth its corners and impose conformal boundary conditions (CBCs) at those corners, labeled $a,b,c$. These CBCs correspond to Cardy states in the framework of boundary conformal field theory (BCFT). For a diagonal rational CFT (RCFT), there are only finitely many conformal families. In such RCFTs, the set of CBCs and the set of conformal families are both in one-to-one correspondence with the primary fields, and can thus be labeled accordingly by the primaries.

\begin{figure}[h]
    \centering
    \subfigure[\label{fig:FP_tensor}]{
    \includegraphics[width=0.2\linewidth]{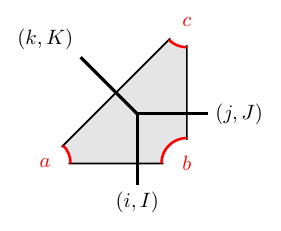}
    }
    \hspace{0.3cm}
    \subfigure[\label{fig:disk_diag}]{
    \includegraphics[width=0.2\linewidth]{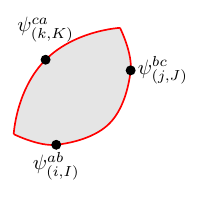}
    }
    \hspace{0.2cm}
    \subfigure[\label{fig:pants_diag}]{
    \includegraphics[width=0.2\linewidth]{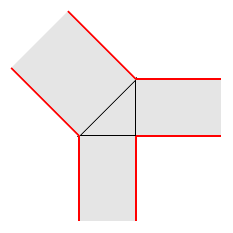}
    }
    \caption{(a) Fixed-point TC structure. (b) Disk diagram. (c) Pants diagram.}
\end{figure}

The edge of a triangle, bounded by CBCs at its two endpoints, can be interpreted as hosting a state created by a boundary-changing operator (BCO). These states can be prepared in different ways that are related by gauge transformations, as discussed in \cite{chengPrecisionReconstructionRational2025}. In one gauge, the triangle is extended into the so-called ``disk diagram" (see Fig.~\ref{fig:disk_diag}),  where BCOs are inserted along boundary arcs via the state-operator correspondence. In an alternative gauge, the triangle is mapped to the ``pants diagram" (See Fig.~\ref{fig:pants_diag}), where the states are prepared from asymptotic infinity and evolved inward to the triangle edges. In both constructions, the BCOs are labeled by primary fields $i,j,k$ (ranging over a finite set in an RCFT), together with their corresponding infinite towers of descendants $I,J,K$. The BCOs satisfy the boundary operator product expansions
\begin{equation}\label{eq:boundaryope}
    \psi^{ab}_i(x)\psi^{bc}_j(y)=\sum_kC^{abc}_{ijk}\abs{x-y}^{h_k-h_i-h_j}\psi_k(y)+\text{descendants},
\end{equation}
where $\psi^{ab}_i(x)$ denotes a BCO that changes the boundary condition from $a$ to $b$, labeled by the primary index $i$. The structure constants $C^{abc}_{ijk}$ encode the fusion data and depend explicitly on the boundary conditions $a,b,c$.  The conformal weights $h_i,h_j, h_k$ correspond to the associated primaries, and the subleading terms represent contributions from descendant fields.

With these ingredients in place, the CFT fixed-point (FP) tensor is constructed as the three-point correlation function of BCOs.   It takes the following \textit{TC} (TC) form:
\begin{equation}\label{eq:sup_FPtensor}
    \mathcal{T}^{abc}_{(i,I)(j,J)(l,K)}=\alpha^{ijk}_{IJK}C^{abc}_{ijk},
\end{equation}
where $\alpha^{ijk}_{IJK}$ denotes the three-point conformal block, fixed entirely by conformal symmetry, and $C^{abc}_{ijk}$ is the structure constant introduced in Eq.~\eqref{eq:boundaryope}. As shown in  \cite{chengPrecisionReconstructionRational2025},  $\mathcal{T}$ can be computed as the three-point correlation function of BCOs inserted along the real axis of the upper half-plane (UHP), upon applying an appropriate conformal map. Crucially, the topological data are contained in the structure constants $C^{abc}_{ijk}$, which depend on the fusion data and the boundary conditions. In contrast, the conformal block $\alpha^{ijk}_{IJK}$ captures the geometric information of the regulated triangle and is sensitive to the gauge choice. For instance, the disk and pants diagrams in Fig.~\ref{fig:disk_diag} and Fig.~\ref{fig:pants_diag} yield different expressions for $\alpha$, related by a gauge transformation. 

\begin{figure}
    \centering
    \subfigure{
    \includegraphics[width=0.45\linewidth]{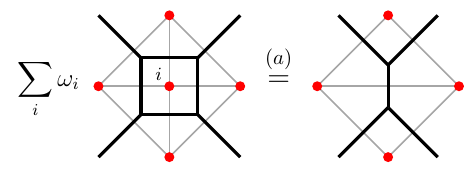}
    }\\[0.5em]
    \subfigure{
    \includegraphics[width=0.37\linewidth]{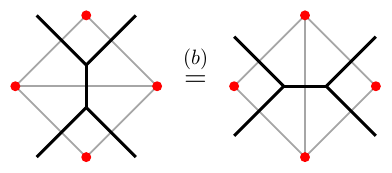}
    }
    \caption{(a) Coarse-graining condition. (b) Crossing-symmetry condition. CBCs are indicated by red dots.}
    \label{fig:FP_conditions}
\end{figure}

The FP tensors defined above can be shown to satisfy both the crossing symmetry and the coarse-graining condition displayed in Fig.~\ref{fig:FP_conditions} when they are sewn together to form four-point correlation functions. In the coarse-graining step, the boundary condition in the center (which forms a closed loop in the triple-line tensor picture) is summed over with a weight $\omega_i$ proportional to the corresponding quantum dimensions. For a diagonal RCFT,
\begin{equation}\label{eq:omega_weight}
    \omega_i=S_{00}^{1/2}S_{i0},
\end{equation}
where $S_{ij}$ is the $S$-matrix element of the modular transformation. 

Both the crossing move and the coarse-graining move are essential in the TC renormalization (TCR) scheme introduced in the main text.

\supsection{Details of the TCR algorithm}\label{sup:rgdetail}

In this section, we outline the technical details of the TCR algorithm, including the SVD procedure, the loop optimization step, and the improvement of initial tensors for loop optimization using QR/LQ decompositions applied to TCs.

\begin{figure}
    \centering
    \includegraphics[width=.75\linewidth]{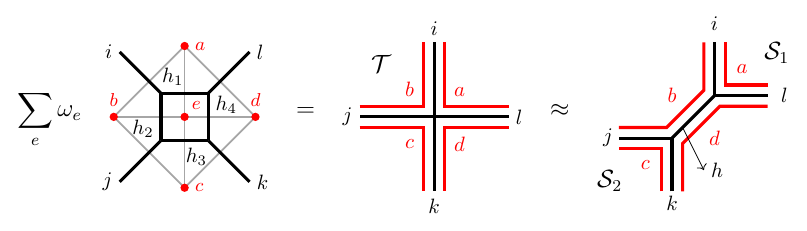}
    \caption{Construction of the initial rank-4 TC $\mathcal{T}$ and its truncated SVD into two rank-3 TCs $\mathcal{S}_1$ and $\mathcal{S}_2$.}
    \label{fig:initialrank4svd}
\end{figure}

\supsubsection{SVD}
As shown in Fig.~\ref{fig:initialrank4svd}, the initial rank-4 tensor is constructed as a weighted contraction of four rank-3 tensors:
\begin{equation}
    \mathcal{T}^{abcd}_{ijkl}
	=\sum_{e} \omega_e \sum_{h_1, h_2, h_3, h_4}
    \mathcal{T}^{bea}_{h_2h_1i}
	\mathcal{T}^{ceb}_{h_3h_2j}
	\mathcal{T}^{dec}_{h_4h_3k}
	\mathcal{T}^{aed}_{h_1h_4l}.	
\end{equation}
For brevity, we use a single index to denote the pair consisting of a primary field and its descendant label (e.g., $i$ is a shorthand for $(i,I)$). In TCs, the ordering of BCO indices ($i,j,\dots,h_1,h_2,\dots$) and CBC indices ($a,b,c,\dots$) is fixed by a chosen convention. As noted in the main text, we can absorb an overall factor of $\omega_a^{1/4}\omega_b^{1/4}\omega_c^{1/4}\omega_d^{1/4}$ into each tensor block at the outset. This removes the need for weighted contractions in the subsequent steps—every tensor contraction that follows is an ordinary contraction with internal CBC indices summed.

To perform an SVD on the rank-4 TC $\mathcal{T}$ shown on the right-hand side of Fig.~\ref{fig:initialrank4svd}, we fix the two CBCs $b,d$, which determine the intermediate leg label $h$, and group the remaining two boundary conditions. As an illustrative example, consider the case $n=3$ (corresponding to the Ising CFT), where $n$ is the number of CBCs. Let the three CBCs be labeled by the set $\{1,2,3\}$. For each pair of boundary conditions in the set $\{(b,d)\,|\,b,d=\{1,2,3\}\}$, we perform a truncated SVD—keeping at most $\chi$ largest singular values—on a block matrix obtained by grouping tensor blocks with different CBCs $a,c$. For a fixed $(b,d)$, each rank-4 tensor block with CBCs $a,b,c,d$ is reshaped into a matrix $N^{abcd}$. In components, it reads (the square bracket $[\cdots]$ denotes a joint index)
\begin{equation}\label{eq:componentsconvention}
    N^{abcd}([l,i],[j,k])=\mathcal{T}^{abcd}(i,j,k,l)\equiv\mathcal{T}^{abcd}_{ijkl}.
\end{equation}
We then assemble the $3\times 3$ block matrix $M^{bd}$ whose $(a,c)$ component is $N^{abcd}$:
\begin{equation}
    M^{bd}=\begin{pmatrix}
    N^{1b1d} & N^{1b2d} & N^{1b3d}\\
    N^{2b1d} & N^{2b2d} & N^{2b3d}\\
    N^{3b1d} & N^{3b2d} & N^{3b3d}
    \end{pmatrix}.
\end{equation}
Performing an SVD of $M^{bd}$ gives
\begin{equation}
	M^{bd}=U\Lambda V^\dagger=\begin{pmatrix}	
		U^{1}\\ U^{2}\\ U^{3}
	\end{pmatrix}\Lambda
    \begin{pmatrix}	
		V^{\dagger\,1} & V^{\dagger\,2} & V^{\dagger\,3}
	\end{pmatrix},
\end{equation}
where $\Lambda$ contains the largest $\chi$ singular values, and the column dimension of $U$ as well as the row dimension of $V^\dagger$ are truncated accordingly. We have omitted the labels $b,d$ in matrices $U^i$ and $V^{\dagger\,j}$ for clarity. The matrices $U$ and $V^\dagger$ still retain their block structure along one dimension because the row index of $U$ and the column index of $V^\dagger$ remain unchanged under the SVD. We then construct two rank-3 TCs $\mathcal{S}_1,\mathcal{S}_2$ such that
\begin{equation}
    \mathcal{T}^{abcd}_{ijkl}\approx{\mathcal{S}_1}^{dab}_{lih}{\mathcal{S}_2}^{bcd}_{jkh},
\end{equation}
where the repeated index $h$ is summed. Denoting the corresponding matrix representations (using the convention in Eq.~\eqref{eq:componentsconvention})
\begin{equation}
    S_1^{dab}([l,i],h)={\mathcal{S}_1}^{dab}(l,i,h) \qand S_2^{bcd}(h,[j,k])={\mathcal{S}_2}^{bcd}(j,k,h),
\end{equation}
the block structure is given by
\begin{equation}
	\begin{pmatrix}	
	 	S_1^{d1b}\\ S_1^{d2b}\\ S_1^{d3b}
	\end{pmatrix}=
	\begin{pmatrix}	
		U^1\\ U^2\\ U^3
	\end{pmatrix}\Lambda^{1/2}
    \qand
	\begin{pmatrix}	
	 	S_2^{b1d} & S_2^{b2d} & S_2^{b3d}
	\end{pmatrix}=\Lambda^{1/2}\begin{pmatrix}	
		V^{\dagger\,1} & V^{\dagger\,2} & V^{\dagger\,3}
    \end{pmatrix}.
\end{equation}
Iterating this SVD procedure over all $(b,d)$ produces all blocks of the rank-3 TCs $\mathcal{S}_1,\mathcal{S}_2$. Similarly, the other kind of SVD shown in the main text
can be done by fixing $a,c$ and grouping $b,d$, yielding another two rank-3 TCs $\mathcal{S}_3,\mathcal{S}_4$.

\begin{figure}
    \centering
    \includegraphics[width=0.55\linewidth]{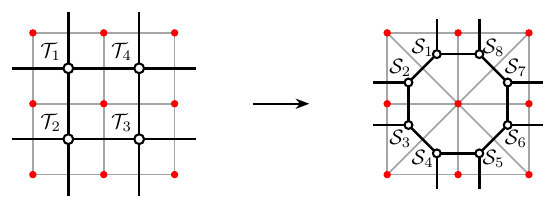}
    \caption{Square-to-octagon transformation via SVD. TCs are labeled counterclockwise.}
\label{fig:squaretooctagon}
\end{figure}

\begin{figure}
    \centering
    \includegraphics[width=.85\linewidth]{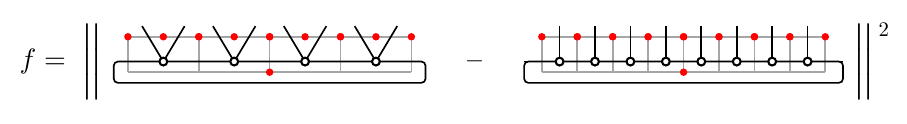}
    \caption{Cost function defined as the distance between two periodic MPS wave functions with TC structure. The leftmost and rightmost CBCs are identified in both MPS states.}
    \label{fig:cost_sup}
\end{figure}

\supsubsection{Loop optimization}
Consider a $2\times 2$ unit cell. After performing the two types of SVD on all four rank-4 TCs in the unit cell, we obtain an octagon consisting of eight rank-3 TCs $\mathcal{S}_1,\dots,\mathcal{S}_8$, see Fig.~\ref{fig:squaretooctagon}.
The cost function $f$ is defined as the 2-norm of the difference between the two diagrams in Fig.~\ref{fig:cost_sup}, viewed as periodic MPS wave functions.

\begin{figure}
    \centering
    \subfigure[\label{fig:sup_fabcd}]{
    \includegraphics[width=\linewidth]{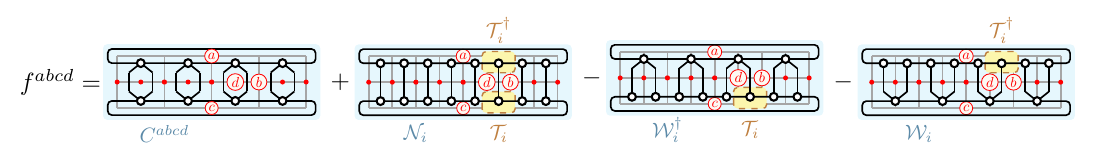}
    }
    \subfigure[\label{fig:sup_NTW}]{
    \includegraphics[width=.55\linewidth]{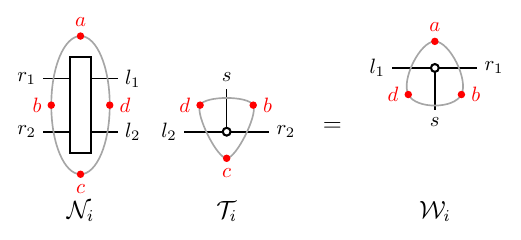}
    }
    \caption{(a) One term in the total cost function with the CBCs around site $i$ specified. For all specified CBCs, the red dots are replaced by their labels; all other unlabeled CBCs are summed over. (b) A block of the linear equation to be solved at site $i$ after reshaping the tensor blocks into matrices, where CBCs $b,d$ are fixed and $a,c$ run over all boundary conditions.}
    \label{fig:fabcd}
\end{figure}

Since we will optimize the rank-3 TCs in the octagon site by site, when optimizing the site $i$, the cost function $f$ decomposes into a sum over CBCs assignments around the site $i$:
\begin{equation}
    f=\sum_{abcd}f^{abcd},
\end{equation}
where each term $f^{abcd}$ is expanded as shown in Fig.~\ref{fig:fabcd}, with all unlabeled CBCs summed.
Adopting the index-labeling convention of Fig.~\ref{fig:sup_NTW}, for each fixed set of boundary conditions $\{a,b,c,d\}$, we first reshape the tensor blocks into matrices. In components,
\begin{equation}\label{eq:NTWmatrix}
\begin{aligned}
    N^{abcd}_i([r_1,l_1],[r_2,l_2])={}&\mathcal{N}^{abcd}_i(r_1,r_2,l_1,l_2),\\
    T^{cdb}_i([r_2,l_2],s)={}&\mathcal{T}^{cdb}_i(l_2,s,r_2),\\
    W^{adb}_i([r_1,l_1],s)={}&\mathcal{W}^{adb}_i(l_1,s,r_1),
\end{aligned}
\end{equation}
where we have used the TC component convention in Eq.~\eqref{eq:componentsconvention}. The specific ordering of CBC and BCO indices in $\mathcal{N}_i,\mathcal{T}_i,\mathcal{W}_i$ chosen on the right-hand side of Eq.~\eqref{eq:NTWmatrix} is fixed once and used throughout the algorithm. The total cost function now can be written as
\begin{equation}
    f=\sum_{abcd}f^{abcd}=\sum_{abcd}\bqty{C^{abcd}+\tr(T^{\dagger\,adb}_iN_i^{abcd}T_i^{cdb})-\tr(W^{\dagger\,cdb}_iT^{cdb}_i)-\tr(T^{\dagger\,adb}_iW^{adb}_i)},
\end{equation}
where
\begin{equation}
    T^{\dagger\,adb}_i=(T^{adb}_i)^\dagger \qand W^{\dagger\,cdb}_i=(W^{cdb}_i)^\dagger.
\end{equation}
The stationarity condition $\pdv*{f}{T_i^{\dagger\,adb}}=0$ yields
\begin{equation}\label{eq:NTWoneblock}
    \sum_c N_i^{abcd}T_i^{cdb}=W^{adb}_i.
\end{equation}
Because Eq.~\eqref{eq:NTWoneblock} couples different boundary conditions $a$ to the same blocks $T^{cdb}$, we group the blocks for different $a,c$ into block matrices $M_{N,i}, M_{T,i}, M_{W,i}$ and optimize all blocks $T^{cdb}$ at the same time. Fixing $b,d$ and grouping $a,c$ produces block matrices $M^{bd}_{N,i}, M^{bd}_{T,i}$, and $M^{bd}_{W,i}$ such that the condition $\pdv*{f}{M^{\dagger\,bd}_{T,i}}$ reduces to the linear equation $M^{bd}_{N,i}M^{bd}_{T,i}=M^{bd}_{W,i}$. For example, in the $n=3$ Ising CFT case:
\begin{equation}\label{eq:NTWblocks}
    M_{N}^{bd}M_{T}^{bd}=M_{W}^{bd}\quad\Longleftrightarrow\quad
    \begin{pmatrix}
        N^{1b1d} & N^{1b2d} & N^{1b3d} \\
        N^{2b1d} & N^{2b2d} & N^{2b3d} \\
        N^{3b1d} & N^{3b2d} & N^{3b3d}
    \end{pmatrix}\begin{pmatrix}
        T^{1db} \\ T^{2db} \\ T^{3db}
    \end{pmatrix}=\begin{pmatrix}
        W^{1db} \\ W^{2db} \\ W^{3db}
    \end{pmatrix},
\end{equation}
where the site index $i$ has been suppressed. Once the solution for $M_T^{bd}$ is found, each block $T^{cdb}$ is extracted and reshaped back into a rank-3 tensor using Eq.~\eqref{eq:NTWmatrix}. Although one could assemble all $(b,d)$ sectors into a single larger block matrix—given that all tensor blocks have already been truncated to the cutoff dimension $\chi$—it is computationally advantageous to optimize each block matrix $M^{bd}_T$ separately (equivalently, to minimize each $f^{bd}$ independently). This blockwise strategy enables full parallelization across all $(b,d)$ sectors and leads to a substantial performance gains. Remarkably, this blockwise optimization procedure already yields highly accurate results.

After optimizing the rank-3 TC on a single site, we sweep through all sites in the octagon (e.g., in the simple order $1\rightarrow8,1\rightarrow8,...$). The loop optimization is terminated either when the number of sweeps reaches a predetermined maximum (typically 20-50 for simple models), or when the change in the total cost function falls below a chosen threshold $\epsilon$. Concretely, the stopping criterion is
\begin{equation}
    \abs{f_i-f_{i+1}}<\epsilon,
\end{equation}
where $f_{i+1}$ denotes the total cost function after one additional full sweep compared to $f_i$. A typical choice is $\epsilon=10^{-12}$.

Finally, as noted in the main text, the weights can be absorbed into the initial tensor by multiplying each block with CBCs $a,b,c,d$ by the factor $\omega_a^{1/4}\omega_b^{1/4}\omega_c^{1/4}\omega_d^{1/4}$. With this simplification, the contractions shown in Fig.~\ref{fig:sup_fabcd} no longer correspond to the geometric gluing of four triangles in the coarse-graining step, where the summed CBC $e$ is weighted by exactly $\omega_e$ (see the relation on the left-hand side of Fig.~\ref{fig:initialrank4svd}). Consequently, the loop optimization performed here should be viewed as a direct optimization of each rank-3 tensor, without any interpretation in terms of geometric gluing of shapes.

\supsubsection{QR/LQ decomposition}
The octagon on the right-hand side of Fig.~\ref{fig:squaretooctagon} is obtained by performing SVDs on matrices constructed from local tensors. The initial tensors for loop optimization can be further improved by applying more global truncations via QR/LQ decompositions \cite{wangClusterUpdateTensor2011}.

We first describe the QR decomposition for a rank-3 TC on a single site; the LQ procedure is analogous.
\begin{figure}
    \centering
    \includegraphics[width=.8\linewidth]{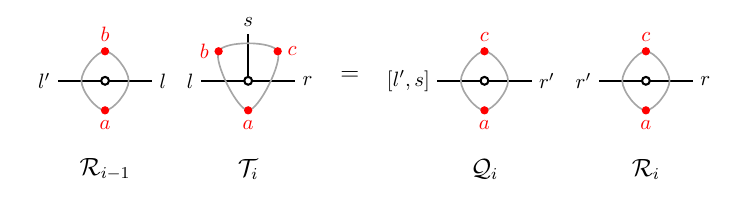}
    \caption{QR decomposition on a rank-3 TC at site $i$. On the right-hand side, the joint index $[l^\prime,s]$ includes components associated with different boundary condition $b$.}
    \label{fig:qr}
\end{figure}
As illustrated in Fig.~\ref{fig:qr}, we begin with a rank-3 TC $\mathcal{T}_i$ that is left-multiplied by a rank-2 TC. The multiplication is straightforward: for each pair of boundary conditions $(b,a)$, we perform ordinary matrix multiplication over the BCO index $c$, where the tensor blocks of $\mathcal{T}_i$ are reshaped to the matrices $T_i(l,[s,r])$. The resulting TC is denoted $\mathcal{T}_i^\prime=\mathcal{R}_{i-1}\mathcal{T}_i$. To perform the QR decomposition on $\mathcal{T}_i^\prime$, we fix boundary conditions $c,a$ and group over boundary condition $b$. For each triple $\{a,b,c\}$, we first reshape the rank-3 tensor block to a matrix:
\begin{equation}
    N^{abc}_i([l^\prime,s],r)=\mathcal{T}^{\prime\,abc}_i(l^\prime,s,r),
\end{equation}
using again the TC component convention of Eq.~\eqref{eq:componentsconvention}. For each fixed $(c,a)$, we then assemble the block matrix $M^{ca}_i$ by stacking the matrices $N^{abc}_i$ for all values of $b$. The QR decomposition is performed on the block matrix $M^{ca}_i$ for each pair of $(c,a)$. For example, in the $n=3$ Ising CFT case,
\begin{equation}
    M^{ca}_i=\begin{pmatrix}
        N^{a1c}_i\\N^{a2c}_i\\N^{a3c}_i
    \end{pmatrix}=Q^{ca}_iR^{ca}_i,
\end{equation}
where $Q^{ca}_i$ and $R^{ca}_i$ are the corresponding blocks of a matrix complex $\mathcal{Q}_i$ and $\mathcal{R}_i$, respectively. Collecting the QR results for all pairs $(c,a)$, we obtain the matrix complex relation
\begin{equation}\label{eq:qr}
    \mathcal{R}_{i-1}\mathcal{T}_{i}=\mathcal{Q}_i\mathcal{R}_i.
\end{equation}
Note that it is unnecessary to separate the index $b$ in $\mathcal{Q}_i$ and reshape it back into a rank-3 TC, since only $\mathcal{R}_i$ (and $\mathcal{L}_i$ from LQ decompositions) are used subsequently.

\begin{figure}
    \centering
    \subfigure{
    \includegraphics[width=0.9\linewidth]{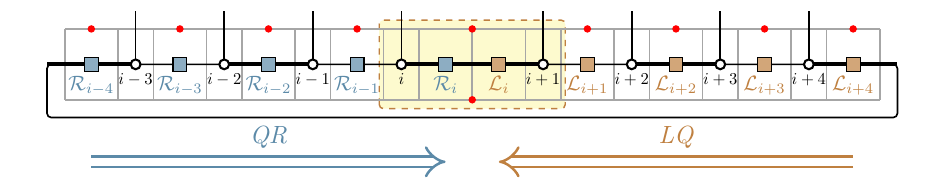}
    }\\[1em]
    \subfigure{
    \includegraphics[width=0.7\linewidth]{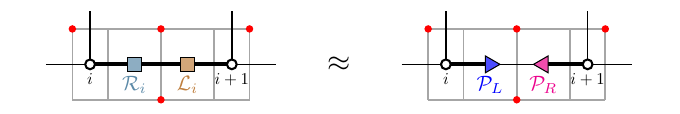}
    }
    \caption{Truncation on the bond between sites $i$ and $i+1$ via successive QR/LQ decompositions. The projectors $\mathcal{P}_L$ and $\mathcal{P}_R$ are constructed from $\mathcal{R}_i$ and $\mathcal{L}_i$. Thickened lines denote legs produced by SVDs prior to truncation. Sites are identified modulo the period 8.}
    \label{fig:qrlqchain}
\end{figure}

The truncation on the bond between site $i$ and $i+1$ is illustrated in Fig.~\ref{fig:qrlqchain}.
We begin by setting $\mathcal{R}_{i-4}=\mathcal{I}$, where $\mathcal{I}$ denotes a matrix complex whose blocks are identity matrices. We then apply successive QR decompositions Eq.~\eqref{eq:qr} to the right until reaching site $i$. Similarly, with $\mathcal{L}_{i+4}=\mathcal{I}$, we perform successive LQ decompositions
\begin{equation}
    \mathcal{T}_{i+1}\mathcal{L}_{i+1}=\mathcal{L}_i\mathcal{Q}_i
\end{equation}
propagating leftward until site $i+1$. On the bond between sites $i$ and $i+1$, we insert an identity matrix complex $\mathcal{I}=\mathcal{R}_i^{-1}\mathcal{R}_i\mathcal{L}_i\mathcal{L}_i^{-1}$, where $\mathcal{R}_i^{-1},\mathcal{L}_i^{-1}$ consist of blockwise inverses of the corresponding blocks in $\mathcal{R}_i,\mathcal{L}_i$. We next perform an SVD with cutoff dimension $\chi$ on the product $\mathcal{R}_i\mathcal{L}_i$:
\begin{equation}\label{eq:RLSVD}
    \mathcal{R}_i\mathcal{L}_i\approx\mathcal{U}_i\mathit{\Lambda}_i\mathcal{V}^\dagger_i.
\end{equation}
This implies the approximate inverse relation
\begin{equation}
    \mathcal{L}_i^{-1}\mathcal{R}_i^{-1}\approx\mathcal{V}_i\frac{1}{\mathit{\Lambda}_i}\mathcal{U}^\dagger_i,
\end{equation}
where singular values below a threshold $\lambda$ (e.g., $10^{-12}$) are discarded to maintain numerical stability. The projectors on this bond are then constructed as $\mathcal{I}\approx\mathcal{P}_L\mathcal{P}_R$, with
\begin{equation}\label{eq:PLPR}
\begin{aligned}
    \mathcal{P}_L={}&\mathcal{R}_i^{-1}\mathcal{U}_i\sqrt{\mathit{\Lambda}_i}=\mathcal{L}_i\mathcal{V}_i\frac{1}{\sqrt{\mathit{\Lambda_i}}},\\
    \mathcal{P}_R={}&\sqrt{\mathit{\Lambda}_i}\mathcal{V}^\dagger_i\mathcal{L}_i^{-1}=\frac{1}{\sqrt{\mathit{\Lambda_i}}}\mathcal{U}^\dagger_i\mathcal{R}_i.
\end{aligned}
\end{equation}
In practice, we construct projectors on all bonds, not only those whose bond dimensions have increased. We also emphasize that all $\mathcal{R}_i, \mathcal{L}_i$ are computed before applying any projectors.

After applying the projectors, the resulting rank-3 TCs serve as the initial variational MPS wavefunction for loop optimization. This improved MPS has a substantially smaller total cost function and thus provides a much better starting point for the optimization procedure.

\supsection{Extracting conformal data from the FP tensor}\label{sup:cftdata}

Once the rank-4 tensors in a $2\times 2$ unit cell are obtained at each RG step, we can extract conformal data—such as scaling dimensions, central charge, and conformal spins—by constructing transfer matrices on a cylinder and analyzing their eigenvalues. In this section, we describe in detail how to compute the scaling dimensions and central charge from the FP tensors.

\begin{figure}[h]
    \centering
    \newsavebox{\mybox}
    \savebox{\mybox}{
    \includegraphics[width=.27\linewidth]{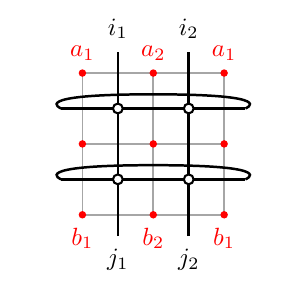}
    }
    \subfigure[]{
        \raisebox{\dimexpr.5\ht\mybox-.5\height}{
    \includegraphics[width=.37\linewidth]{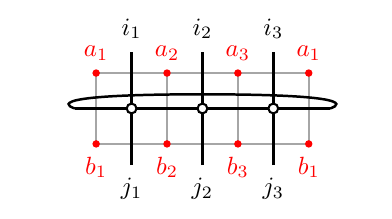}
    }}
    \subfigure[\label{fig:T_2x2}]{
        \usebox{\mybox}
    }
    \caption{Transfer matrix constructed from rank-4 TCs. All decendant labels are suppressed for simplicity. (a) Example of a $1\times 3$ transfer matrix, with boundary conditions at the two ends identified. (b) A $2\times 2$ transfer matrix used in the Loop-TNR algorithm, where the unit cell is $2\times 2$. All unlabeled intermediate boundary conditions are summed.}
    \label{fig:transfermatrix}
\end{figure}

Consider an $m\times n$ transfer matrix with $m$ rows and $n$ columns. The conformal data are encoded in its eigenvalues $\lambda_i$, which satisfy \cite{cardyOperatorContentTwodimensional1986a,guTensorentanglementfilteringRenormalizationApproach2009}
\begin{equation}\label{eq:eigenvalues_transfermatrix}
    \lambda_i=e^{-2\pi\Im(\tau)(\Delta_i-\frac{c}{12})+2\pi i\Re(\tau)s_i-\epsilon mn}\quad (i=0,1,2,...)
\end{equation}
where $\Delta_i=h_i+\bar{h}_i$ is the scaling dimension, $c$ is the central charge, and $s_i=h_i-\bar{h}_i$ is the conformal spin. The parameter $\tau$ is the modular parameter of the torus obtained by wrapping the tensor network. For the transfer matrices used here (Fig. \ref{fig:transfermatrix}), the modular parameter takes the purely imaginary form $\tau=im/n$, so that $\Re(\tau)=0$. The constant $\epsilon$ denotes the ground-state energy density of the infinite system, and appears as an overall normalization factor for the tensor network. 

One way to eliminate the normalization factor $\epsilon$ in Eq.~\eqref{eq:eigenvalues_transfermatrix} is to construct two transfer matrices of different lengths. For example, when all tensors are uniform (as in the Levin-Nave TRG scheme), we may build two transfer matrices of size $1\times 1$ and $1\times 2$, with eigenvalues $\lambda_{1i}$ and $\lambda_{2i}$, respectively. Assuming the lowest scaling dimension is $\Delta_0=0$ (otherwise we simply shift the spectrum so that the smallest $\Delta_i$ is set to $0$ and extract the effective central charge and scaling dimensions), the central charge follows from
\begin{equation}
    c=-\frac{4}{\pi}\pqty{\ln\lambda_{20}-2\ln\lambda_{10}}.
\end{equation}

\begin{figure}
    \centering
    \includegraphics[width=0.32\linewidth]{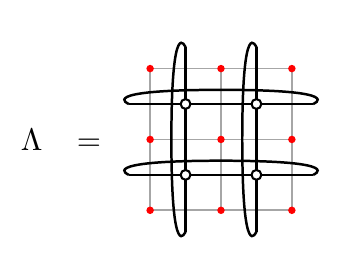}
    \caption{Full contraction of a $2\times 2$ unit cell. All CBCs, indicated by red dots, are summed.}
    \label{fig:normalization}
\end{figure}
A more convenient way to remove the normalization factor is to properly normalize the tensors at each RG step \cite{guTensorentanglementfilteringRenormalizationApproach2009}. Along an RG flow, the scale-invariant tensors
\begin{equation}
    \mathcal{T}_{\text{inv}}^{(i)}\rightarrow\mathcal{T}_{\text{inv}}^{(i+1)},
\end{equation}
satisfy the defining property
\begin{equation}
    \text{tTr}(\otimes \mathcal{T}_{\text{inv}}^{(i)})^{N_i}=\text{tTr}(\otimes \mathcal{T}_{\text{inv}}^{(i+1)})^{N_{i+1}},
\end{equation}
where $\text{tTr}$ denotes full contraction of the tensor (-complex) network, and $N_i$ and $N_{i+1}$ are the numbers of tensors at RG steps $i$ and $i+1$. Within TCR, after each RG step we normalize all tensors in a $2\times 2$ unit cell by a common factor $\Lambda^{1/4}$ to maintain numerical stability:
\begin{equation}
    \mathcal{T}\rightarrow \mathcal{T}^\prime=\mathcal{T}/\Lambda^{1/4},
\end{equation}
where $\Lambda$ is the full contraction shown in Fig.~\ref{fig:normalization}. The properly normalized scale-invariant tensors $\mathcal{T}_{\text{inv}}$ used to construct the transfer matrices (so that their eigenvalues no longer contain the factor $\exp(-\epsilon mn)$) are then given by
\begin{equation}
    \mathcal{T}_{\text{inv}}=\mathcal{T}^\prime/\Lambda^{1/4}.
\end{equation}
Using the $2\times 2$ transfer matrix of Fig.~\ref{fig:T_2x2}, the central charge and scaling dimensions are obtained from
\begin{equation}
    c=\frac{6}{\pi}\frac{\ln\lambda_0}{\Im(\tau)},\qquad \Delta_i=-\frac{1}{2\pi\Im(\tau)}\ln\frac{\lambda_i}{\lambda_0}.
\end{equation}

In constructing the transfer matrix, we must group all matrix blocks associated with different boundary conditions on the open legs into a single large block matrix. For the $2\times 2$ transfer matrix, the row index of this matrix corresponds to the collective boundary conditions $a=(a_1,a_2)$, and the column index corresponds to $b=(b_1,b_2)$, while all intermediate boundary conditions are summed over. As an example, again consider the $n=3$ Ising CFT. Let $N^{a_1a_2,\,b_1b_2}$ denote the matrix block associated with boundary conditions $(a_1,a_2)$ and $(b_1,b_2)$, with components $N^{a_1a_2,\,b_1b_2}([i_1,i_2],[j_1,j_2])$. The full block matrix to be diagonalized is
\begin{equation}
    M=\begin{pmatrix}
      N^{11,11} & N^{11,21} & \cdots & N^{11,33}\\
      N^{21,11} & N^{21,21} & \cdots & N^{21,33}\\
      \vdots & \vdots &\ddots &\vdots\\
      N^{33,11} & N^{33,21} & \cdots & N^{33,33}
    \end{pmatrix}.
\end{equation}

\begin{figure}
    \centering
    \includegraphics[width=0.75\linewidth]{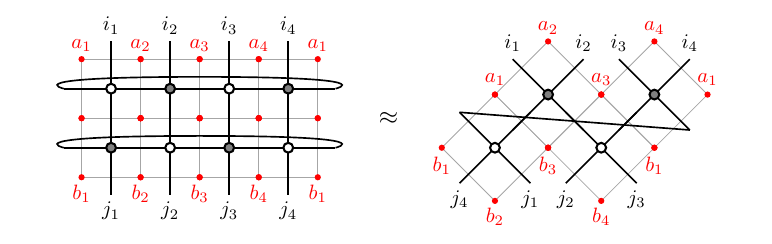}
    \caption{Approximating a $2\times 4$ transfer matrix using tensors from the next RG step. All unlabeled CBCs are summed. The two distinct rank-4 tensors in the unit cell are shown in different colors.}
    \label{fig:approxT2x4}
\end{figure}
To resolve higher scaling dimensions more accurately, we can likewise construct a $2\times 4$ transfer matrix, as illustrated in Fig.~\ref{fig:approxT2x4}. For clarity, the two distinct rank-4 tensors inside each unit cell are represented in different colors. Since an exact construction of the $2\times 4$ transfer matrix requires significantly computational effort, we approximate it using tensors from the next RG step, following Ref.~\cite{chenfengbaoLoopOptimizationTensor2019}.

For matrices of this size, direct diagonalization is impractical. Instead, we extract the leading eigenvalues using iterative methods such as the Arnoldi algorithm, bringing the computational cost of this step to the same order as that of the main TCR procedure implemented via Loop-TNR.

\supsection{Spectra in the pants gauge}

We now present the TCR results obtained in an alternative gauge for the three-point function—the pants gauge, illustrated in Fig.~\ref{fig:pants_diag}. In this gauge, the leading behavior of the three-point conformal block $\alpha^{ijk}_{IJK}$, restricted to primary fields, takes the form
\begin{equation}\label{eq:alpha_pants}
    \alpha^{ijk}_{000} \approx 0.450^{h_i+h_j}0.820^{h_k}.
\end{equation}

\begin{figure}[h]
    \includegraphics[width=.4\linewidth]{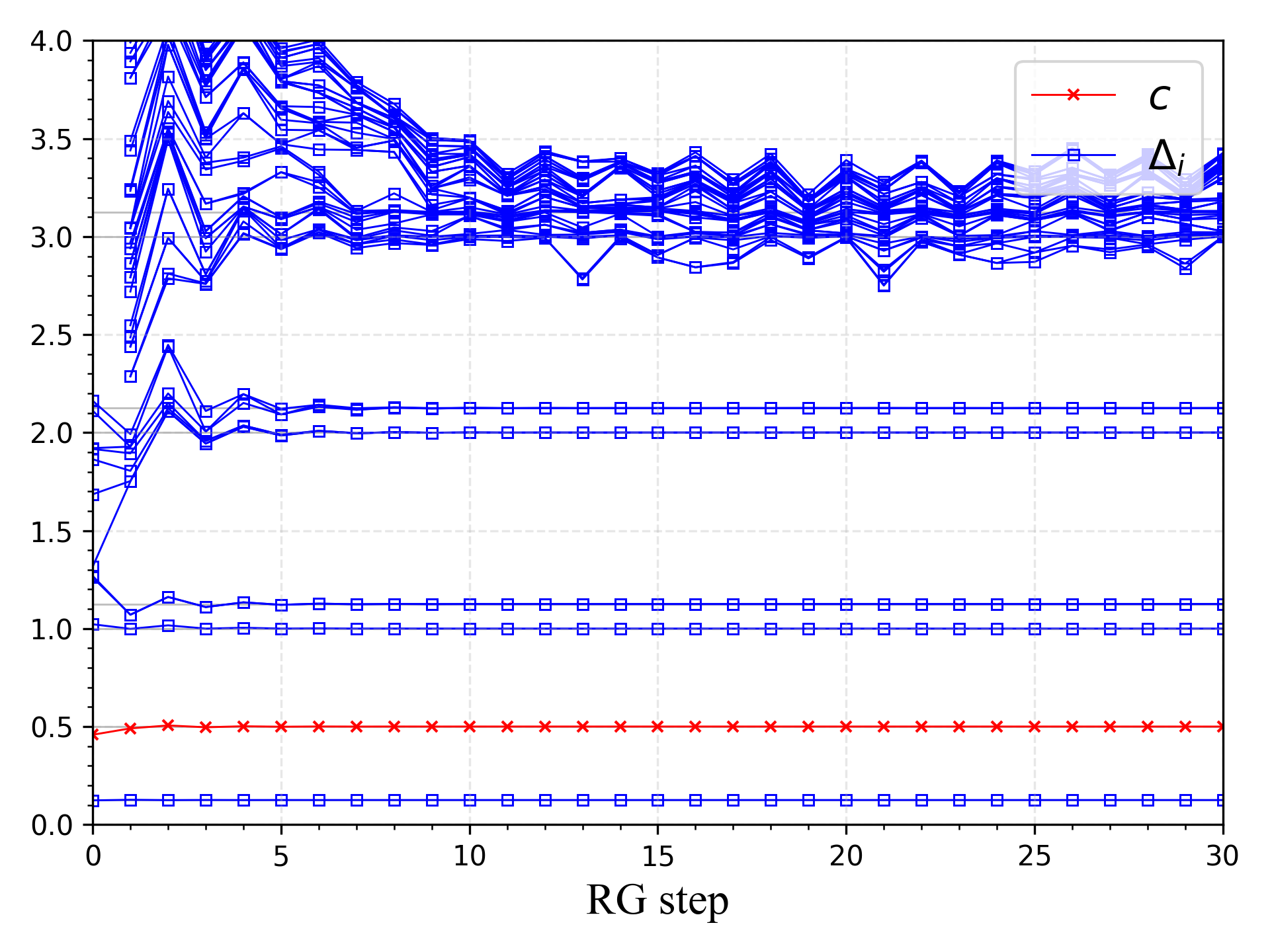}
    \includegraphics[width=.4\linewidth]{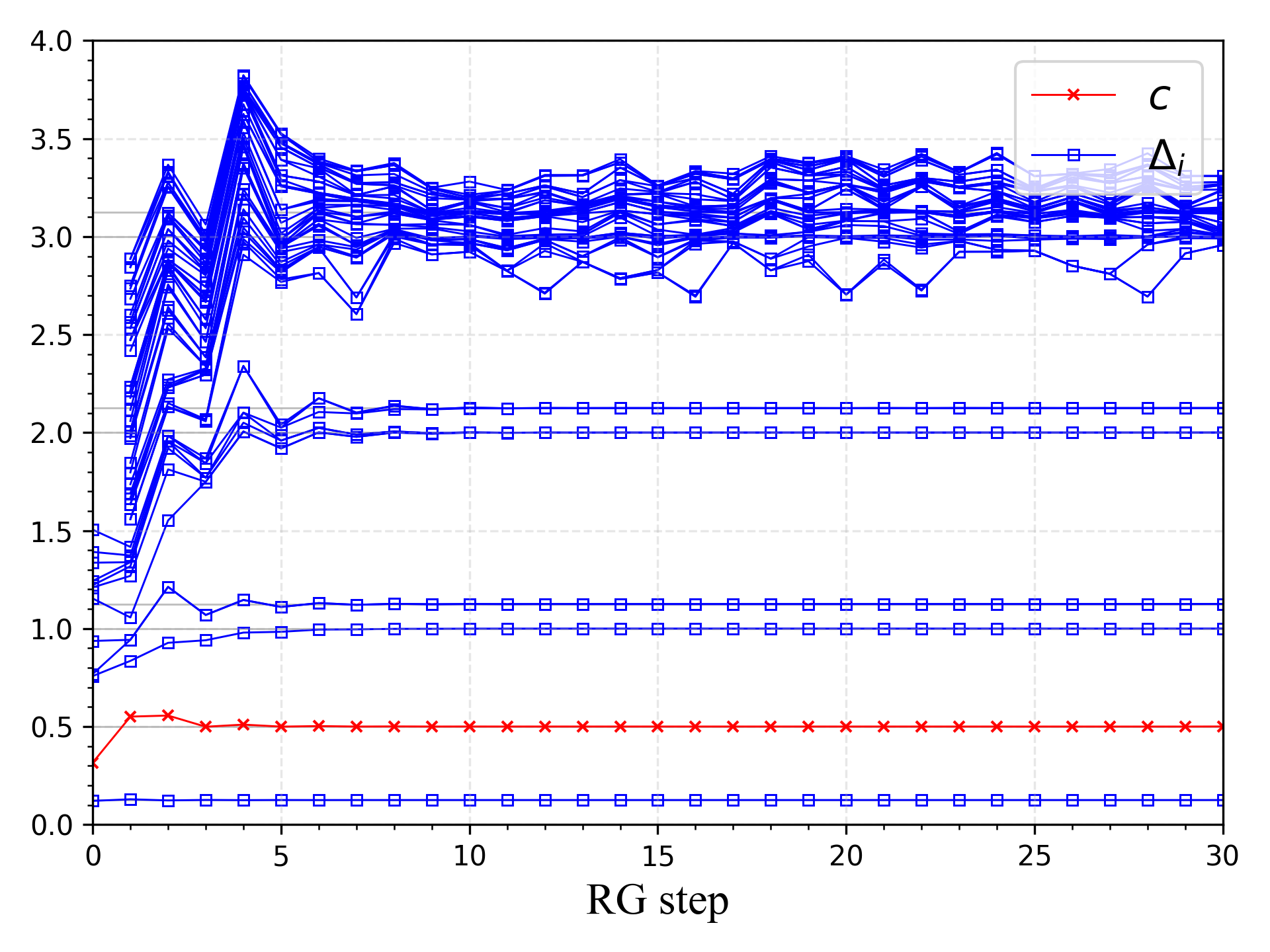}
    \caption{Ising CFT spectrum in the pants gauge, for $\mathcal{N}=\mathcal{N}_{ijk}$ (left) and $\mathcal{N}=1$ (right), $\chi=16$.}
    \label{fig:ising_pants}
\end{figure}

\begin{figure}[h]
    \includegraphics[width=.4\linewidth]{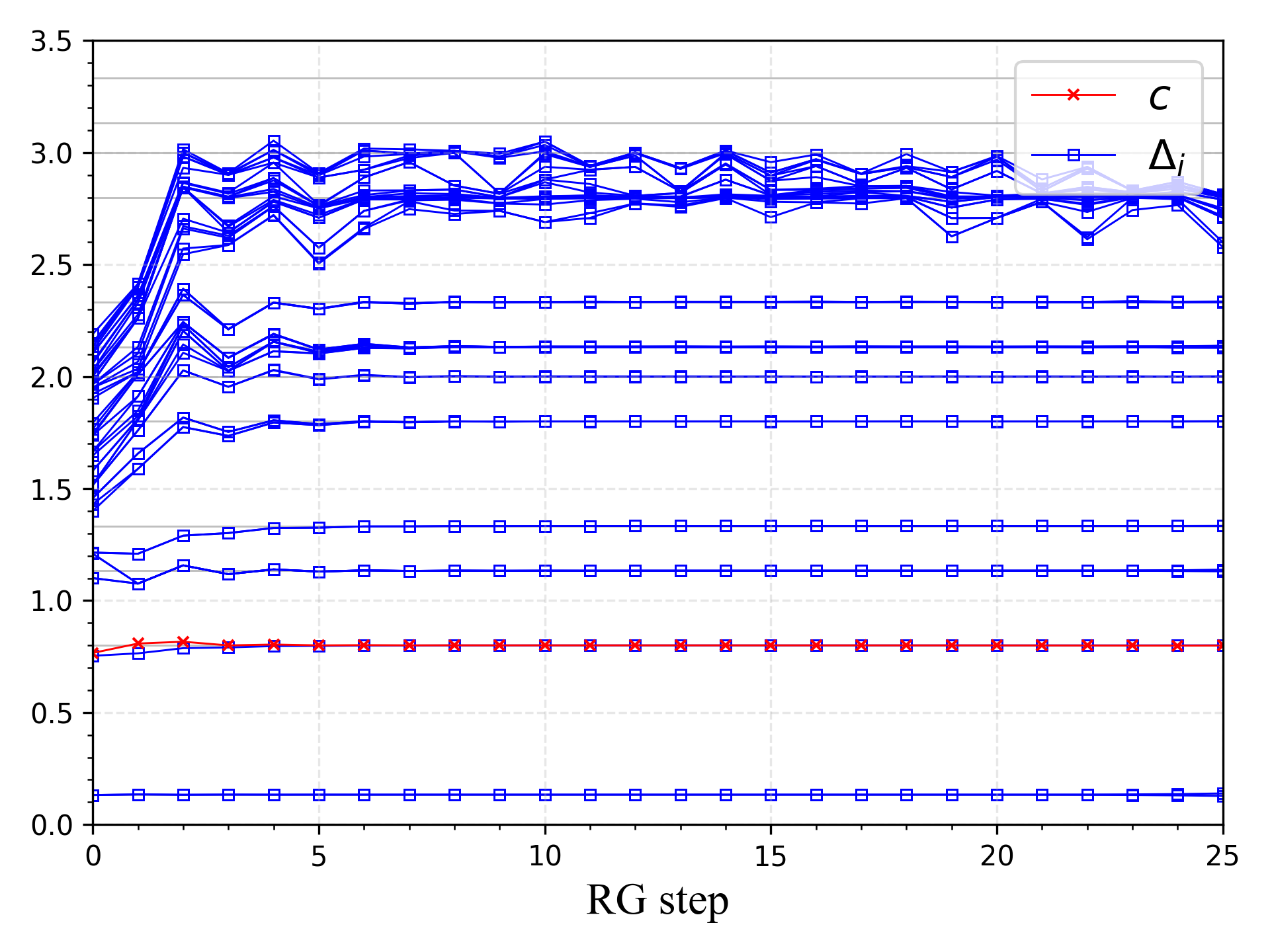}
    \includegraphics[width=.4\linewidth]{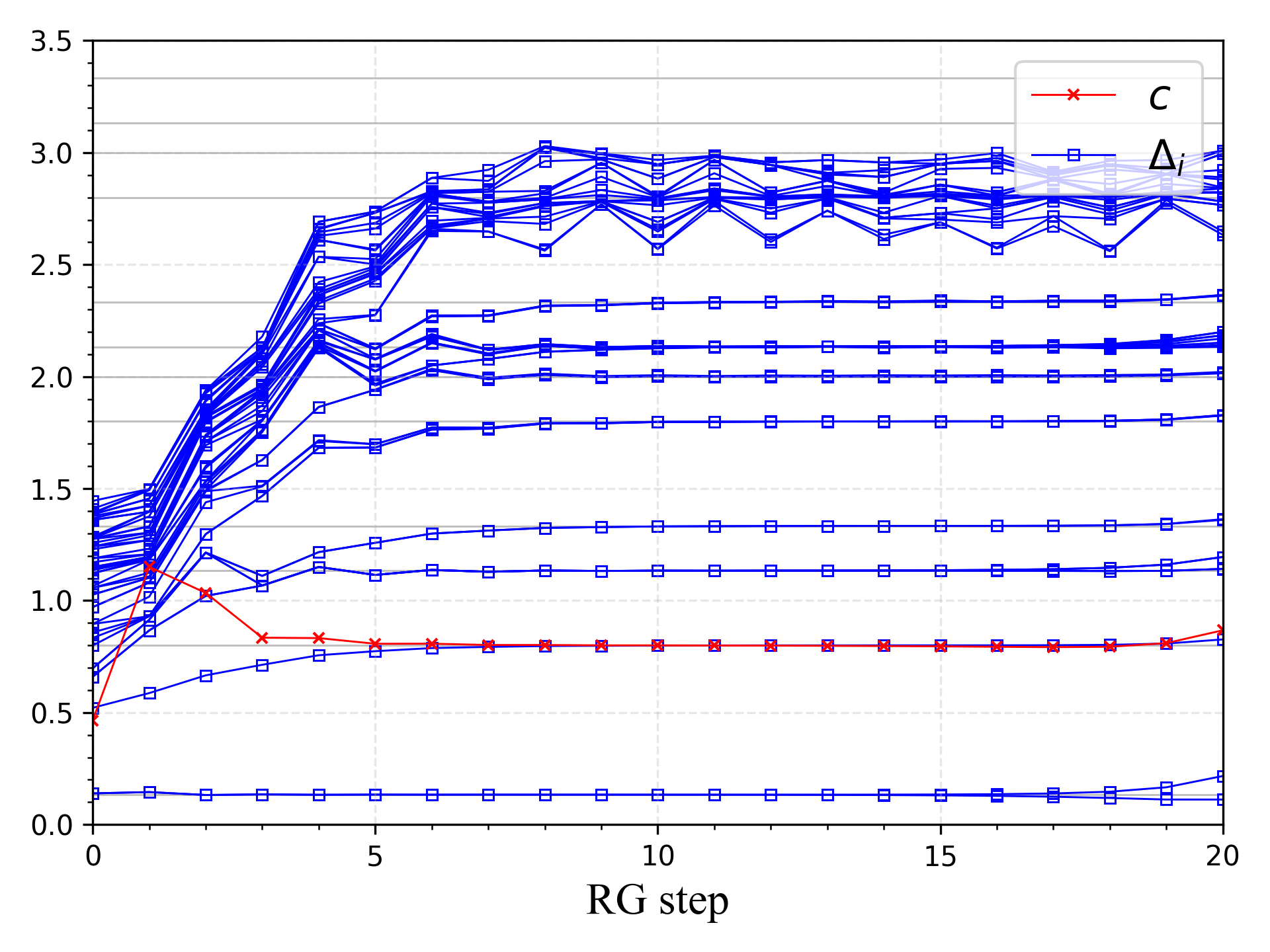}
    \caption{3-state Potts CFT spectrum in the pants gauge, for $\mathcal{N}=\mathcal{N}_{ijk}$ (left) and $\mathcal{N}=1$ (right), $\chi=16$.}
    \label{fig:3potts_pants}
\end{figure}

As in the main text, we consider two distinct cases.
In the first case, we use $\mathcal{N}_{ijk}$ as defined in Eq.~\eqref{eq:N}, which incorporates the bulk OPE data of the CFT. The input tensors are then chosen as
\begin{equation}
\mathcal{T}^{abc}_{ijk} = \alpha^{ijk}_{000} C^{abc}_{ijk},\qquad (\mathcal{N}=\mathcal{N}_{ijk}).
\end{equation}

In the second case, we set $\mathcal{N}_{ijk}=1$, so that the tensors contain only the topological data. This choice reduces $C_{ijk}^{abc}$ to its topological counterpart $\hat{C}^{abc}_{ijk}$, and the corresponding input tensors become
\begin{equation}
\mathcal{T}^{abc}_{ijk} = \alpha^{ijk}_{000} \hat{C}^{abc}_{ijk},\qquad (\mathcal{N}=1).
\end{equation}

Figures~\ref{fig:ising_pants} and \ref{fig:3potts_pants} display the spectra of the Ising CFT $\mathcal{M}(4,3)$ in the A-series and the 3-state Potts CFT $\mathcal{M}(6,5)$ in the $D$-series, respectively. We observe that, in the pants gauge, primary-field input data remain sufficient to generate stable RG flows with the correct spectra. Remarkably, the topological bootstrap corresponding to $\mathcal{N}_{ijk}=1$ continues to succeed in this gauge, indicating that the scheme is general and robust, and largely insensitive to the gauge choice in the CFT FP tensor construction.

\supsection{More minimal models}

Next, we apply the TCR algorithm to two additional minimal models: the tricritical Ising CFT $\mathcal{M}(5,4)$ and the tetracritical Ising CFT $\mathcal{M}(6,5)$, both in the $A$-series. 

\begin{figure}[h]
    \includegraphics[width=.4\linewidth]{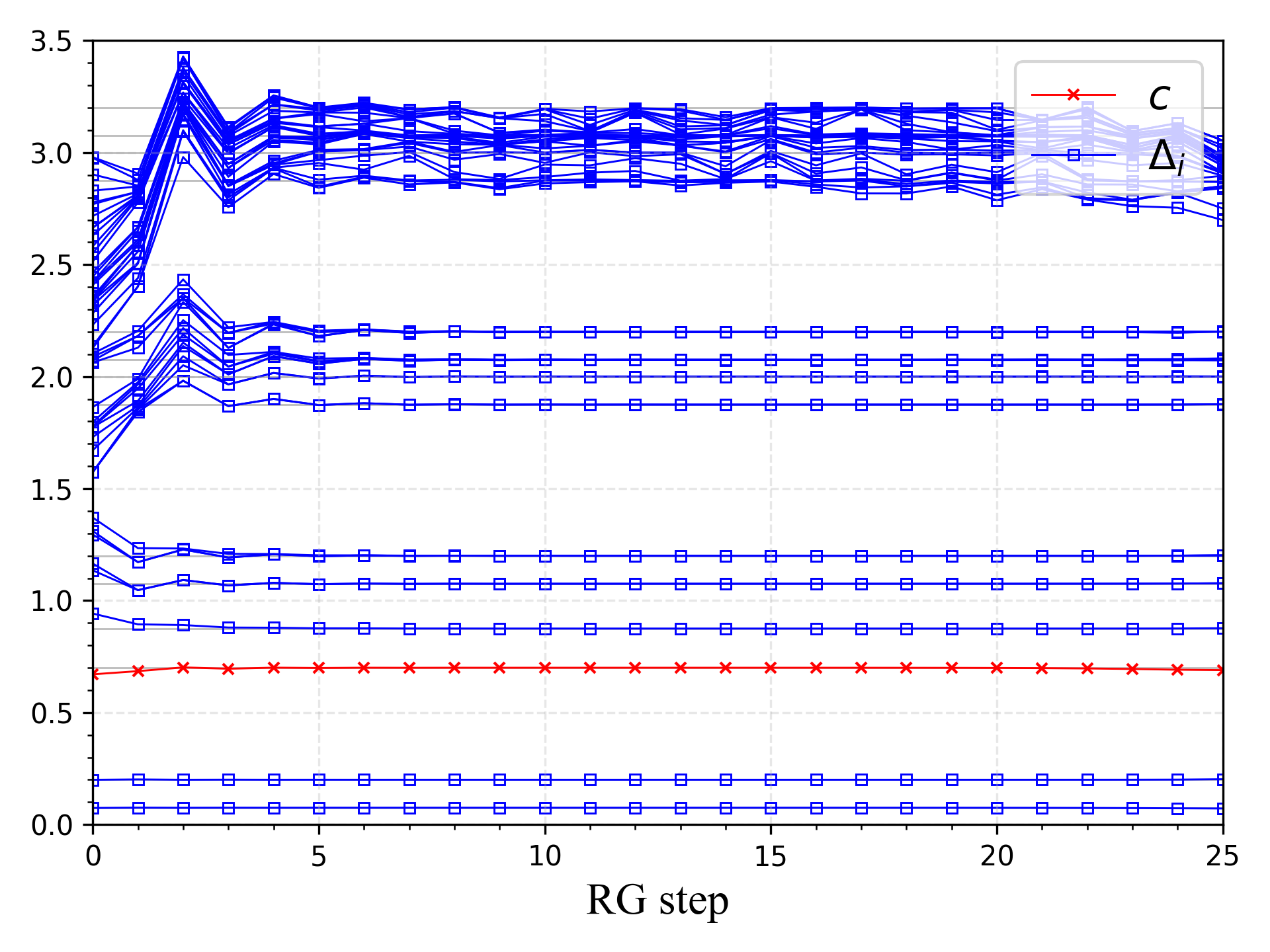}
    \includegraphics[width=.4\linewidth]{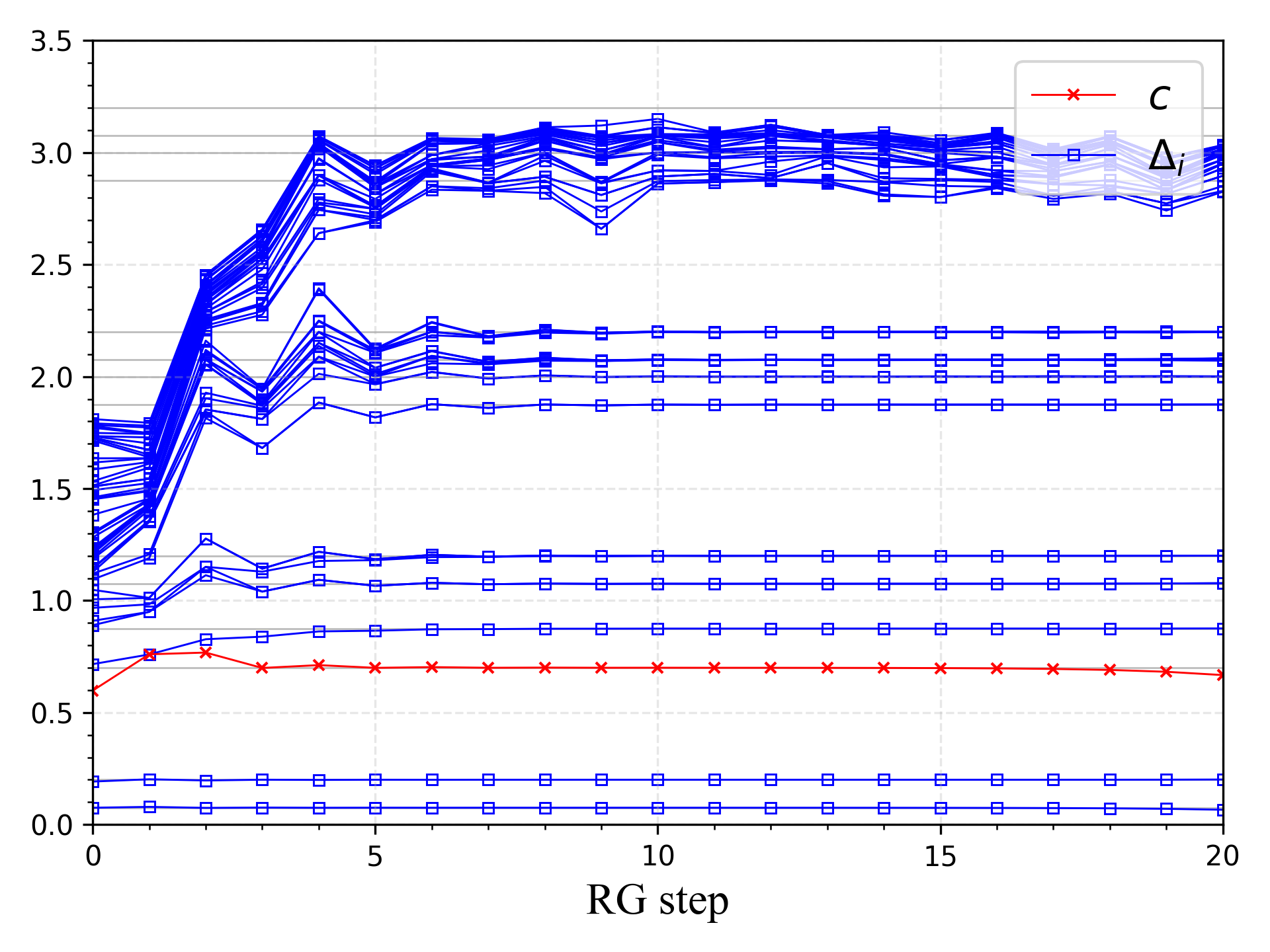}
    \caption{Tricritical Ising CFT spectrum in the disk gauge, for $\mathcal{N}=\mathcal{N}_{ijk}$ (left) and $\mathcal{N}=1$ (right), $\chi=16$.}
    \label{fig:triising_disk}
\end{figure}

\begin{figure}[h]
    \includegraphics[width=.4\linewidth]{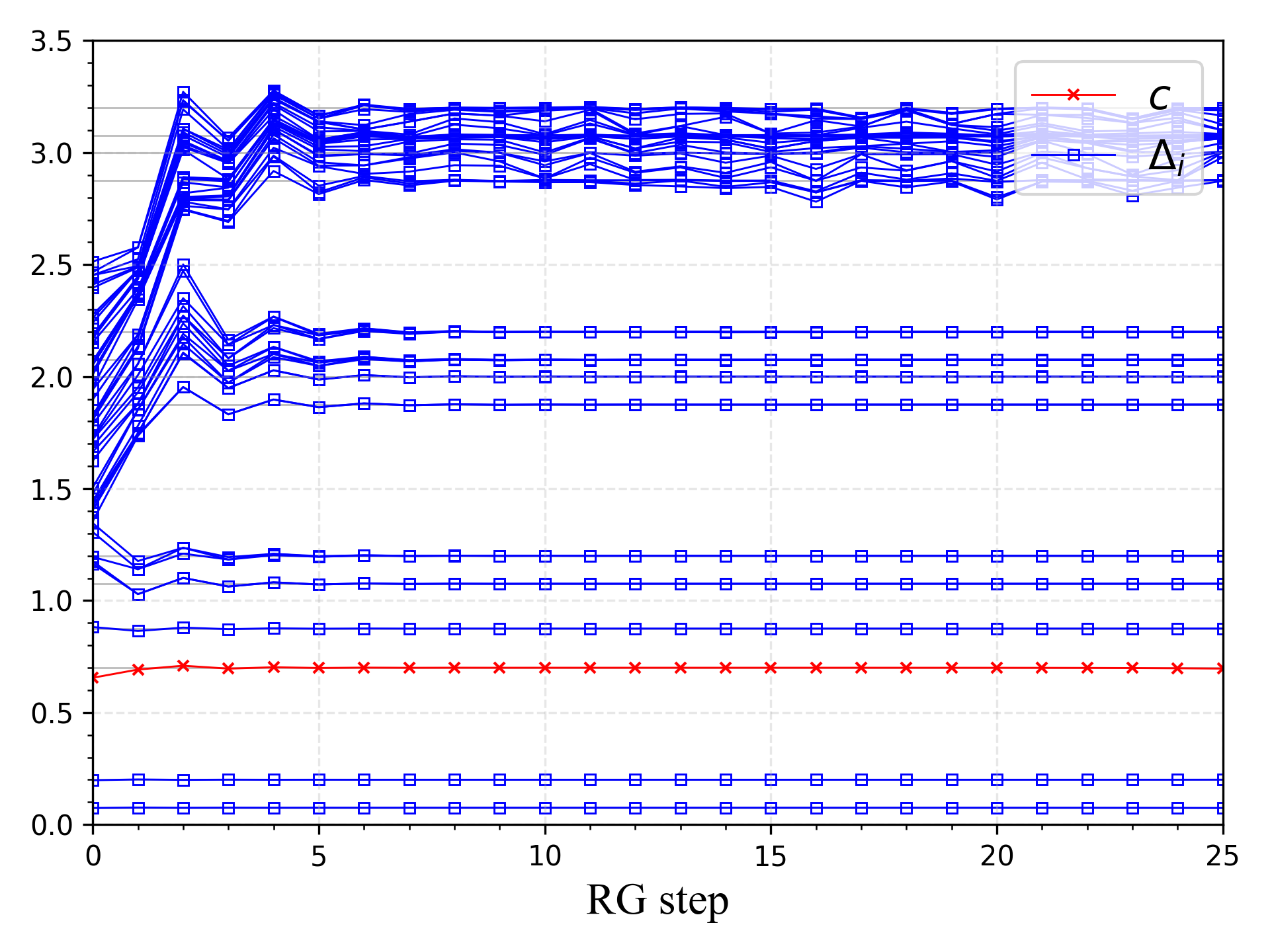}
    \includegraphics[width=.4\linewidth]{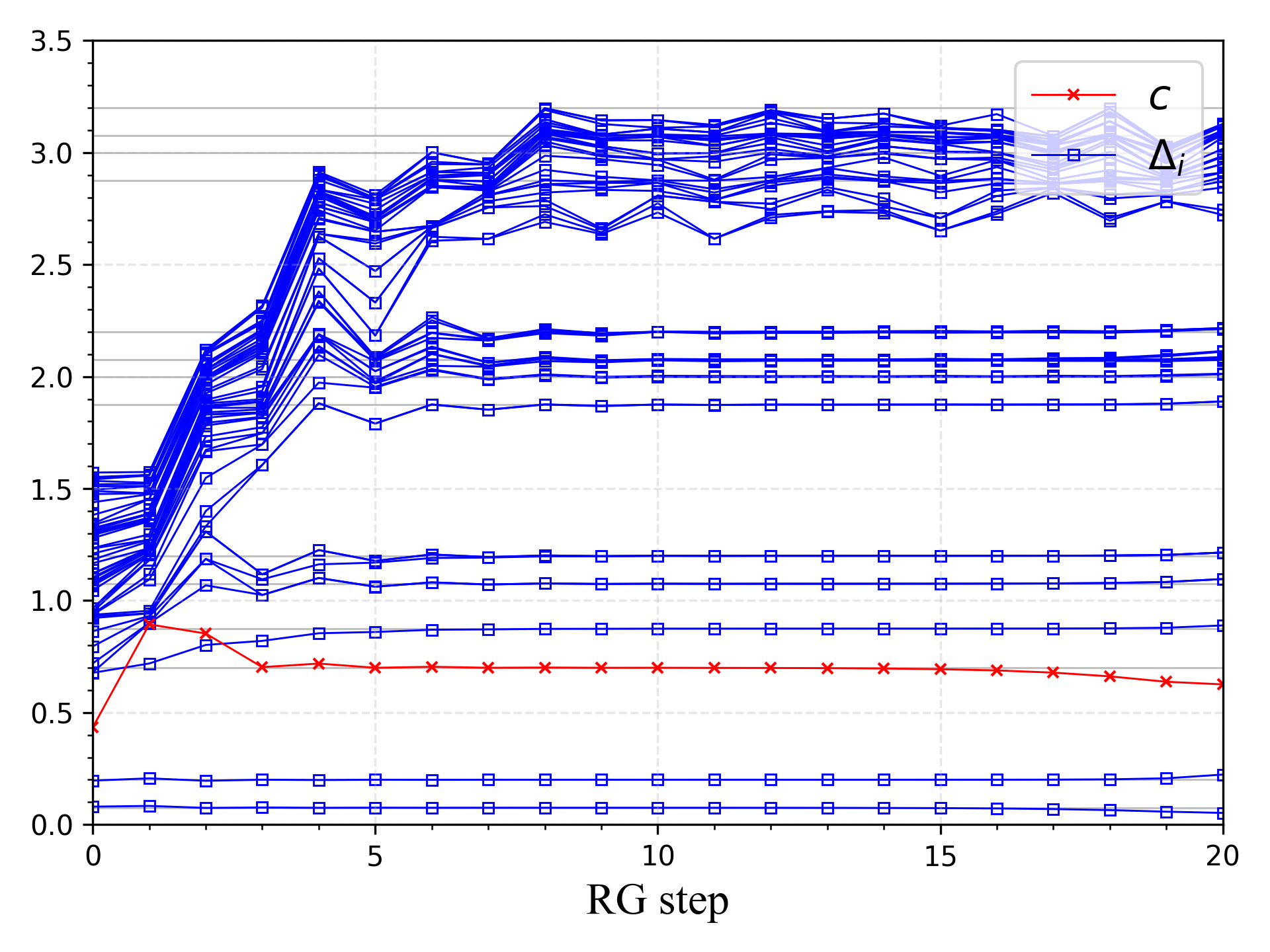}
    \caption{Tricritical Ising CFT spectrum in the pants gauge, for $\mathcal{N}=\mathcal{N}_{ijk}$ (left) and $\mathcal{N}=1$ (right), $\chi=16$.}
    \label{fig:triising_pants}
\end{figure}

The tricritical Ising CFT has six primary fields with scaling dimensions
\begin{equation}
    \Delta=\Bqty{0,\frac{3}{40},\frac{1}{5},\frac{7}{8},\frac{6}{5},3}
\end{equation}
and has central charge $c=0.7$. The conformal boundary conditions correspond one-to-one with the primaries. The fusion rules can be found, for example, in Ref.~\cite{difrancescoConformalFieldTheory1997}, and a complete list of structure constants is provided in \cite{chengPrecisionReconstructionRational2025}. The spectra of the tricritical Ising CFT in both the disk and pants gauges are shown in Fig.~\ref{fig:triising_disk} and Fig.~\ref{fig:triising_pants}, respectively. We find that the $\mathcal{N}=1$ topological bootstrap yields stable RG flows in both gauges and reproduces the correct spectra.

\begin{figure}[h]
    \includegraphics[width=.4\linewidth]{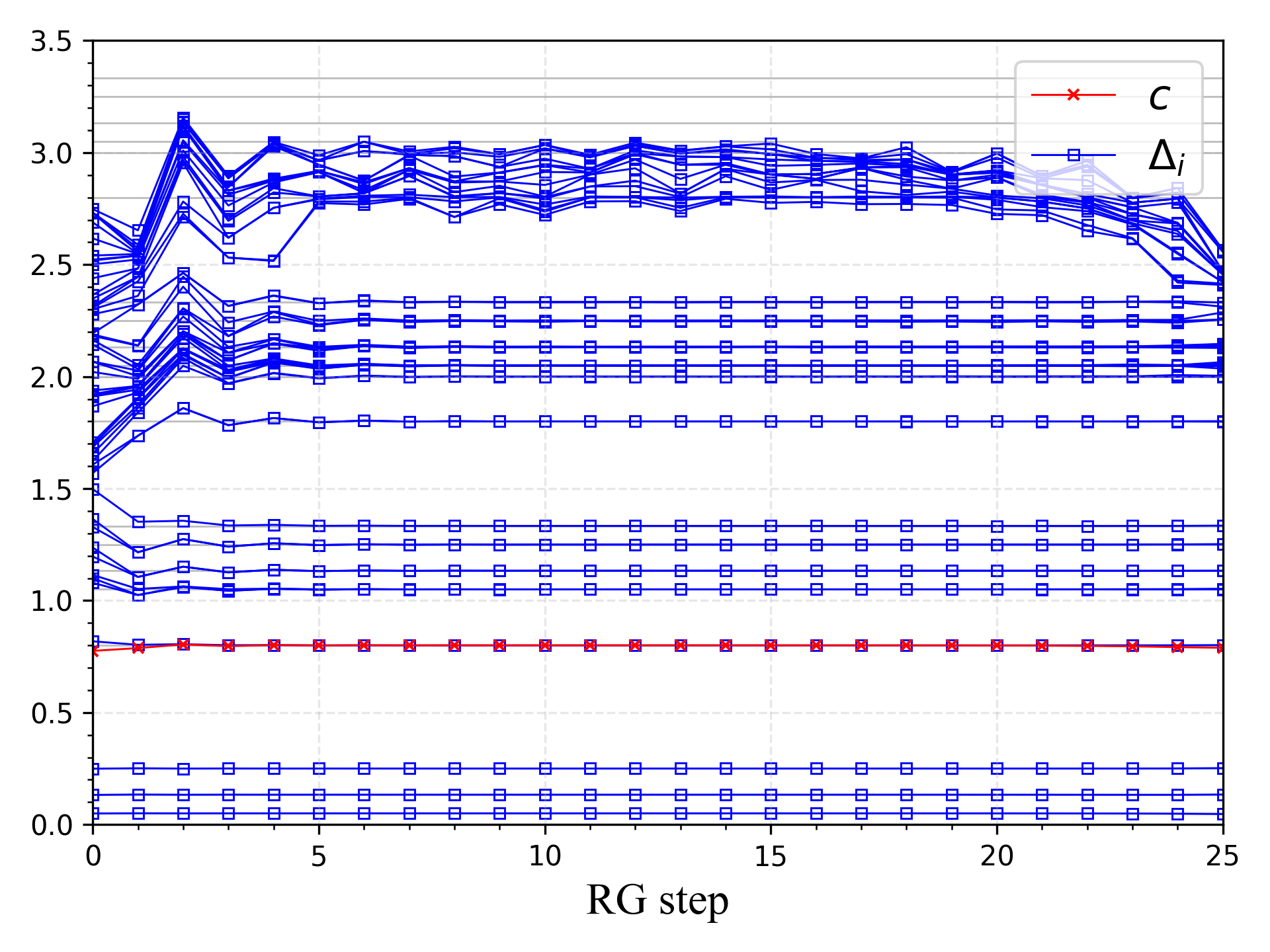}
    \includegraphics[width=.4\linewidth]{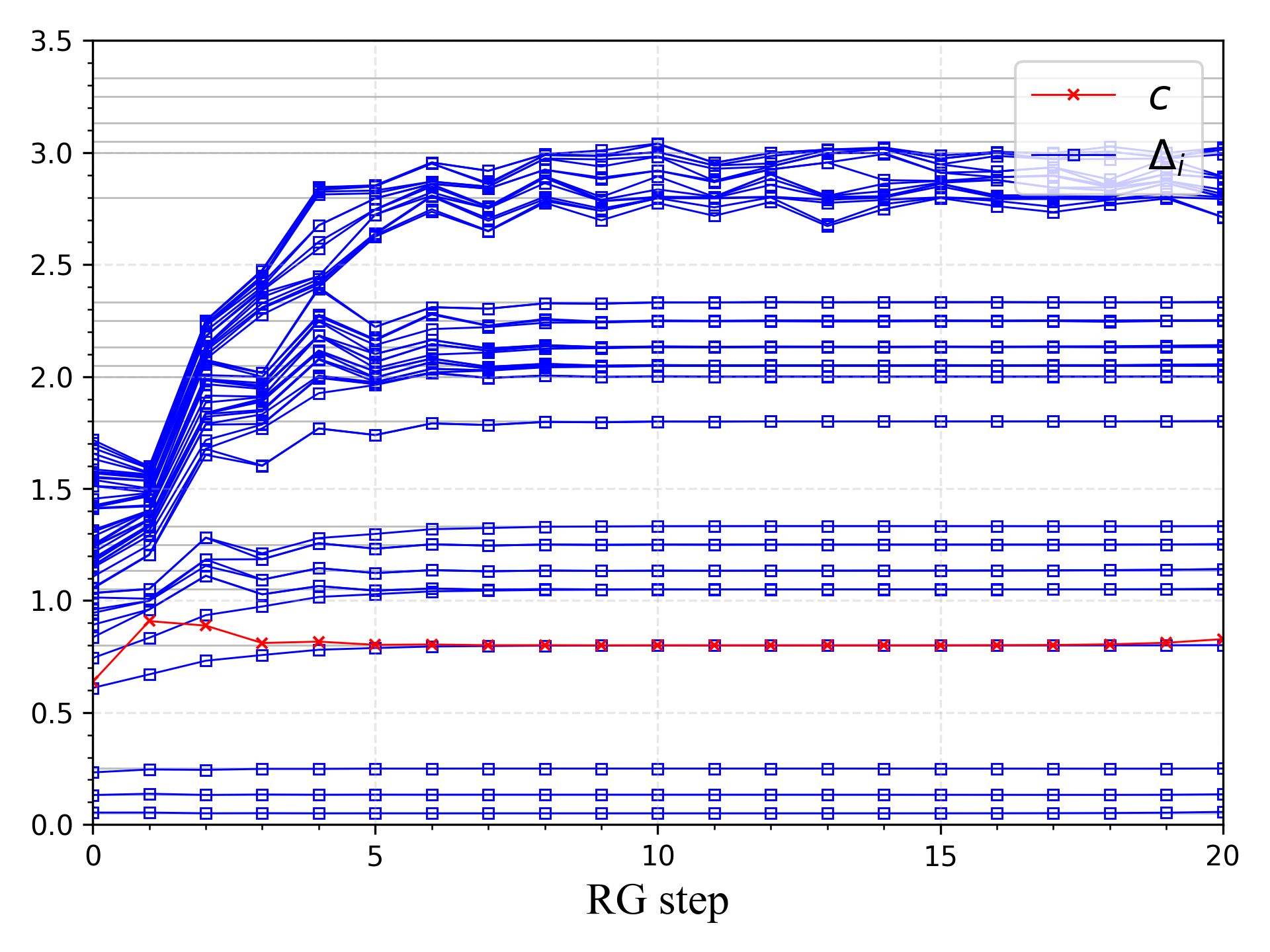}
    \caption{Tetracritical Ising CFT spectrum in the disk gauge, for $\mathcal{N}=\mathcal{N}_{ijk}$ (left) and $\mathcal{N}=1$ (right), $\chi=16$.}
    \label{fig:tetraising_disk}
\end{figure}
\begin{figure}[h]
    \includegraphics[width=.4\linewidth]{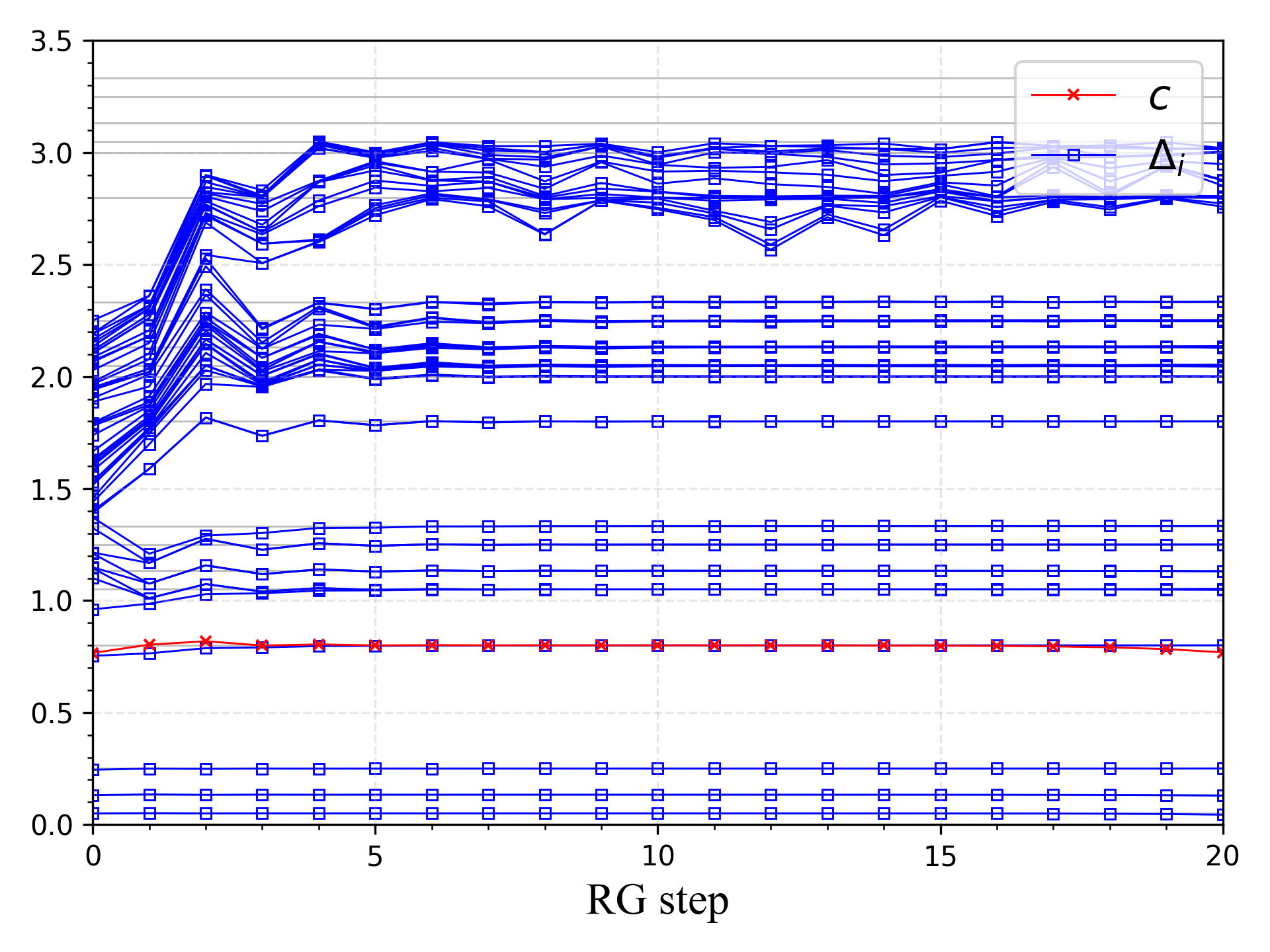}
    \includegraphics[width=.4\linewidth]{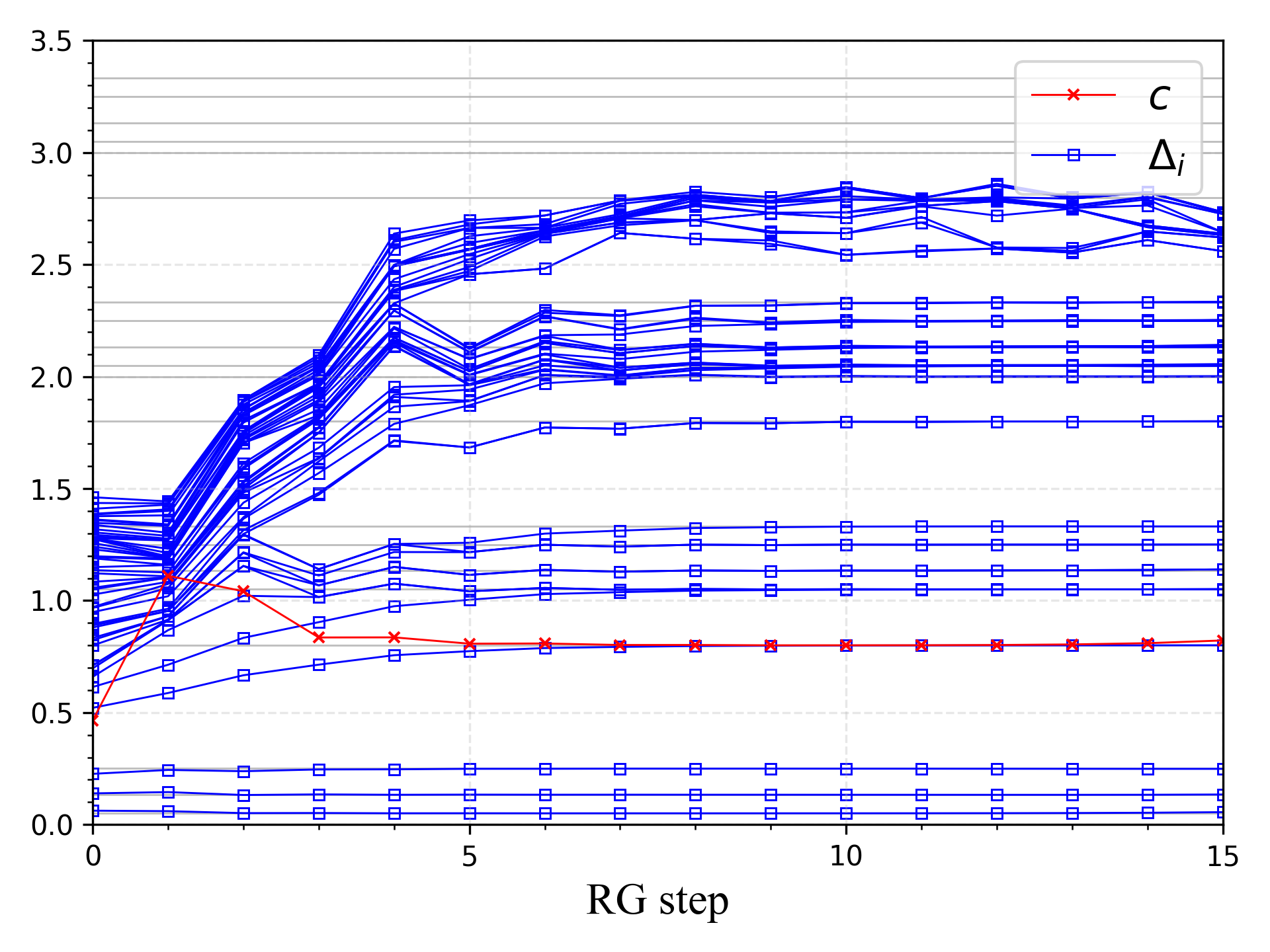}
    \caption{Tetracritical Ising CFT spectrum in the pants gauge, for $\mathcal{N}=\mathcal{N}_{ijk}$ (left) and $\mathcal{N}=1$ (right), $\chi=16$.}
    \label{fig:tetraising_pants}
\end{figure}

The tetracritical Ising CFT has ten primary fields, whose holomorphic conformal dimensions are
\begin{equation}
    h=\Bqty{0,\frac{1}{40},\frac{1}{15},\frac{1}{8},\frac{2}{5},\frac{21}{40},\frac{2}{3},\frac{7}{5},\frac{13}{8},3}
\end{equation}
and it has central charge $c=0.8$. A list of structure constants for this model is provided in the final section of the Supplementary Material. The spectra of the tetracritical Ising CFT in both the disk and pants gauges are shown in Fig.~\ref{fig:tetraising_disk} and Fig.~\ref{fig:tetraising_pants}, respectively. In practice, we find that only a subset of six primaries are needed to obtain stable RG flows. Denoting these six fields by $\{\mathbf{I}, \sigma, \epsilon, Z, X, Y\}$ (following the convention of Ref.~\cite{difrancescoConformalFieldTheory1997}), they have conformal dimensions
\begin{equation}
    h^\prime=\Bqty{0,\frac{1}{15},\frac{2}{5},\frac{2}{3},\frac{7}{5},3}\subset h,
\end{equation}
and they form a fusion subcategory of the full theory. The sufficiency of this reduced set can be understood from their modular data.
Recall that in the FP tensor construction \cite{chengPrecisionReconstructionRational2025}, the hole-shrinking condition is verified by evaluating the weighted sum of Cardy states $\ket{i}_c$:
\begin{equation}
\begin{split}
    \sum_{i\in h'}S_{00}^{1/2}S_{0i}\ket{i}_c=&\sum_{i\in h'}\sum_{j\in h} S_{00}^{1/2}S_{0i}S_{ij}/S_{0j}^{1/2}\ket{j}\rangle\\
    =&\frac{1}{2}\left(\ket{0}\rangle+\ket{3}\rangle\right).
\end{split}
\end{equation}
where $\ket{j}\rangle$ denotes the Ishibashi state. This linear combination indeed creates a shrinkable boundary because the state $\ket{3}\rangle$ produces only an irrelevant operator, so its presence does not obstruct the RG flow towards the fixed point. 
The FP tensors obtained from this reduced set correctly reproduce the scaling dimensions of all ten bulk primary fields of the tetracritical Ising CFT:
\begin{equation}
    \Delta=\Bqty{0,\frac{1}{20},\frac{2}{15},\frac{1}{4},\frac{4}{5},\frac{21}{20},\frac{4}{3},\frac{14}{5},\frac{13}{4},6}.
\end{equation}
We also find that the $\mathcal{N}=1$ topological bootstrap continues to work well in both gauges for this more complicated theory.

\supsection{Conformal spins}

In this section, we describe how to compute the conformal spins from the FP tensors and present the resulting conformal spins for the $\mathcal{N}=1$ topological bootstrap example. 

\begin{figure}[h]
    \centering
    \subfigure[\label{fig:spina1}]{
    \includegraphics[width=0.58\linewidth]{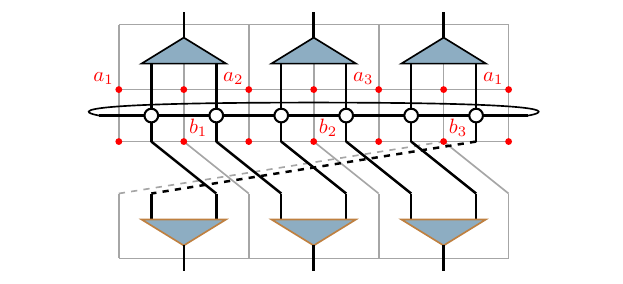}
    }\\[1em]
    \subfigure[\label{fig:spina2}]{
    \includegraphics[width=0.7\linewidth]{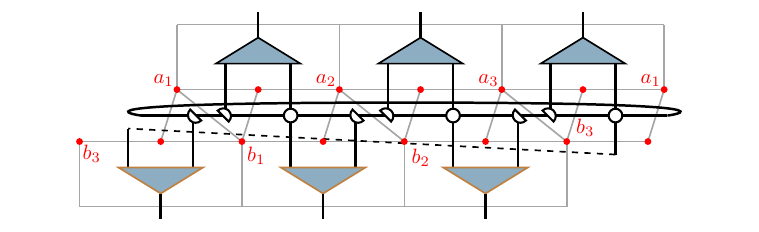}
    }\\[1em]
    \newsavebox{\myboxa}
    \savebox{\myboxa}{
    \includegraphics[width=0.5\linewidth]{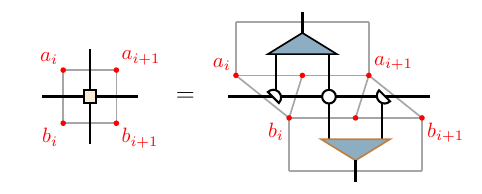}
    }
    \subfigure[\label{fig:spina3}]{
        \raisebox{\dimexpr.5\ht\myboxa-.5\height}{
    \includegraphics[width=0.4\linewidth]{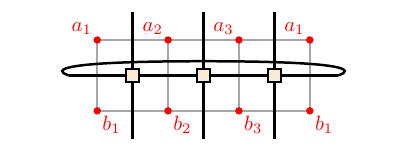}
    }
    }\hspace{0.5em}
    \subfigure[\label{fig:spina4}]{
        \usebox{\myboxa}
    }
    \caption{Constructing a $L=6$ twisted transfer matrix and approximating it by an $L=3$ transfer matrix. Blue triangles (with black or brown borders) denote projectors acting on each pair of double legs, reducing them to single legs with cutoff dimension $\chi$. All unlabeled CBCs are summed, and the CBCs on the two horizontal ends are identified. (a) A block of the $L=6$ twisted transfer matrix with boundary conditions $a_1,a_2,a_3,b_1,b_2,b_3$. (b) SVD is performed on tensors at odd sites, followed by the application of the projectors. (c) The resulting approximate transfer matrix corresponding to (a) now has length three. (d) Each square in (c) represents a new rank-4 TC.}
    \label{fig:spina}
\end{figure}
To extract conformal spins, we require a transfer matrix with a nonzero real part of the modular parameter $\tau$. This can be achieved by introducing a twist in the transfer matrix. Here we follow the method of Ref.~\cite{hauruTopologicalConformalDefects2016}, adapted to TCs. We begin by performing one additional step of Levin-Nave TRG (on TCs, without loop optimization) to render the tensors uniform. As shown in Fig.~\ref{fig:spina1}, we then construct a transfer matrix of length six using these uniform tensors, with the lower legs shifted one site to the right. Since diagonalizing a length-six transfer matrix is computationally demanding, we approximate it by a shorter transfer matrix of length three. This is accomplished by inserting projectors (depicted as blue triangles) on the double legs, which reduce the bond dimension in each block from $\chi^2$ to $\chi$. Next, as illustrated in Fig.~\ref{fig:spina2}, we perform an SVD on the rank-4 tensors located at odd sites and interpret the portion shown in Fig.~\ref{fig:spina4} (with all unlabeled CBCs summed over) as a new rank-4 tensor. The resulting transfer matrix—constructed from these new rank-4 tensors, represented as squares in Fig.~\ref{fig:spina3}—has length three. Its leading eigenvalues $\lambda_i$ can then be efficiently computed using iterative methods such as the Arnoldi algorithm.

\begin{figure}[h]
    \centering
    \newsavebox{\myboxb}
    \savebox{\myboxb}{
    \includegraphics[width=0.5\linewidth]{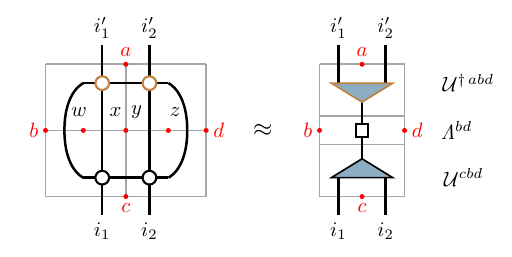}
    }
    \subfigure[\label{fig:spinb1}]{
        \raisebox{\dimexpr.5\ht\myboxb-.5\height}{
        \includegraphics[width=0.24\linewidth]{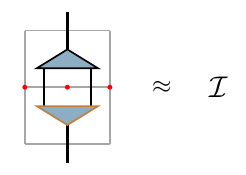}
    }}\hspace{1.5em}
    \subfigure[\label{fig:spinb2}]{
        \usebox{\myboxb}
    }\\[1em]
    \subfigure[\label{fig:spinb3}]{
    \includegraphics[width=0.75\linewidth]{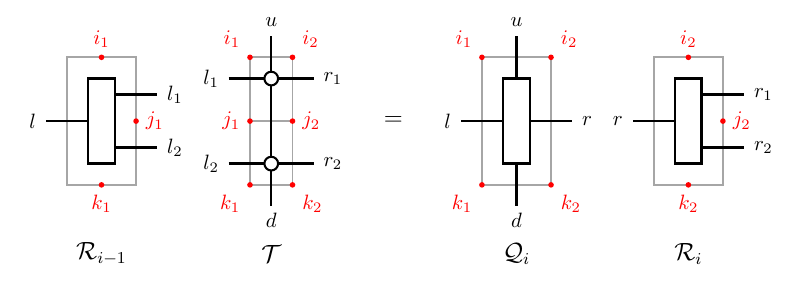}
    }\\[1em]
    \subfigure[\label{fig:spinb4}]{
    \includegraphics[width=0.75\linewidth]{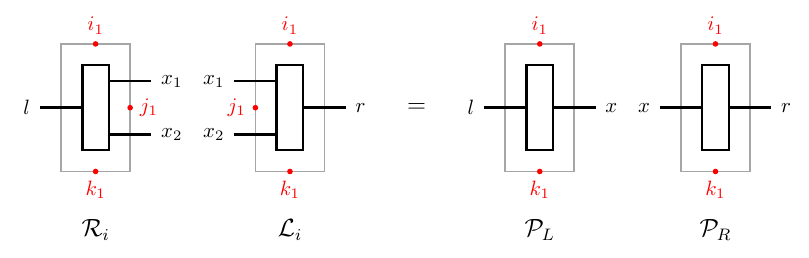}
    }
    \caption{Two methods for constructing projectors. (a) The two projectors must compose to the identity matrix complex $\mathcal{I}$, with the internal CBC summed. (b) Projectors found via HOSVD. All indices $i_1,i_2,i^\prime_1,i^\prime_2,w,x,y,z$ label tensor legs. Tensors in brown circles are complex conjugates of the tensors in black circles. (c) QR decomposition applied to a double-layer tensor $\mathcal{T}$; on the left-hand side, the CBC $j_1$ is summed. (d) Projectors constructed from $\mathcal{R}$ and $\mathcal{L}$; again, the CBC $j_1$ is summed on the left-hand side.}
    \label{fig:spinb}
\end{figure}
The projectors can be obtained either via HOSVD \cite{xieCoarsegrainingRenormalizationHigherorder2012} or by using successive QR/LQ decompositions, as illustrated in Fig.~\ref{fig:spinb}. Concretely, in the HOSVD approach, the rank-3 TC $\mathcal{U}$ (which becomes unitary after reshaping into a matrix complex) for the joint index $[i_1,i_2]$ (see Fig.~\ref{fig:spinb2}) is identical to the rank-3 TC $\mathcal{U}$ in the SVD of the matrix complex $\mathcal{M}([i_1,i_2],[w,x,y,z])$. The blocks $\mathcal{U}^{cbd}$ are obtained from the decomposition
\begin{equation}
    M^{cbd\dots}=U^{cbd}\lambda^{bd} V^{\dagger\,\dots},
\end{equation}
where the ellipsis $\dots$ denotes the unlabeled CBCs on the left-hand side of Fig.~\ref{fig:spinb2}. This is equivalent to finding the unitary matrix in the eigenvalue problem shown in Fig.~\ref{fig:spinb2}, which has substantially lower computational cost. For each fixed set of boundary conditions $\{a,b,c,d\}$, we form a tensor block by contracting all internal legs (and summing over all internal CBCs):
\begin{equation}
    N^{abcd}\equiv(MM^\dagger)^{abcd}\equiv M^{cbd}(M^{abd})^\dagger.
\end{equation}
This eigenvalue problem is solved exactly as in the SVD procedure described earlier: we fix $b,d$ and group over $a,c$. After grouping $a,c$, the resulting rank-4 TC $\mathcal{N}$ can be regarded as a matrix complex, and it is Hermitian. Thus, we obtain
\begin{equation}\label{eq:hosvdu}
    N^{abcd}=U^{cbd}\Lambda^{bd}U^{\dagger\,abd}\quad\text{with}\quad U^{\dagger\,abd}=(U^{abd})^\dagger,
\end{equation}
where $\Lambda^{bd}=(\lambda^{bd})^2$. Restoring the index $c$ then yields a rank-3 TC $\mathcal{U}$ whose blocks $\mathcal{U}^{cbd}$ are given by Eq.~\eqref{eq:hosvdu}. The projector with the black border is simply $\mathcal{U}$, while the projector with the brown border is its Hermitian conjugate $\mathcal{U}^\dagger$.

An alternative general method for determining the projectors is to use successive QR/LQ decompositions applied to double-layer TCs. The QR decomposition procedure is illustrated in Fig.~\ref{fig:spinb3}. Since the tensors are uniform, we suppress the site index $i$ in $\mathcal{T}$, and the index $i$ can in fact be viewed as the label for the iteration step of the QR/LQ process. On the left-hand side of Fig.~\ref{fig:spinb3}, we first left-multiply $\mathcal{T}$ by another matrix complex $\mathcal{R}_{i-1}$, summing over the internal CBC $j_1$. During this multiplication, each block of $\mathcal{R}_{i-1}$ is reshaped into the matrix $R_{i-1}(l,[l_1,l_2])$, while each block of $\mathcal{T}$ is reshaped into $T([l_1,l_2],[u,r_1,r_2,d])$. The product yields a new TC $\mathcal{T}^\prime$ with blocks $\mathcal{T}^{\prime\,i_1k_1i_2j_2k_2}(l,[u,r_1,r_2,d])$. To perform the QR decomposition, we reshape each block
\begin{equation}
    T^{\prime\,i_1k_1i_2j_2k_2}(l,[u,r_1,r_2,d])\rightarrow T^{\prime\prime\,i_1k_1i_2j_2k_2}([u,l,d],[r_1,r_2]).
\end{equation}
For each fixed pair $(i_2,k_2)$, we then construct large block matrices $T^{\prime\prime\,i_2 k_2}$ by grouping the CBCs $(i_1,k_1)$ along the row index and $j_2$ along the column index. QR decomposition of these block matrices gives
\begin{equation}
    \mathcal{R}_{i-1}\mathcal{T}=\mathcal{Q}_i\mathcal{R}_i.
\end{equation}
The LQ decomposition proceeds analogously, yielding
\begin{equation}
    \mathcal{T}\mathcal{L}_{i-1}=\mathcal{L}_i\mathcal{Q}_i.
\end{equation}
With $\mathcal{R}_0=\mathcal{L}_0=\mathcal{I}$, the projectors are then constructed as in Fig.~\ref{fig:spinb4}, using Eq.~\eqref{eq:RLSVD} and \eqref{eq:PLPR}. We note that HOSVD works efficiently for all unitary models, while for non-unitary models the successive QR/LQ method with a small number of iterations ($i=1\text{–}2$) is more reliable.
\begin{figure}[h]
    \centering
    \includegraphics[width=.4\linewidth]{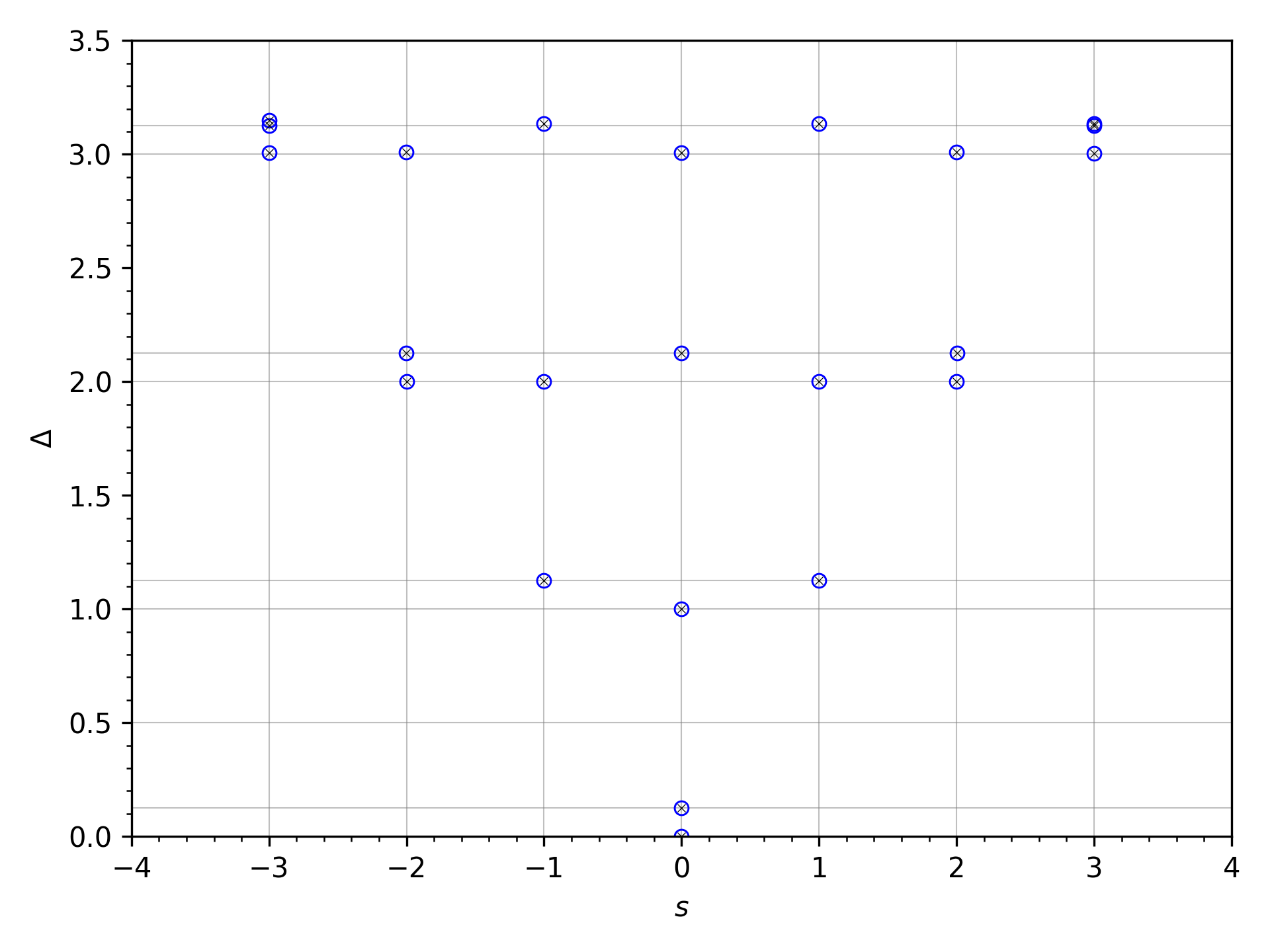}
    \includegraphics[width=.4\linewidth]{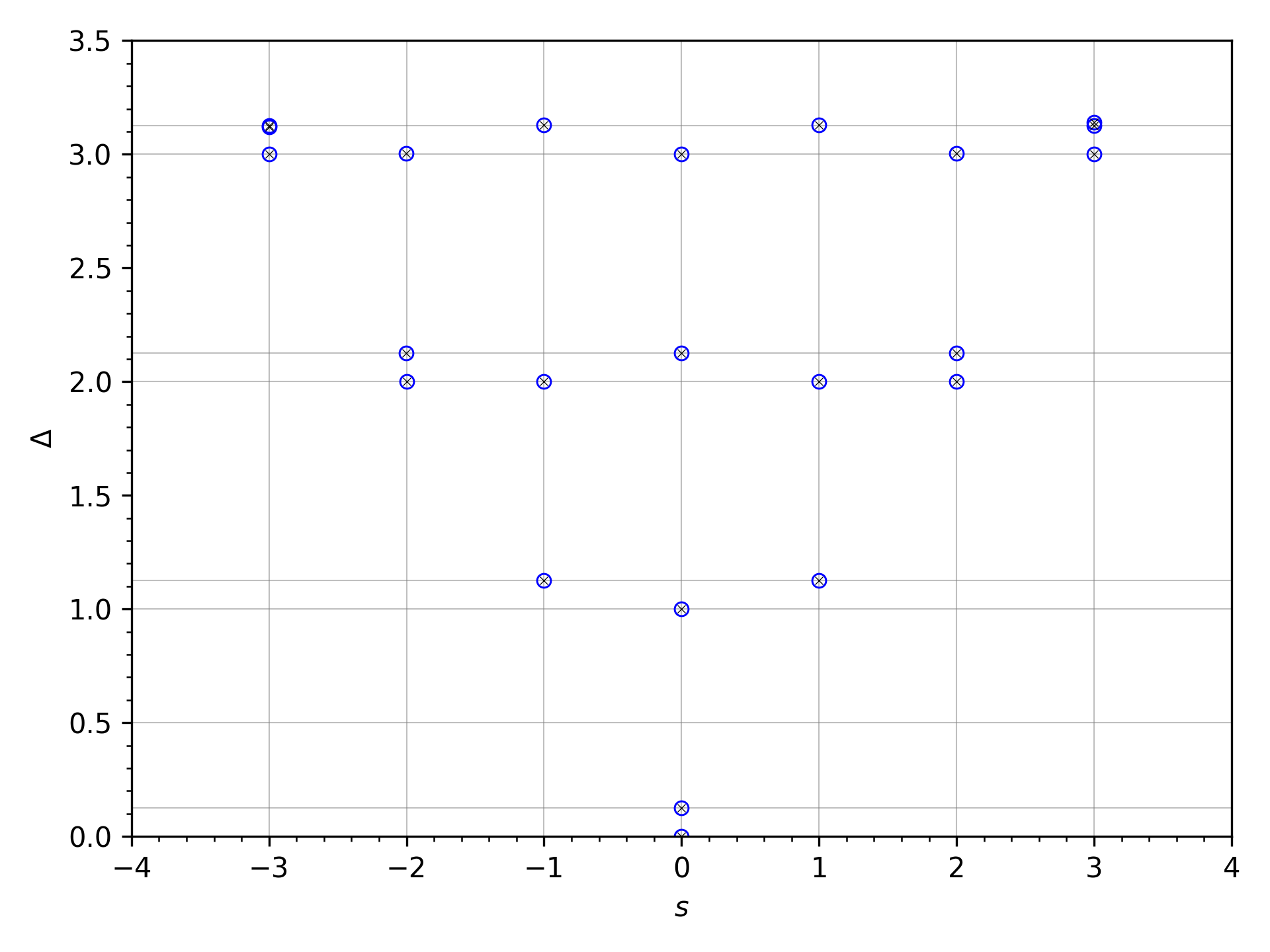}
    \caption{Ising CFT conformal spins for $\mathcal{N}=1$, $\chi=16$. Left: step 20 in the disk gauge (shown in the main text).
    Right: step 20 in the pants gauge (right panel of Fig.~\ref{fig:ising_pants}).
    }
    \label{fig:ising_N=1_spin}
\end{figure}
\begin{figure}[h]
    \centering
    \includegraphics[width=.4\linewidth]{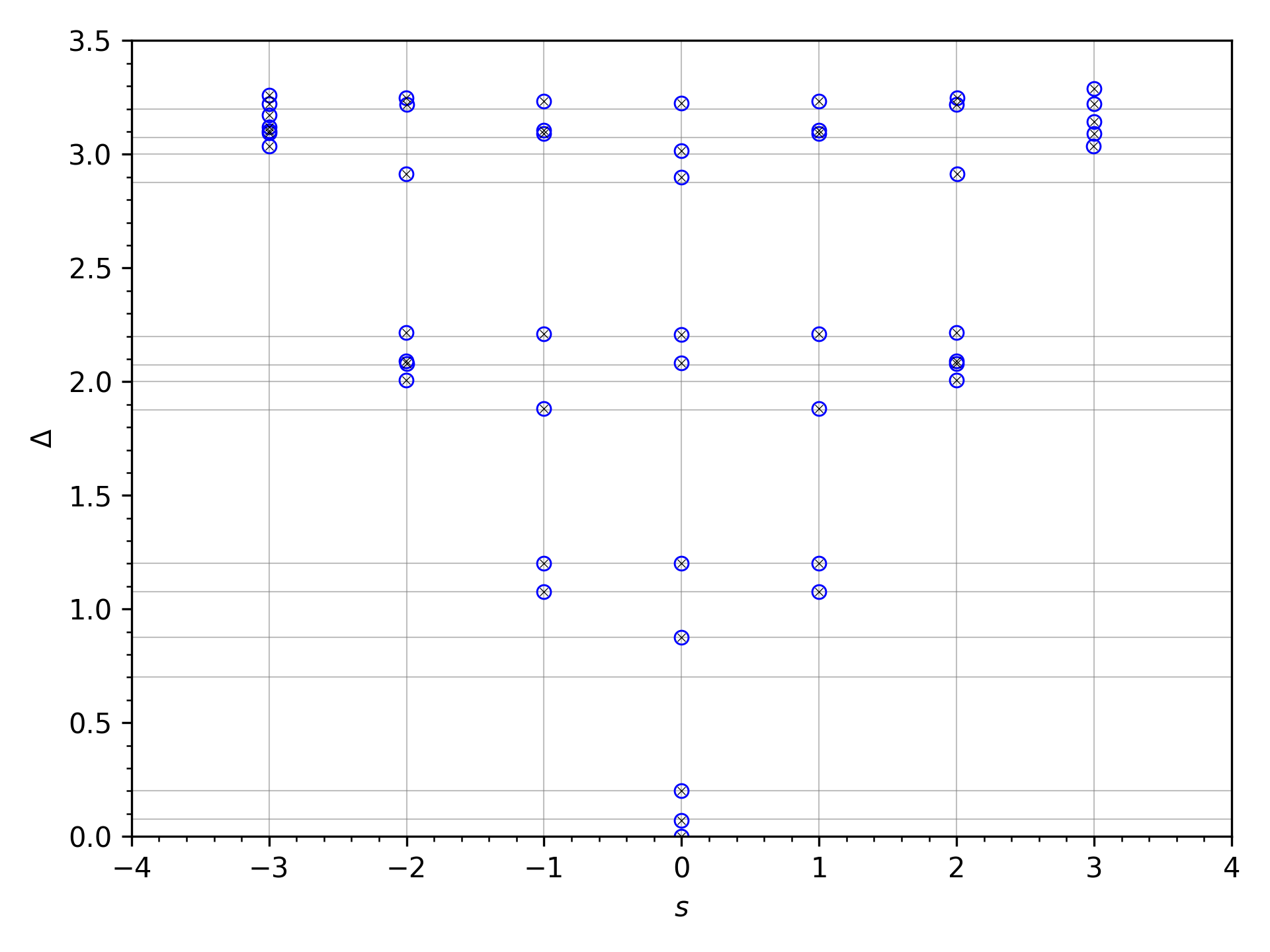}
    \includegraphics[width=.4\linewidth]{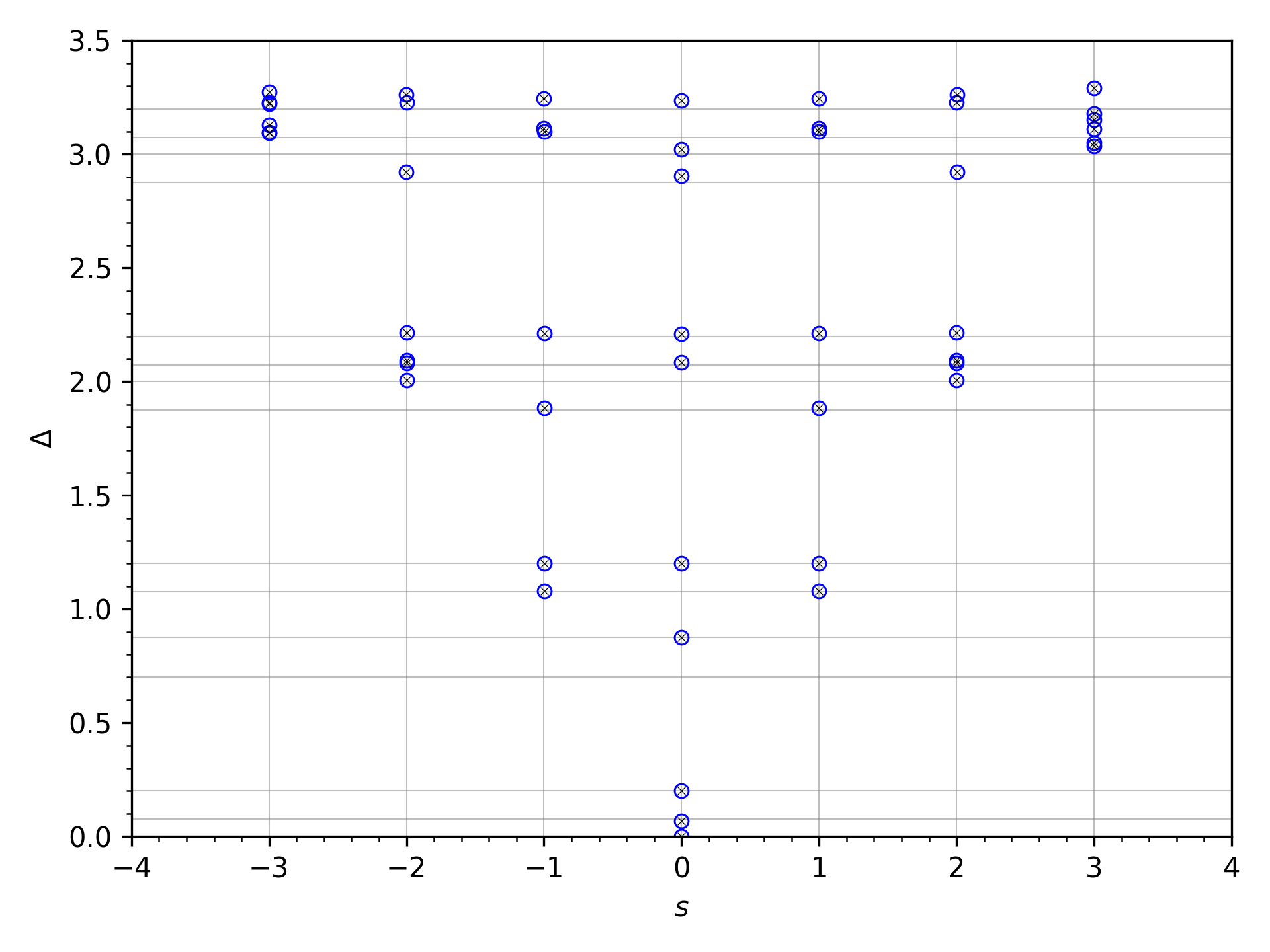}
    \caption{Tricritical Ising CFT conformal spins for $\mathcal{N}=1$, $\chi=16$. Left: step 15 in the disk gauge (right panel of Fig.~\ref{fig:triising_disk}).
    Right: step 13 in the pants gauge (right panel of Fig.~\ref{fig:triising_pants}).
    }
    \label{fig:triising_N=1_spin}
\end{figure}
\begin{figure}[h]
    \centering
    \includegraphics[width=.4\linewidth]{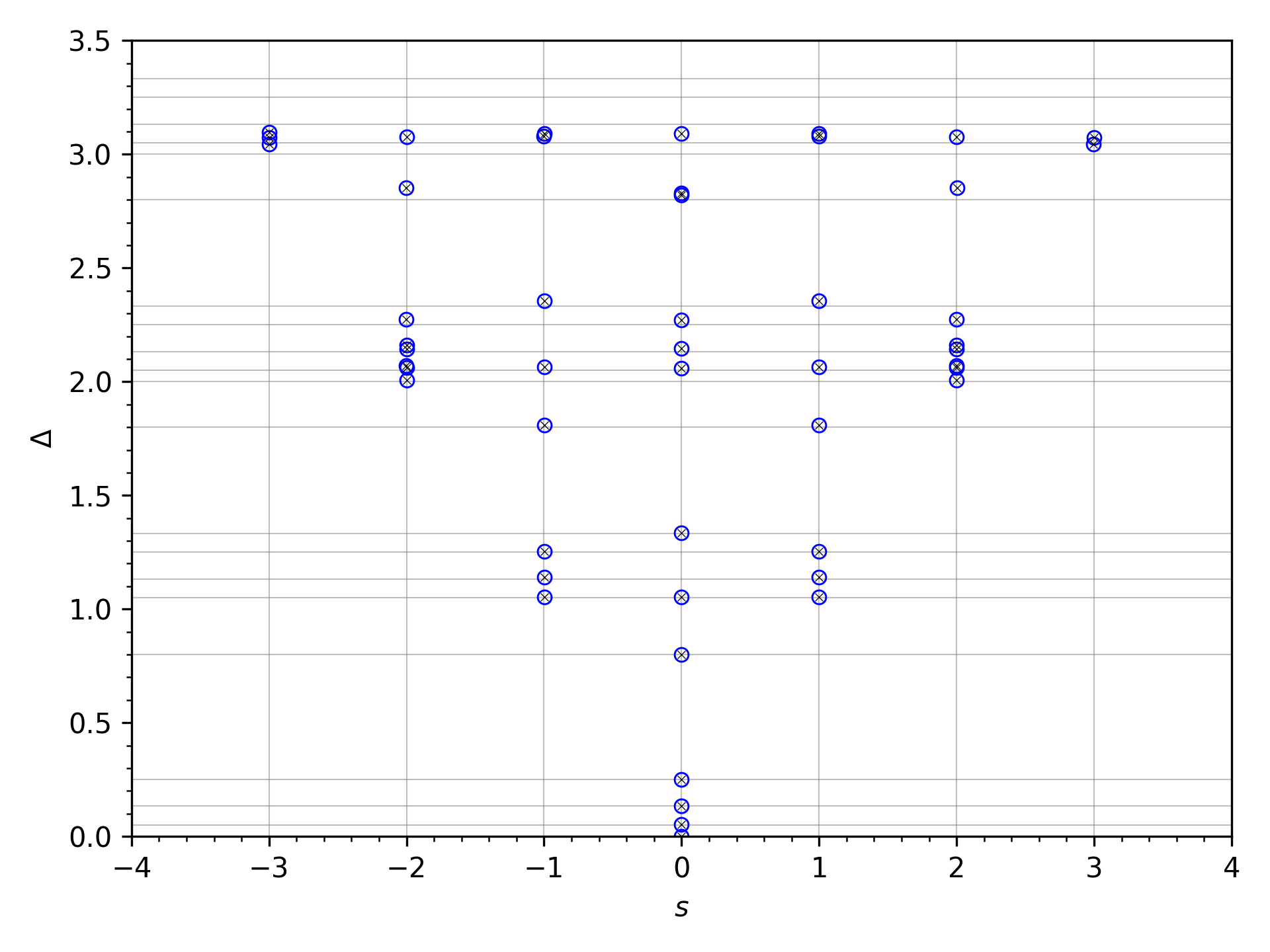}
    \includegraphics[width=.4\linewidth]{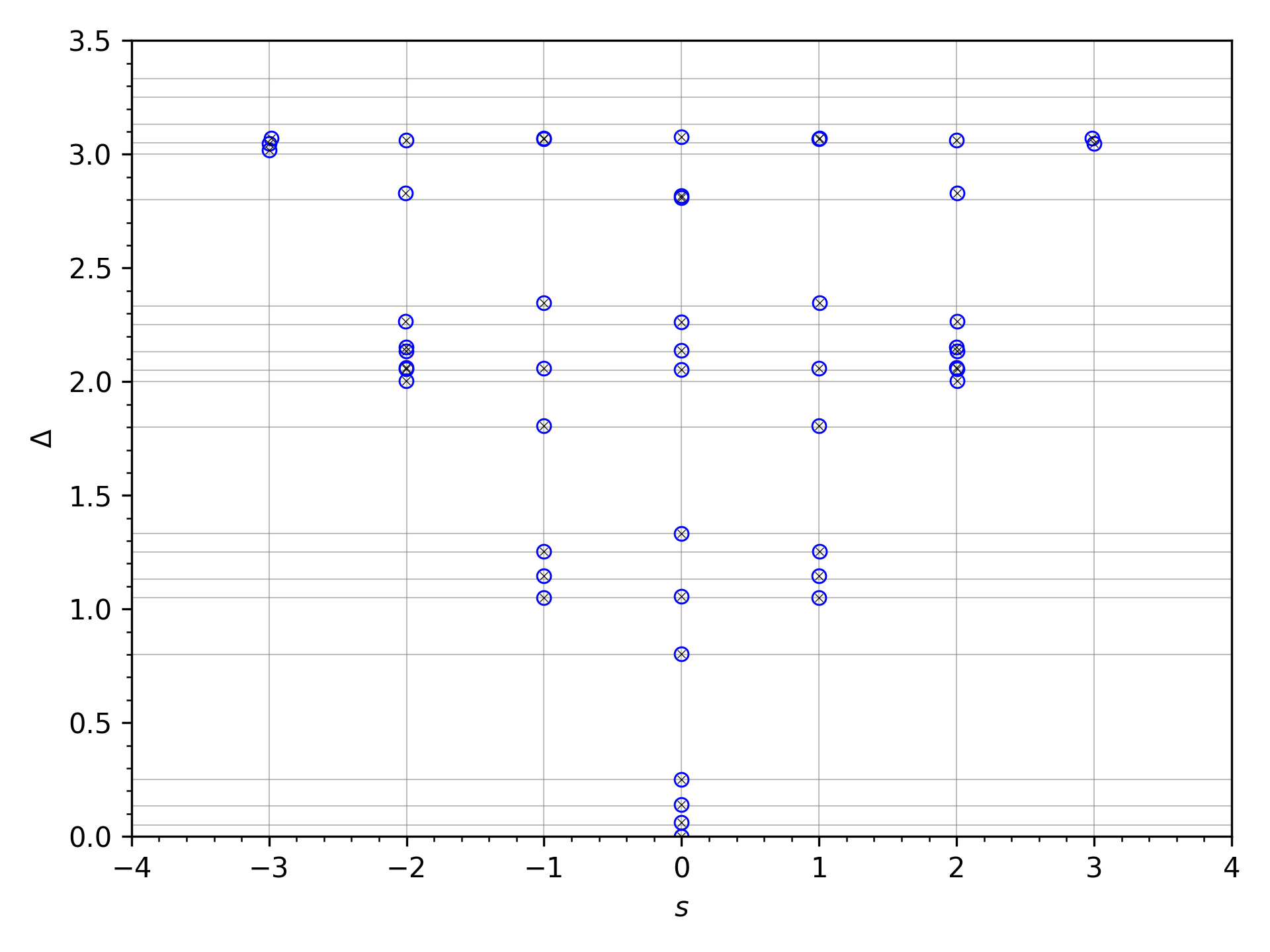}
    \caption{Tetracritical Ising CFT conformal spins for $\mathcal{N}=1$, $\chi=16$. Left: step 15 in the disk gauge (right panel of Fig.~\ref{fig:tetraising_disk}). 
    Right: step 12 in the pants gauge (right panel of Fig.~\ref{fig:tetraising_pants}).
    }
    \label{fig:tetraising_N=1_spin}
\end{figure}
\begin{figure}[h]
    \centering
    \includegraphics[width=.4\linewidth]{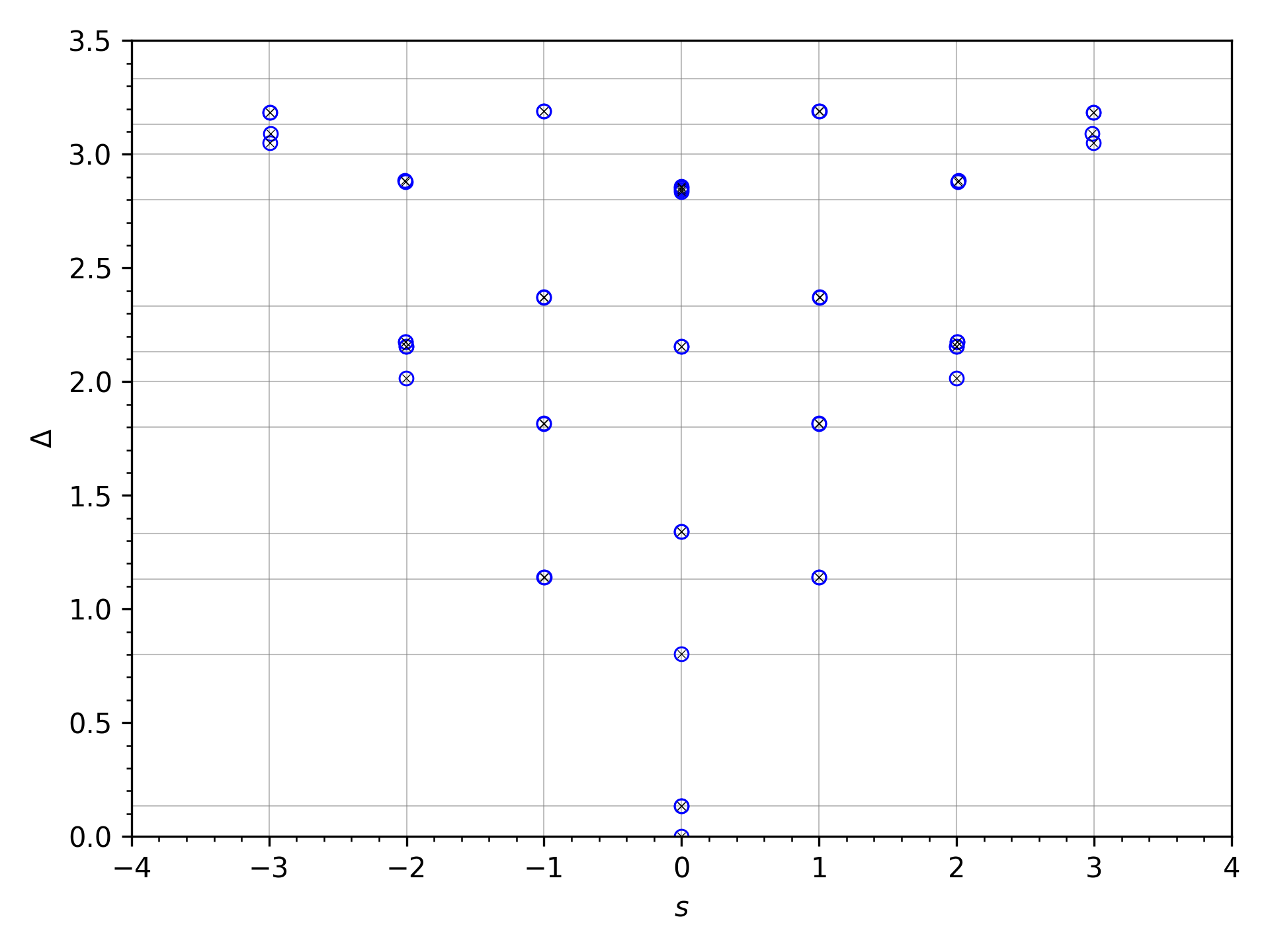}
    \includegraphics[width=.4\linewidth]{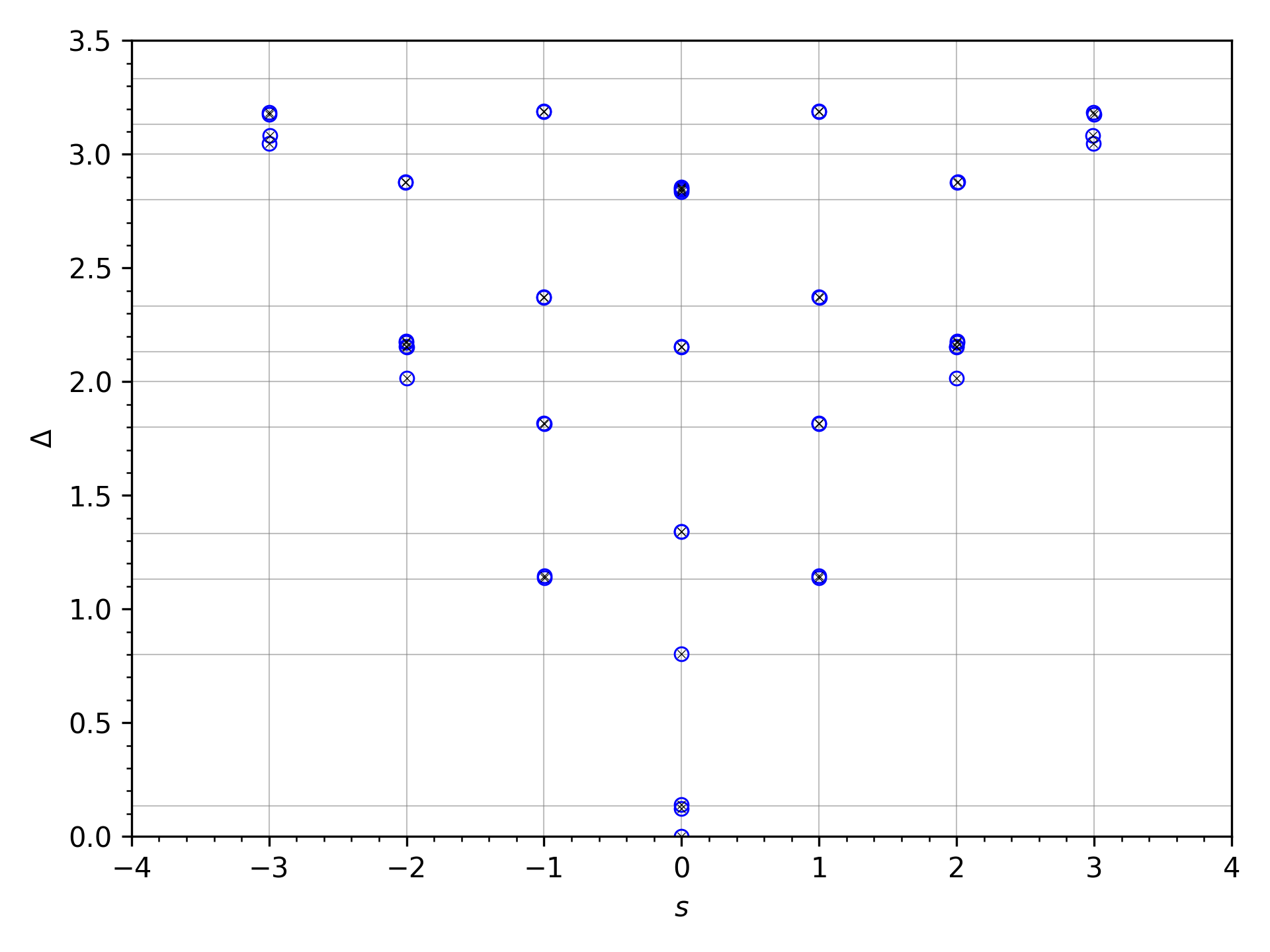}
    \caption{3-state Potts CFT conformal spins for $\mathcal{N}=1$, $\chi=16$. Left: step 15 in the disk gauge (shown in the main text). 
    Right: step 13 in the pants gauge (right panel of Fig.~\ref{fig:3potts_pants}).
    }
    \label{fig:3potts_N=1_spin}
\end{figure}

The transfer matrix shown in Fig.~\ref{fig:spina} has modular parameter $\tau=(1+i)/6$, so $\Re(\tau)=1/6$. The conformal spins $s_i$ (defined modulo $1/\Re(\tau)=6$) are therefore extracted via
\begin{equation}
    s_i=\frac{\Im(\ln\lambda_i)}{2\pi\Re(\tau)}=\frac{3\Im(\ln\lambda_i)}{\pi}.
\end{equation}

The conformal-spin spectra for the Ising, tricritical Ising, tetracritical Ising, and 3-state Potts CFTs in the $\mathcal{N}=1$ topological bootstrap examples are shown in Fig.~\ref{fig:ising_N=1_spin}--\ref{fig:3potts_N=1_spin}, presented in both the disk and pants gauges.
All projectors are obtained using HOSVD. In every case, the resulting spectra exhibit the correct spin values and degeneracies expected from the corresponding CFTs. This provides strong additional evidence that the FP tensors produced by the $\mathcal{N}=1$ topological bootstrap procedure faithfully realize the genuine FP tensors of the underlying CFTs.

\supsection{Non-unitary CFT}

Finally, we examine the simplest non-unitary case by applying TCR algorithm to the Yang–Lee CFT $\mathcal{M}(5,2)$. This theory contains two primary operators $\Bqty{1,\tau}$, with scaling dimensions
\begin{equation}
    \Delta=\Bqty{0,-\frac{2}{5}},
\end{equation}
and a negative central charge $c=-22/5$. In practical calculations, it is convenient to shift the lowest scaling dimension $\Delta_0=-2/5$ to zero. Using the eigenvalue expression in Eq.~\eqref{eq:eigenvalues_transfermatrix}—which assumes the lowest scaling dimension is treated as $\Delta_0=0$—we extract the effective scaling dimensions $\Delta_{\text{eff}}$ and effective central charge $c_{\text{eff}}$:
\begin{equation}
    \Delta_{\text{eff}}=\Delta_i-\Delta_0\qand \frac{c_{\text{eff}}}{12}=\frac{c}{12}-\Delta_0.
\end{equation}
For the two primaries $\Bqty{1,\tau}$, this yields the effective spectrum
\begin{equation}
    \Delta_{\text{eff}}=\Bqty{\frac{2}{5},0},
\end{equation}
with effective central charge $c_{\text{eff}}=0.4$. The only nontrivial fusion rule of the model is $\tau\times\tau=1+\tau$. The conformal boundary conditions are likewise labeled by $\Bqty{1,\tau}$. The corresponding quantum dimensions are
\begin{equation}
    d_1=1,\qquad d_\tau=-\frac{1}{p}\quad (p=\frac{1+\sqrt{5}}{2}).
\end{equation}
The sole nontrivial bulk OPE coefficient is $C^{\text{bulk}}_{\tau\tau\tau}\approx 1.91131i$ \cite{cardyConformalInvarianceYangLee1985}. The quantum $6j$-symbols for Yang-Lee CFT are listed below:
\begin{equation}
    \bmqty{1&1&1\\1&1&1}=1,\quad\bmqty{\tau&\tau&1\\1&1&\tau}=-\sqrt{p}i,\quad \bmqty{\tau&\tau&1\\\tau&\tau&1}=\bmqty{\tau&\tau&\tau\\\tau&\tau&1}=-p,\quad\bmqty{\tau&\tau&\tau\\\tau&\tau&\tau}=-1-p.
\end{equation}
Using these data, the structure constants $C^{abc}_{ijk}$ are computed using Eq.~\eqref{eq:C},\eqref{eq:hatC},and \eqref{eq:N}.

\begin{figure}[h]
    \centering
    \newsavebox{\myboxc}
    \savebox{\myboxc}{
    \includegraphics[width=0.2\linewidth]{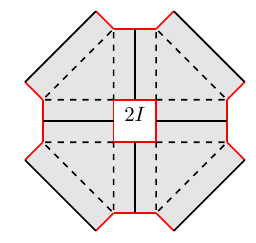}
    }
    \subfigure[\label{fig:decay1}]{
    \raisebox{\dimexpr.5\ht\myboxc-.5\height}{
    \includegraphics[width=0.16\linewidth]{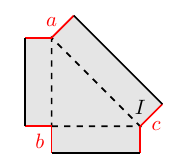}
    }}\hspace{5em}
    \subfigure[\label{fig:decay2}]{
        \usebox{\myboxc}
    }
    \caption{(a) Adding a small exponential factor $\exp(-Ih_i)$ to each component on each tensor leg in the pants gauge. This corresponds to evolving the state over the shaded rectangular region. (b) The decay length $I$ controls the effective size of the hole in the coarse-graining step.}
\end{figure}

In the pants gauge with input $\mathcal{N}=\mathcal{N}_{ijk}$, a slight modification of the tensor is required to obtain a stable spectrum. Specifically, we introduce a small exponential factor $\exp(-Ih_i)$ on each tensor leg, where $h_i$ is the holomorphic conformal dimension of the corresponding BCO. Physically, this corresponds to allowing the state on each edge of the triangle to evolve over a narrow rectangular strip (see Fig.~\ref{fig:decay1}). The parameter $I$ therefore controls the effective ``hole size" introduced in the coarse-graining step, as illustrated in Fig.~\ref{fig:decay2}. Concretely, for a tensor leg specified by CBC $(a,b)$, we multiply each block associated with that pair by the factor $\exp(-Ih_i)$, where $i$ labels the BCO types permitted by $(a,b)$. We find that choosing $I$ not too small leads to a stable RG flow for inputs $\mathcal{N}=\mathcal{N}_{ijk}$ in the pants gauge.

\begin{figure}[h]
    \centering
    \includegraphics[width=.4\linewidth]{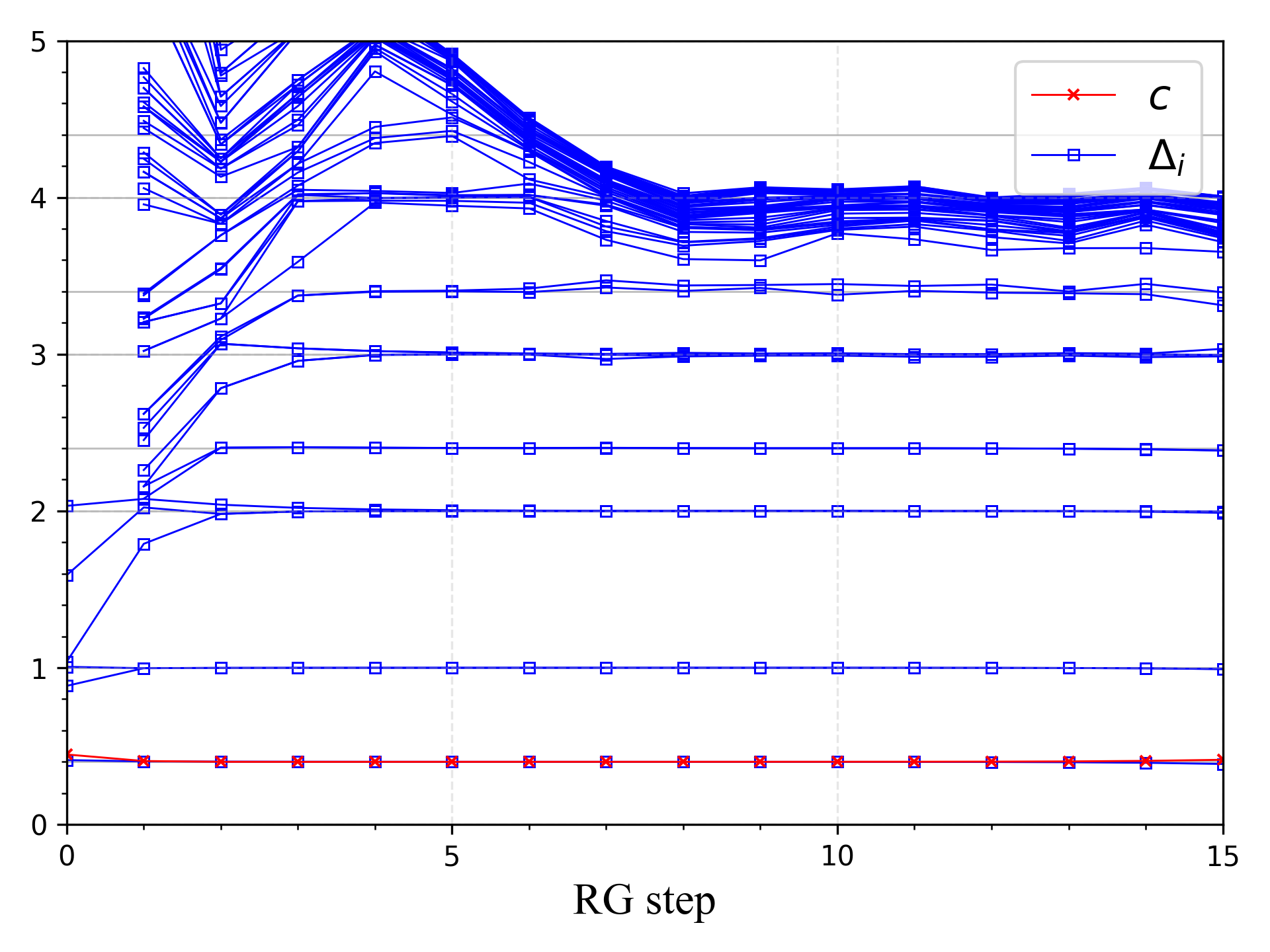}
    \includegraphics[width=.4\linewidth]{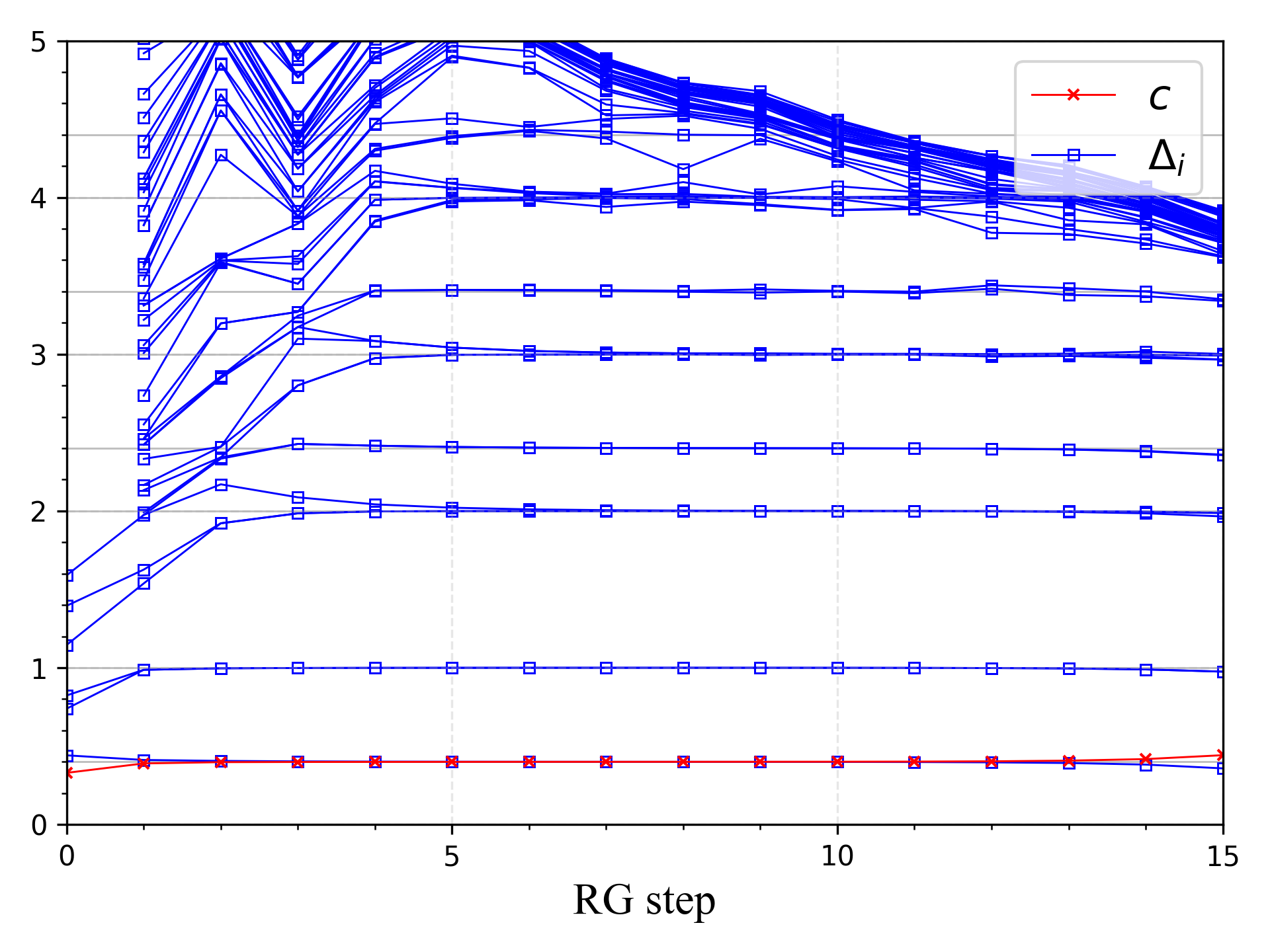}
    \caption{Yang-Lee CFT spectrum in the disk gauge, for $\mathcal{N}=\mathcal{N}_{ijk}$ (left) and $\mathcal{N}=1$ (right), $\chi=24$, obtained using $2\times 4$ transfer matrices.}
    \label{fig:yanglee_disk}
\end{figure}

\begin{figure}[h!]
    \centering
    \includegraphics[width=.4\linewidth]{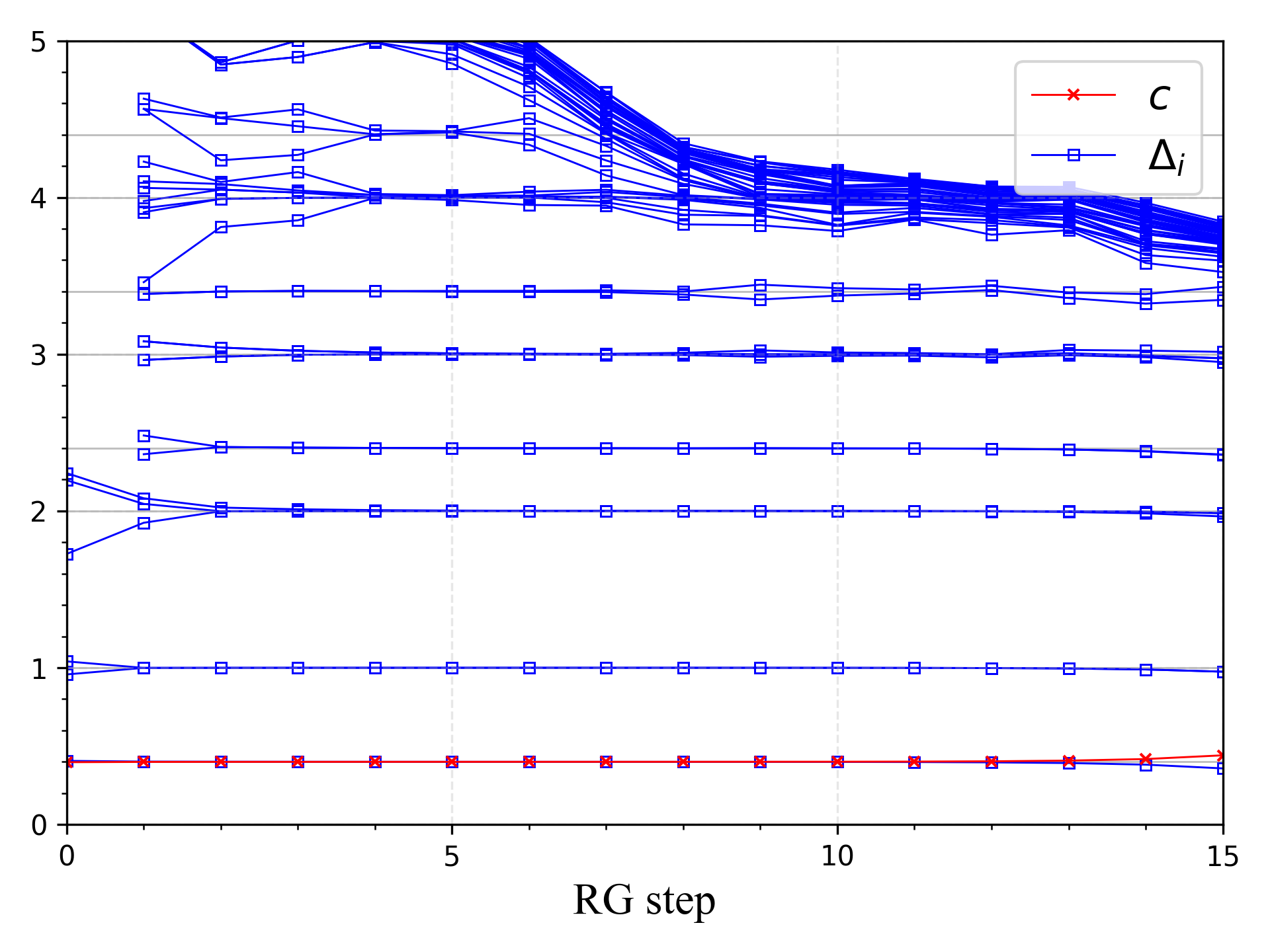}
    \includegraphics[width=.4\linewidth]{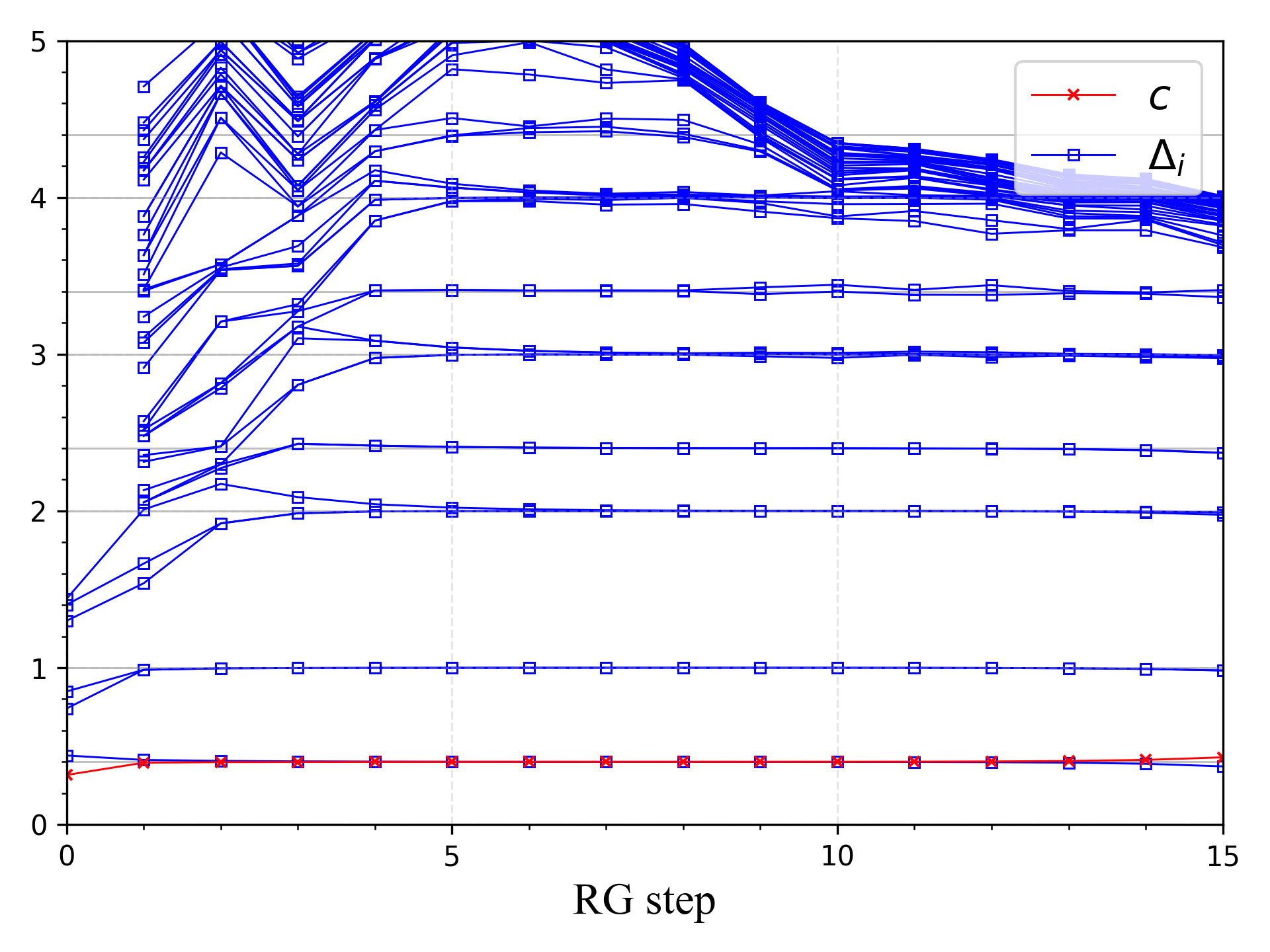}
    \caption{Yang-Lee CFT spectrum in the pants gauge, for $\mathcal{N}=\mathcal{N}_{ijk}$ (left) and $\mathcal{N}=1$ (right), $\chi=24$, obtained using $2\times 4$ transfer matrices. An exponential decay factor $\exp(-Ih_i)$ is applied to each component on each tensor leg at initialization, as described in the text. For $\mathcal{N}=\mathcal{N}_{ijk}$, we use $I=0.8$.}
    \label{fig:yanglee_pants}
\end{figure}

The spectra in the disk and pants gauges are shown in Fig.~\ref{fig:yanglee_disk} and Fig.~\ref{fig:yanglee_pants}, respectively. All spectra are obtained using $2\times 4$ transfer matrices. These results confirm that the $\mathcal{N}=1$ topological bootstrap continues to function reliably even for this non-unitary model.
\begin{figure}
    \centering
    \includegraphics[width=0.65\linewidth]{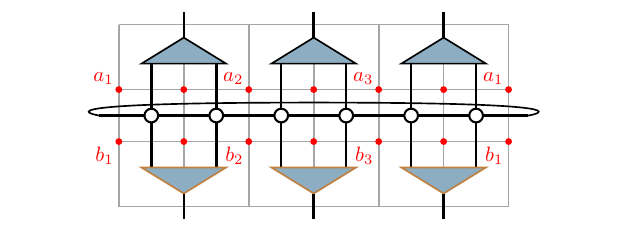}
    \caption{An $L=6$ untwisted transfer matrix using the same projectors as in the twisted case.}
    \label{fig:spinc}
\end{figure}

\begin{figure}
    \centering
    \includegraphics[width=.4\linewidth]{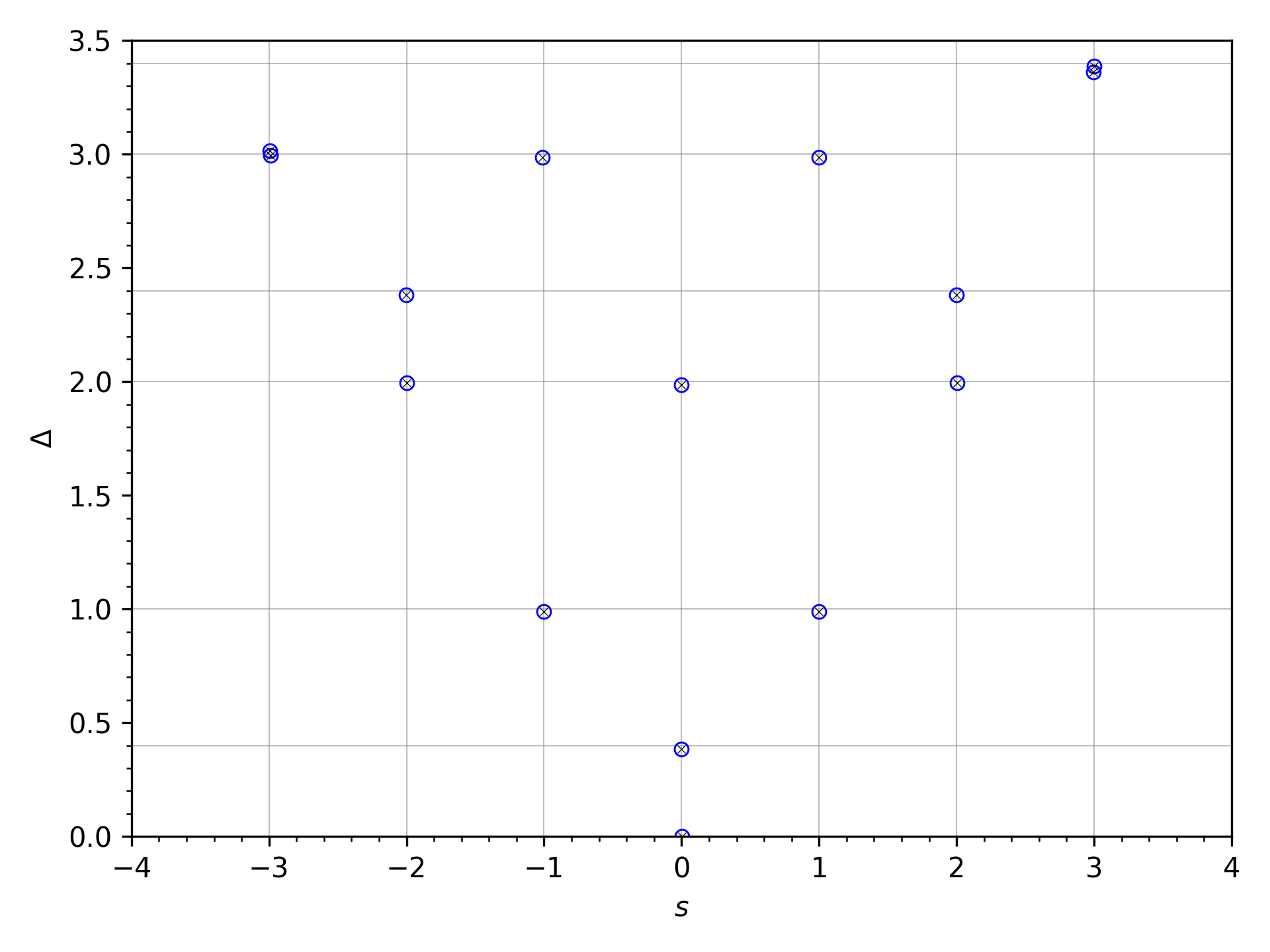}
    \includegraphics[width=.4\linewidth]{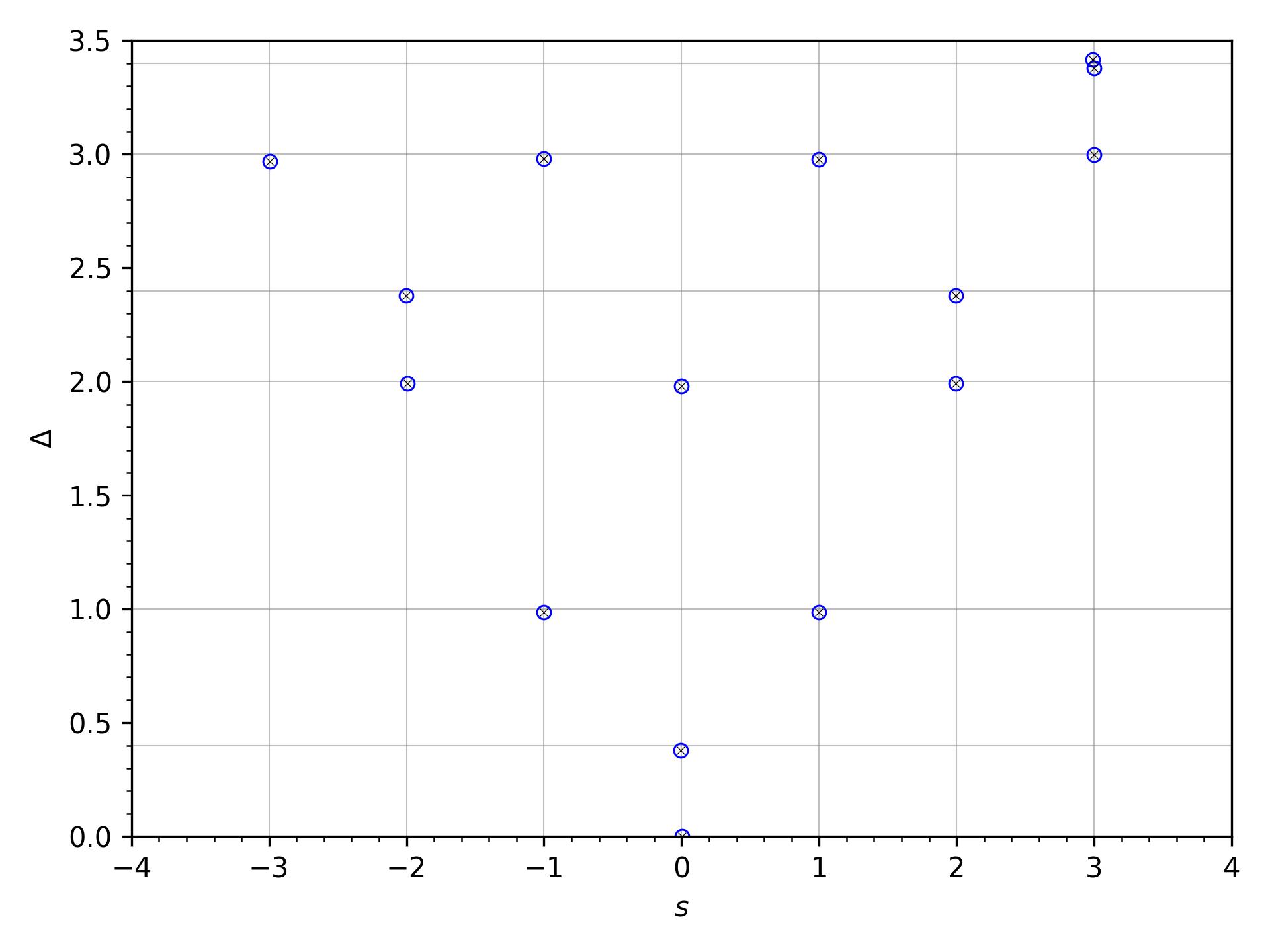}
    \caption{Yang-Lee CFT conformal spins for $\mathcal{N}=1$, $\chi=24$. Left: step 10 in the disk gauge (right panel of Fig.~\ref{fig:yanglee_disk}). Right: step 11 in the pants gauge (right panel of Fig.~\ref{fig:yanglee_pants}).}
    \label{fig:yanglee_spin}
\end{figure}

To extract the conformal spins, we also construct an untwisted $L=6$ transfer matrix (Fig.~\ref{fig:spinc}) using the same projectors as in Fig.~\ref{fig:spina1} for the twisted case. The resulting $L=3$ transfer matrix—serving as an approximation to the original untwisted $L=6$ transfer matrix—can be diagonalized to obtain eigenvalues $\lambda$. Let $\lambda^t$ denote the eigenvalues of the twisted $L=6$ transfer matrix. The conformal spins are then extracted from the imaginary part of $\ln(\lambda^t/\lambda)$, which removes the contribution from the projectors. In this Yang–Lee case, the projectors are obtained via successive QR/LQ decompositions on double-layer tensors, as illustrated in Figs.~\ref{fig:spinb3} and \ref{fig:spinb4}. Consequently, the spins—defined modulo 6—are given by
\begin{equation}
    s_i=\frac{3}{\pi}\Im(\ln\frac{\lambda^t_i}{\lambda_i}).
\end{equation}
The resulting spin spectra for the Yang–Lee CFT in the $\mathcal{N}=1$ topological bootstrap construction are shown in Fig.~\ref{fig:yanglee_spin}. The spins exhibit the correct integer values and the correct degeneracies expected from the theory.

\supsection{Computation of the structure constants}

We summarize below the computation of the boundary structure constants for the minimal models. For the $A$-series minimal models, the conformal boundary conditions are in one-to-one correspondence with the primary fields. Denoting boundary conditions by ${a,b,c}$ and primaries by ${i,j,k}$, the topological structure constants can be expressed as~\cite{FUCHS2005539}
\begin{equation}\label{eq:hatC}
    \hat{C}_{ijk}^{abc}=(d_id_jd_k)^{1/4} \bmqty{i&j&k\\c&a&b}.
\end{equation}
where the bracket denotes the quantum $6j$-symbols, which can be computed using the formulas in Refs.~\cite{Furlan:1989ra,kirillovRepresentationsAlgebraUqsl21989}.
The corresponding CFT structure constants differ from this purely topological quantity by a rescaling factor,
\begin{equation}\label{eq:C}
    C_{ijk}^{abc}=\hat{C}_{ijk}^{abc}/\mathcal{N}_{ijk},
\end{equation}
where the rescaling factors $\mathcal{N}_{ijk}$'s are determined by the bulk OPE coefficients:
\begin{equation}\label{eq:N}
\mathcal{N}_{ijk}=1/\sqrt{C_{ijk}^{\text{bulk}}}.
\end{equation}
The latter can be computed using the general formula of Ref.~\cite{DOTSENKO1985291}. 

The fusion matrix associated with the CFT fusion category is also related to the quantum $6j$-symbol through
\begin{equation}
[F^{ijk}_l]_{mn}=\sqrt{d_md_n} \bmqty{i&j&m\\k&l&n}.
\end{equation}
From this, one obtains the fusion matrix for the conformal blocks,
\begin{equation}
[F^{ijk}_l]_{mn}^{block}=\frac{\mathcal{N}_{ijm}\mathcal{N}_{klm}}{\mathcal{N}_{iln}\mathcal{N}_{jkn}}[F^{ijk}_l]_{mn}.
\end{equation}

For the $D$-series models, the boundary conditions are labeled by $a=(\alpha,\beta)\in(A,D)$, where $\alpha$ and $\beta$ denote the nodes on the respective $A$- and $D$-type Dynkin diagrams~\cite{Behrend_1998}. Using the fusion matrices of the $A$-series model, one can determine the $D$-series boundary structure constants from the crossing equation
\begin{equation}
    \sum_m[F^{ijk}_l]^{\textrm{blocks}}_{mn} C_{ijm}^{abc}C_{mkl}^{acd}=C_{inl}^{abd}C_{jkn}^{bcd},
\end{equation}
following the general algorithm described in Ref.~\cite{Runkel:1999dz}. Note that the BCO's are normalized such that their two point structure constants are equal to one.


We list below the calculated boundary structure constants for the 3-state Potts CFT and the tetracritical Ising CFT.  

For the 3-state Potts CFT, we follow the convention of Ref.~\cite{CARDY1989581} in labeling the conformal boundary conditions as $\{A,B,C,AB,AC,BC\}$, and we follow Ref.~\cite{difrancescoConformalFieldTheory1997} in labeling the six Virasoro primary fields as $\{\mathbf{I}, \sigma, \epsilon, Z, X, Y\}$, with corresponding conformal dimensions $\Bqty{0,\frac{1}{15},\frac{2}{5},\frac{2}{3},\frac{7}{5},3}$. The conformal boundary conditions satisfy the following fusion rules: 
\begin{equation}
\begin{aligned}
    &A\times A= \mathbf{I}+Y, \qquad A\times B= Z, \qquad A\times AB=\sigma, \qquad A\times BC=\epsilon+ X\\
    &AB\times AB=\mathbf{I}+\epsilon+X+Y,\qquad AB\times AC=Z+\sigma. 
\end{aligned}
\end{equation}
The remaining rules can be inferred from the $S_3$ symmetry acting on $\{A, B, C\}$ and $\{AB, AC, BC\}$. The boundary structure constants for the 3-state Potts CFT are listed in Table~\ref{tab:structure}.

For the tetracritical Ising CFT, the structure constants are listed in Table~\ref{tab:tetra}. In this case, both the conformal boundary conditions and the primary fields are labeled by $\{\mathbf{I}, \sigma, \epsilon, Z, X, Y\}$, obeying the following fusion rules,
\begin{equation}
\begin{aligned}
&\epsilon \times \epsilon = \mathbf{I} + X, \qquad
\epsilon \times \sigma = \sigma + Z, \quad
\epsilon \times X = \epsilon + Y,  \quad
\epsilon \times Y = X, \\
&\sigma \times X = \sigma + Z, \quad
\sigma \times Y = \sigma,  \quad
\sigma \times Z = \epsilon + \sigma + X, \quad \epsilon \times Z = \sigma \\
&X \times X = \mathbf{I} + X, \quad
X \times Y = \epsilon,  \quad
X \times Z = \sigma, \quad Z \times Z = \mathbf{I} + Y + Z ,  \\
&Y \times Z = Z, \quad Y \times Y = \mathbf{I}, \quad \sigma \times \sigma = \mathbf{I}  + \epsilon + \sigma + X + Y + Z. 
\end{aligned}
\end{equation}

All listed structure constants are understood to obey permutation symmetry, and entries related by such permutations are therefore omitted.

\begin{table}[h]
\centering
\small
\setlength{\tabcolsep}{2pt}
\renewcommand{\arraystretch}{1.6}
\resizebox{.9\linewidth}{!}{
\begin{tabular}{|l|l|l|l|l|l|l|l|l|l|l|l|}
\hline
$C^{A|A|A}_{III}$ & 1.0 & $C^{A|A|A}_{IYY}$ & 1.0 & $C^{A|A|BC}_{I{\epsilon}{\epsilon}}$ & 1.0 & $C^{A|A|BC}_{IXX}$ & 1.0 & $C^{A|A|BC}_{Y{\epsilon}X}$ & 0.8987 & $C^{A|A|B}_{IZZ}$ & 1.0 \\
\hline
$C^{A|A|B}_{YZZ}$ & 0.5666 & $C^{A|A|C}_{IZZ}$ & 1.0 & $C^{A|A|C}_{YZZ}$ & -0.5666 & $C^{A|A|AC}_{I{\sigma}{\sigma}}$ & 1.0 & $C^{A|A|AC}_{Y{\sigma}{\sigma}}$ & -0.0218 & $C^{A|A|AB}_{I{\sigma}{\sigma}}$ & 1.0 \\
\hline
$C^{A|A|AB}_{Y{\sigma}{\sigma}}$ & 0.0218 & $C^{A|BC|BC}_{{\epsilon}I{\epsilon}}$ & 0.7862 & $C^{A|BC|BC}_{{\epsilon}{\epsilon}X}$ & 0.8579 & $C^{A|BC|BC}_{{\epsilon}X{\epsilon}}$ & 0.8579 & $C^{A|BC|BC}_{{\epsilon}YX}$ & 0.7065 & $C^{A|BC|BC}_{XIX}$ & 0.7862 \\
\hline
$C^{A|BC|BC}_{XXX}$ & -1.2869 & $C^{A|BC|B}_{{\epsilon}{\sigma}Z}$ & 0.8165 & $C^{A|BC|B}_{X{\sigma}Z}$ & 0.6236 & $C^{A|BC|C}_{{\epsilon}{\sigma}Z}$ & 0.8165 & $C^{A|BC|C}_{X{\sigma}Z}$ & -0.6236 & $C^{A|BC|AC}_{{\epsilon}Z{\sigma}}$ & 0.6419 \\
\hline
$C^{A|BC|AC}_{{\epsilon}{\sigma}{\sigma}}$ & 0.6553 & $C^{A|BC|AC}_{XZ{\sigma}}$ & 0.4903 & $C^{A|BC|AC}_{X{\sigma}{\sigma}}$ & -0.143 & $C^{A|BC|AB}_{{\epsilon}Z{\sigma}}$ & 0.6419 & $C^{A|BC|AB}_{{\epsilon}{\sigma}{\sigma}}$ & 0.6553 & $C^{A|BC|AB}_{XZ{\sigma}}$ & -0.4903 \\
\hline
$C^{A|BC|AB}_{X{\sigma}{\sigma}}$ & 0.143 & $C^{A|B|B}_{ZIZ}$ & 1.0 & $C^{A|B|B}_{ZYZ}$ & 0.5666 & $C^{A|B|C}_{ZZZ}$ & 1.1547 & $C^{A|B|AC}_{Z{\epsilon}{\sigma}}$ & 0.8165 & $C^{A|B|AC}_{ZX{\sigma}}$ & 0.6236 \\
\hline
$C^{A|B|AB}_{Z{\sigma}{\sigma}}$ & 0.5774 & $C^{A|C|C}_{ZIZ}$ & 1.0 & $C^{A|C|C}_{ZYZ}$ & 0.5666 & $C^{A|C|AC}_{Z{\sigma}{\sigma}}$ & 0.5774 & $C^{A|C|AB}_{Z{\epsilon}{\sigma}}$ & 0.8165 & $C^{A|C|AB}_{ZX{\sigma}}$ & 0.6236 \\
\hline
$C^{A|AC|AC}_{{\sigma}I{\sigma}}$ & 0.7862 & $C^{A|AC|AC}_{{\sigma}{\epsilon}{\sigma}}$ & 0.6553 & $C^{A|AC|AC}_{{\sigma}X{\sigma}}$ & 0.143 & $C^{A|AC|AC}_{{\sigma}Y{\sigma}}$ & 0.0171 & $C^{A|AC|AB}_{{\sigma}Z{\sigma}}$ & 0.4539 & $C^{A|AC|AB}_{{\sigma}{\sigma}{\sigma}}$ & 0.9267 \\
\hline
$C^{A|AB|AB}_{{\sigma}I{\sigma}}$ & 0.7862 & $C^{A|AB|AB}_{{\sigma}{\epsilon}{\sigma}}$ & 0.6553 & $C^{A|AB|AB}_{{\sigma}X{\sigma}}$ & 0.143 & $C^{A|AB|AB}_{{\sigma}Y{\sigma}}$ & 0.0171 & $C^{BC|BC|BC}_{III}$ & 0.7862 & $C^{BC|BC|BC}_{I{\epsilon}{\epsilon}}$ & 0.7862 \\
\hline
$C^{BC|BC|BC}_{IXX}$ & 0.7862 & $C^{BC|BC|BC}_{IYY}$ & 0.7862 & $C^{BC|BC|BC}_{X{\epsilon}{\epsilon}}$ & -0.5302 & $C^{BC|BC|BC}_{{\epsilon}XY}$ & 0.7065 & $C^{BC|BC|BC}_{XXX}$ & 0.7954 & $C^{BC|BC|B}_{I{\sigma}{\sigma}}$ & 0.7862 \\
\hline
$C^{BC|BC|B}_{{\epsilon}{\sigma}{\sigma}}$ & 0.6553 & $C^{BC|BC|B}_{X{\sigma}{\sigma}}$ & -0.143 & $C^{BC|BC|B}_{Y{\sigma}{\sigma}}$ & -0.0171 & $C^{BC|BC|C}_{I{\sigma}{\sigma}}$ & 0.7862 & $C^{BC|BC|C}_{{\epsilon}{\sigma}{\sigma}}$ & -0.6553 & $C^{BC|BC|C}_{X{\sigma}{\sigma}}$ & -0.143 \\
\hline
$C^{BC|BC|C}_{Y{\sigma}{\sigma}}$ & 0.0171 & $C^{BC|BC|AC}_{IZZ}$ & 0.7862 & $C^{BC|BC|AC}_{I{\sigma}{\sigma}}$ & 0.7862 & $C^{BC|BC|AC}_{{\epsilon}Z{\sigma}}$ & 0.6419 & $C^{BC|BC|AC}_{{\epsilon}{\sigma}{\sigma}}$ & -0.405 & $C^{BC|BC|AC}_{XZ{\sigma}}$ & 0.4903 \\
\hline
$C^{BC|BC|AC}_{X{\sigma}{\sigma}}$ & 0.0884 & $C^{BC|BC|AC}_{YZZ}$ & 0.4454 & $C^{BC|BC|AC}_{Y{\sigma}{\sigma}}$ & -0.0171 & $C^{BC|BC|AB}_{IZZ}$ & 0.7862 & $C^{BC|BC|AB}_{I{\sigma}{\sigma}}$ & 0.7862 & $C^{BC|BC|AB}_{{\epsilon}Z{\sigma}}$ & -0.6419 \\
\hline
$C^{BC|BC|AB}_{{\epsilon}{\sigma}{\sigma}}$ & 0.405 & $C^{BC|BC|AB}_{XZ{\sigma}}$ & 0.4903 & $C^{BC|BC|AB}_{X{\sigma}{\sigma}}$ & 0.0884 & $C^{BC|BC|AB}_{YZZ}$ & -0.4454 & $C^{BC|BC|AB}_{Y{\sigma}{\sigma}}$ & 0.0171 & $C^{BC|B|B}_{{\sigma}I{\sigma}}$ & 1.0 \\
\hline
$C^{BC|B|B}_{{\sigma}Y{\sigma}}$ & -0.0218 & $C^{BC|B|C}_{{\sigma}Z{\sigma}}$ & 0.5774 & $C^{BC|B|AC}_{{\sigma}{\epsilon}Z}$ & 0.6419 & $C^{BC|B|AC}_{{\sigma}{\epsilon}{\sigma}}$ & 0.6553 & $C^{BC|B|AC}_{{\sigma}XZ}$ & 0.4903 & $C^{BC|B|AC}_{{\sigma}X{\sigma}}$ & -0.143 \\
\hline
$C^{BC|B|AB}_{{\sigma}{\sigma}Z}$ & -0.4539 & $C^{BC|B|AB}_{{\sigma}{\sigma}{\sigma}}$ & 0.9267 & $C^{BC|C|C}_{{\sigma}I{\sigma}}$ & 1.0 & $C^{BC|C|C}_{{\sigma}Y{\sigma}}$ & -0.0218 & $C^{BC|C|AC}_{{\sigma}{\sigma}Z}$ & -0.4539 & $C^{BC|C|AC}_{{\sigma}{\sigma}{\sigma}}$ & 0.9267 \\
\hline
$C^{BC|C|AB}_{{\sigma}{\epsilon}Z}$ & 0.6419 & $C^{BC|C|AB}_{{\sigma}{\epsilon}{\sigma}}$ & 0.6553 & $C^{BC|C|AB}_{{\sigma}XZ}$ & 0.4903 & $C^{BC|C|AB}_{{\sigma}X{\sigma}}$ & -0.143 & $C^{BC|AC|AC}_{ZIZ}$ & 0.7862 & $C^{BC|AC|AC}_{Z{\epsilon}{\sigma}}$ & 0.6419 \\
\hline
$C^{BC|AC|AC}_{ZX{\sigma}}$ & -0.4903 & $C^{BC|AC|AC}_{ZYZ}$ & -0.4454 & $C^{BC|AC|AC}_{{\sigma}I{\sigma}}$ & 0.7862 & $C^{BC|AC|AC}_{{\sigma}{\epsilon}{\sigma}}$ & -0.405 & $C^{BC|AC|AC}_{{\sigma}X{\sigma}}$ & -0.0884 & $C^{BC|AC|AC}_{{\sigma}Y{\sigma}}$ & 0.0171 \\
\hline
$C^{BC|AC|AB}_{ZZZ}$ & 0.9078 & $C^{BC|AC|AB}_{Z{\sigma}{\sigma}}$ & 0.4539 & $C^{BC|AC|AB}_{{\sigma}Z{\sigma}}$ & -0.4539 & $C^{BC|AC|AB}_{{\sigma}{\sigma}Z}$ & 0.4539 & $C^{BC|AC|AB}_{{\sigma}{\sigma}{\sigma}}$ & 0.5727 & $C^{BC|AB|AB}_{ZIZ}$ & 0.7862 \\
\hline
$C^{BC|AB|AB}_{Z{\epsilon}{\sigma}}$ & 0.6419 & $C^{BC|AB|AB}_{ZX{\sigma}}$ & -0.4903 & $C^{BC|AB|AB}_{ZYZ}$ & -0.4454 & $C^{BC|AB|AB}_{{\sigma}I{\sigma}}$ & 0.7862 & $C^{BC|AB|AB}_{{\sigma}{\epsilon}{\sigma}}$ & -0.405 & $C^{BC|AB|AB}_{{\sigma}X{\sigma}}$ & -0.0884 \\
\hline
$C^{BC|AB|AB}_{{\sigma}Y{\sigma}}$ & 0.0171 & $C^{B|B|B}_{III}$ & 1.0 & $C^{B|B|B}_{IYY}$ & 1.0 & $C^{B|B|C}_{IZZ}$ & 1.0 & $C^{B|B|C}_{YZZ}$ & -0.5666 & $C^{B|B|AC}_{I{\epsilon}{\epsilon}}$ & 1.0 \\
\hline
$C^{B|B|AC}_{IXX}$ & 1.0 & $C^{B|B|AC}_{Y{\epsilon}X}$ & 0.8987 & $C^{B|B|AB}_{I{\sigma}{\sigma}}$ & 1.0 & $C^{B|B|AB}_{Y{\sigma}{\sigma}}$ & 0.0218 & $C^{B|C|C}_{ZIZ}$ & 1.0 & $C^{B|C|C}_{ZYZ}$ & -0.5666 \\
\hline
$C^{B|C|AC}_{Z{\sigma}{\epsilon}}$ & 0.8165 & $C^{B|C|AC}_{Z{\sigma}X}$ & -0.6236 & $C^{B|C|AB}_{Z{\epsilon}{\sigma}}$ & 0.8165 & $C^{B|C|AB}_{ZX{\sigma}}$ & -0.6236 & $C^{B|AC|AC}_{{\epsilon}I{\epsilon}}$ & 0.7862 & $C^{B|AC|AC}_{{\epsilon}{\epsilon}X}$ & 0.8579 \\
\hline
$C^{B|AC|AC}_{{\epsilon}X{\epsilon}}$ & -0.8579 & $C^{B|AC|AC}_{{\epsilon}YX}$ & -0.7065 & $C^{B|AC|AC}_{XIX}$ & 0.7862 & $C^{B|AC|AC}_{XXX}$ & 1.2869 & $C^{B|AC|AB}_{{\epsilon}Z{\sigma}}$ & -0.6419 & $C^{B|AC|AB}_{{\epsilon}{\sigma}{\sigma}}$ & 0.6553 \\
\hline
$C^{B|AC|AB}_{XZ{\sigma}}$ & 0.4903 & $C^{B|AC|AB}_{X{\sigma}{\sigma}}$ & 0.143 & $C^{B|AB|AB}_{{\sigma}I{\sigma}}$ & 0.7862 & $C^{B|AB|AB}_{{\sigma}{\epsilon}{\sigma}}$ & -0.6553 & $C^{B|AB|AB}_{{\sigma}X{\sigma}}$ & 0.143 & $C^{B|AB|AB}_{{\sigma}Y{\sigma}}$ & -0.0171 \\
\hline
$C^{C|C|C}_{III}$ & 1.0 & $C^{C|C|C}_{IYY}$ & 1.0 & $C^{C|C|AC}_{I{\sigma}{\sigma}}$ & 1.0 & $C^{C|C|AC}_{Y{\sigma}{\sigma}}$ & 0.0218 & $C^{C|C|AB}_{I{\epsilon}{\epsilon}}$ & 1.0 & $C^{C|C|AB}_{IXX}$ & 1.0 \\
\hline
$C^{C|C|AB}_{Y{\epsilon}X}$ & 0.8987 & $C^{C|AC|AC}_{{\sigma}I{\sigma}}$ & 0.7862 & $C^{C|AC|AC}_{{\sigma}{\epsilon}{\sigma}}$ & -0.6553 & $C^{C|AC|AC}_{{\sigma}X{\sigma}}$ & 0.143 & $C^{C|AC|AC}_{{\sigma}Y{\sigma}}$ & -0.0171 & $C^{C|AC|AB}_{{\sigma}Z{\epsilon}}$ & -0.6419 \\
\hline
$C^{C|AC|AB}_{{\sigma}ZX}$ & 0.4903 & $C^{C|AC|AB}_{{\sigma}{\sigma}{\epsilon}}$ & 0.6553 & $C^{C|AC|AB}_{{\sigma}{\sigma}X}$ & 0.143 & $C^{C|AB|AB}_{{\epsilon}I{\epsilon}}$ & 0.7862 & $C^{C|AB|AB}_{{\epsilon}{\epsilon}X}$ & 0.8579 & $C^{C|AB|AB}_{{\epsilon}X{\epsilon}}$ & -0.8579 \\
\hline
$C^{C|AB|AB}_{{\epsilon}YX}$ & -0.7065 & $C^{C|AB|AB}_{XIX}$ & 0.7862 & $C^{C|AB|AB}_{XXX}$ & 1.2869 & $C^{AC|AC|AC}_{III}$ & 0.7862 & $C^{AC|AC|AC}_{I{\epsilon}{\epsilon}}$ & 0.7862 & $C^{AC|AC|AC}_{IXX}$ & 0.7862 \\
\hline
$C^{AC|AC|AC}_{IYY}$ & 0.7862 & $C^{AC|AC|AC}_{X{\epsilon}{\epsilon}}$ & 0.5302 & $C^{AC|AC|AC}_{{\epsilon}XY}$ & 0.7065 & $C^{AC|AC|AC}_{XXX}$ & -0.7954 & $C^{AC|AC|AB}_{IZZ}$ & 0.7862 & $C^{AC|AC|AB}_{I{\sigma}{\sigma}}$ & 0.7862 \\
\hline
$C^{AC|AC|AB}_{{\epsilon}Z{\sigma}}$ & 0.6419 & $C^{AC|AC|AB}_{{\epsilon}{\sigma}{\sigma}}$ & 0.405 & $C^{AC|AC|AB}_{XZ{\sigma}}$ & 0.4903 & $C^{AC|AC|AB}_{X{\sigma}{\sigma}}$ & -0.0884 & $C^{AC|AC|AB}_{YZZ}$ & 0.4454 & $C^{AC|AC|AB}_{Y{\sigma}{\sigma}}$ & -0.0171 \\
\hline
$C^{AC|AB|AB}_{ZIZ}$ & 0.7862 & $C^{AC|AB|AB}_{Z{\epsilon}{\sigma}}$ & 0.6419 & $C^{AC|AB|AB}_{ZX{\sigma}}$ & 0.4903 & $C^{AC|AB|AB}_{ZYZ}$ & 0.4454 & $C^{AC|AB|AB}_{{\sigma}I{\sigma}}$ & 0.7862 & $C^{AC|AB|AB}_{{\sigma}{\epsilon}{\sigma}}$ & 0.405 \\
\hline
$C^{AC|AB|AB}_{{\sigma}X{\sigma}}$ & -0.0884 & $C^{AC|AB|AB}_{{\sigma}Y{\sigma}}$ & -0.0171 & $C^{AB|AB|AB}_{III}$ & 0.7862 & $C^{AB|AB|AB}_{I{\epsilon}{\epsilon}}$ & 0.7862 & $C^{AB|AB|AB}_{IXX}$ & 0.7862 & $C^{AB|AB|AB}_{IYY}$ & 0.7862 \\
\hline
$C^{AB|AB|AB}_{X{\epsilon}{\epsilon}}$ & 0.5302 & $C^{AB|AB|AB}_{{\epsilon}XY}$ & 0.7065 & $C^{AB|AB|AB}_{XXX}$ & -0.7954 & & & & & &\\
\hline
\end{tabular}}
\caption{Structure constants of the 3-state Potts CFT.}\label{tab:structure}
\end{table}

\begin{table}[h]
\centering
\scriptsize
\setlength{\tabcolsep}{2pt}
\renewcommand{\arraystretch}{1.6}
\resizebox{.9\linewidth}{!}{
\begin{tabular}{|l|l|l|l|l|l|l|l|l|l|l|l|}
\hline
$C^{III}_{III}$ & 1.0 & $C^{II{\epsilon}}_{I{\epsilon}{\epsilon}}$ & 1.0 & $C^{II{\sigma}}_{I{\sigma}{\sigma}}$ & 1.0 & $C^{IIX}_{IXX}$ & 1.0 & $C^{IIY}_{IYY}$ & 1.0 & $C^{IIZ}_{IZZ}$ & 1.0 \\
\hline
$C^{I{\epsilon}{\epsilon}}_{{\epsilon}I{\epsilon}}$ & 0.7862 & $C^{I{\epsilon}X}_{{\epsilon}{\epsilon}X}$ & 0.8579 & $C^{I{\epsilon}{\sigma}}_{{\epsilon}{\sigma}{\sigma}}$ & 0.6553 & $C^{I{\epsilon}Z}_{{\epsilon}{\sigma}Z}$ & 0.8165 & $C^{I{\epsilon}{\epsilon}}_{{\epsilon}X{\epsilon}}$ & 0.8579 & $C^{I{\epsilon}Y}_{{\epsilon}XY}$ & 0.8987 \\
\hline
$C^{I{\epsilon}X}_{{\epsilon}YX}$ & 0.7065 & $C^{I{\epsilon}{\sigma}}_{{\epsilon}Z{\sigma}}$ & 0.6419 & $C^{I{\sigma}{\sigma}}_{{\sigma}I{\sigma}}$ & 0.5559 & $C^{I{\sigma}{\sigma}}_{{\sigma}{\epsilon}{\sigma}}$ & 0.4633 & $C^{I{\sigma}Z}_{{\sigma}{\epsilon}Z}$ & 0.5774 & $C^{I{\sigma}{\sigma}}_{{\sigma}{\sigma}{\sigma}}$ & 0.6553 \\
\hline
$C^{I{\sigma}X}_{{\sigma}{\sigma}X}$ & 0.143 & $C^{I{\sigma}Y}_{{\sigma}{\sigma}Y}$ & 0.0218 & $C^{I{\sigma}Z}_{{\sigma}{\sigma}Z}$ & 0.4082 & $C^{I{\sigma}{\sigma}}_{{\sigma}X{\sigma}}$ & 0.1011 & $C^{I{\sigma}Z}_{{\sigma}XZ}$ & 0.441 & $C^{I{\sigma}{\sigma}}_{{\sigma}Y{\sigma}}$ & 0.0121 \\
\hline
$C^{I{\sigma}{\sigma}}_{{\sigma}Z{\sigma}}$ & 0.3209 & $C^{I{\sigma}X}_{{\sigma}ZX}$ & 0.4903 & $C^{IXX}_{XIX}$ & 0.7862 & $C^{IXY}_{X{\epsilon}Y}$ & 0.8987 & $C^{IXZ}_{X{\sigma}Z}$ & 0.6236 & $C^{IXX}_{XXX}$ & 1.2869 \\
\hline
$C^{IYY}_{YIY}$ & 1.0 & $C^{IYZ}_{YZZ}$ & 0.5666 & $C^{IZZ}_{ZIZ}$ & 0.7071 & $C^{IZZ}_{ZYZ}$ & 0.4006 & $C^{IZZ}_{ZZZ}$ & 0.8165 & $C^{{\epsilon}{\epsilon}{\epsilon}}_{III}$ & 0.7862 \\
\hline
$C^{{\epsilon}{\epsilon}X}_{I{\epsilon}{\epsilon}}$ & 0.7862 & $C^{{\epsilon}{\epsilon}{\sigma}}_{I{\sigma}{\sigma}}$ & 0.7862 & $C^{{\epsilon}{\epsilon}Z}_{I{\sigma}{\sigma}}$ & 0.7862 & $C^{{\epsilon}{\epsilon}{\epsilon}}_{IXX}$ & 0.7862 & $C^{{\epsilon}{\epsilon}Y}_{IXX}$ & 0.7862 & $C^{{\epsilon}{\epsilon}X}_{IYY}$ & 0.7862 \\
\hline
$C^{{\epsilon}{\epsilon}{\sigma}}_{IZZ}$ & 0.7862 & $C^{{\epsilon}XX}_{{\epsilon}I{\epsilon}}$ & 0.7862 & $C^{{\epsilon}{\epsilon}X}_{X{\epsilon}{\epsilon}}$ & -0.5302 & $C^{{\epsilon}XY}_{{\epsilon}{\epsilon}X}$ & 0.8579 & $C^{{\epsilon}X{\sigma}}_{{\epsilon}{\sigma}{\sigma}}$ & 0.405 & $C^{{\epsilon}X{\sigma}}_{{\epsilon}{\sigma}Z}$ & -0.6419 \\
\hline
$C^{{\epsilon}XZ}_{{\epsilon}{\sigma}{\sigma}}$ & 0.6553 & $C^{{\epsilon}XX}_{{\epsilon}X{\epsilon}}$ & 0.5302 & $C^{{\epsilon}XX}_{{\epsilon}XY}$ & -0.7065 & $C^{{\epsilon}X{\epsilon}}_{{\epsilon}YX}$ & 0.7065 & $C^{{\epsilon}X{\sigma}}_{{\epsilon}Z{\sigma}}$ & 0.6419 & $C^{{\epsilon}{\sigma}{\sigma}}_{{\sigma}I{\sigma}}$ & 0.5559 \\
\hline
$C^{{\epsilon}{\sigma}{\sigma}}_{{\sigma}{\epsilon}{\sigma}}$ & -0.2864 & $C^{{\epsilon}{\sigma}{\sigma}}_{{\sigma}{\epsilon}Z}$ & 0.4539 & $C^{{\epsilon}{\sigma}Z}_{{\sigma}{\epsilon}{\sigma}}$ & 0.4633 & $C^{{\epsilon}{\epsilon}{\sigma}}_{X{\sigma}{\sigma}}$ & 0.0884 & $C^{{\epsilon}{\sigma}{\sigma}}_{{\sigma}{\sigma}{\sigma}}$ & 0.405 & $C^{{\epsilon}{\sigma}{\sigma}}_{{\sigma}{\sigma}Z}$ & 0.3209 \\
\hline
$C^{{\epsilon}{\sigma}X}_{{\sigma}{\sigma}Y}$ & 0.0171 & $C^{{\epsilon}{\sigma}Y}_{{\sigma}{\sigma}X}$ & 0.143 & $C^{{\epsilon}{\sigma}Z}_{{\sigma}{\sigma}{\sigma}}$ & 0.6553 & $C^{{\epsilon}{\sigma}{\sigma}}_{{\sigma}X{\sigma}}$ & -0.0625 & $C^{{\epsilon}{\sigma}{\sigma}}_{{\sigma}XZ}$ & -0.3467 & $C^{{\epsilon}{\sigma}Z}_{{\sigma}X{\sigma}}$ & -0.1011 \\
\hline
$C^{{\epsilon}{\sigma}{\sigma}}_{{\sigma}Y{\sigma}}$ & 0.0121 & $C^{{\epsilon}{\sigma}{\epsilon}}_{{\sigma}ZX}$ & 0.4903 & $C^{{\epsilon}{\sigma}{\sigma}}_{{\sigma}Z{\sigma}}$ & -0.3209 & $C^{{\epsilon}ZZ}_{{\sigma}I{\sigma}}$ & 0.7071 & $C^{{\epsilon}Z{\sigma}}_{{\sigma}{\epsilon}Z}$ & 0.4539 & $C^{{\epsilon}{\epsilon}Z}_{X{\sigma}{\sigma}}$ & -0.143 \\
\hline
$C^{{\epsilon}Z{\sigma}}_{{\sigma}{\sigma}Z}$ & -0.3209 & $C^{{\epsilon}ZX}_{{\sigma}{\sigma}Y}$ & -0.0171 & $C^{{\epsilon}Z{\sigma}}_{{\sigma}XZ}$ & 0.3467 & $C^{{\epsilon}ZZ}_{{\sigma}Y{\sigma}}$ & -0.0154 & $C^{{\epsilon}ZY}_{{\sigma}ZX}$ & 0.6236 & $C^{{\epsilon}ZZ}_{{\sigma}Z{\sigma}}$ & 0.4082 \\
\hline
$C^{{\epsilon}{\epsilon}{\epsilon}}_{XXX}$ & 0.7954 & $C^{{\epsilon}{\epsilon}Y}_{XXX}$ & -1.2869 & $C^{{\epsilon}YY}_{XIX}$ & 1.0 & $C^{{\epsilon}YX}_{X{\epsilon}Y}$ & 0.7065 & $C^{{\epsilon}Y{\sigma}}_{X{\sigma}Z}$ & -0.4903 & $C^{{\epsilon}XX}_{YIY}$ & 0.7862 \\
\hline
$C^{{\epsilon}X{\sigma}}_{YZZ}$ & 0.4454 & $C^{{\epsilon}{\sigma}{\sigma}}_{ZIZ}$ & 0.5559 & $C^{{\epsilon}{\sigma}{\sigma}}_{ZYZ}$ & -0.3149 & $C^{{\epsilon}{\sigma}{\sigma}}_{ZZZ}$ & 0.6419 & $C^{{\sigma}{\sigma}{\sigma}}_{III}$ & 0.5559 & $C^{{\sigma}{\sigma}{\sigma}}_{I{\epsilon}{\epsilon}}$ & 0.5559 \\
\hline
$C^{{\sigma}{\sigma}Z}_{I{\epsilon}{\epsilon}}$ & 0.5559 & $C^{{\sigma}{\sigma}{\sigma}}_{I{\sigma}{\sigma}}$ & 0.5559 & $C^{{\sigma}{\sigma}X}_{I{\sigma}{\sigma}}$ & 0.5559 & $C^{{\sigma}{\sigma}Y}_{I{\sigma}{\sigma}}$ & 0.5559 & $C^{{\sigma}{\sigma}Z}_{I{\sigma}{\sigma}}$ & 0.5559 & $C^{{\sigma}{\sigma}{\sigma}}_{IXX}$ & 0.5559 \\
\hline
$C^{{\sigma}{\sigma}Z}_{IXX}$ & 0.5559 & $C^{{\sigma}{\sigma}{\sigma}}_{IYY}$ & 0.5559 & $C^{{\sigma}{\sigma}{\sigma}}_{IZZ}$ & 0.5559 & $C^{{\sigma}{\sigma}X}_{IZZ}$ & 0.5559 & $C^{{\sigma}{\sigma}{\sigma}}_{X{\epsilon}{\epsilon}}$ & 0.3749 & $C^{{\sigma}{\sigma}Z}_{{\epsilon}{\epsilon}X}$ & 0.6067 \\
\hline
$C^{{\sigma}{\sigma}{\sigma}}_{{\epsilon}{\sigma}{\sigma}}$ & 0.2864 & $C^{{\sigma}{\sigma}{\sigma}}_{{\epsilon}{\sigma}Z}$ & 0.4539 & $C^{{\sigma}{\sigma}X}_{{\epsilon}{\sigma}{\sigma}}$ & -0.2864 & $C^{{\sigma}{\sigma}X}_{{\epsilon}{\sigma}Z}$ & -0.4539 & $C^{{\sigma}{\sigma}Y}_{{\epsilon}{\sigma}{\sigma}}$ & 0.4633 & $C^{{\sigma}{\sigma}Z}_{{\epsilon}{\sigma}{\sigma}}$ & -0.4633 \\
\hline
$C^{{\sigma}{\sigma}{\sigma}}_{{\epsilon}XY}$ & 0.4996 & $C^{{\sigma}ZZ}_{{\epsilon}I{\epsilon}}$ & 0.7071 & $C^{{\sigma}{\sigma}Z}_{X{\epsilon}{\epsilon}}$ & -0.6067 & $C^{{\sigma}Z{\sigma}}_{{\epsilon}{\sigma}{\sigma}}$ & 0.4633 & $C^{{\sigma}Z{\sigma}}_{{\epsilon}{\sigma}Z}$ & -0.4539 & $C^{{\sigma}ZX}_{{\epsilon}{\sigma}{\sigma}}$ & -0.4633 \\
\hline
$C^{{\sigma}ZX}_{{\epsilon}{\sigma}Z}$ & 0.4539 & $C^{{\sigma}Z{\sigma}}_{{\epsilon}XY}$ & -0.4996 & $C^{{\sigma}ZZ}_{{\epsilon}YX}$ & 0.6355 & $C^{{\sigma}ZY}_{{\epsilon}Z{\sigma}}$ & -0.5774 & $C^{{\sigma}ZZ}_{{\epsilon}Z{\sigma}}$ & 0.5774 & $C^{{\sigma}{\sigma}{\sigma}}_{X{\sigma}{\sigma}}$ & -0.0625 \\
\hline
$C^{{\sigma}{\sigma}{\sigma}}_{Y{\sigma}{\sigma}}$ & -0.0121 & $C^{{\sigma}{\sigma}X}_{{\sigma}{\sigma}{\sigma}}$ & -0.405 & $C^{{\sigma}{\sigma}X}_{{\sigma}{\sigma}Z}$ & 0.3209 & $C^{{\sigma}{\sigma}Y}_{{\sigma}{\sigma}{\sigma}}$ & -0.6553 & $C^{{\sigma}{\sigma}Z}_{{\sigma}{\sigma}X}$ & 0.1011 & $C^{{\sigma}{\sigma}{\sigma}}_{{\sigma}XZ}$ & 0.3467 \\
\hline
$C^{{\sigma}XX}_{{\sigma}I{\sigma}}$ & 0.7862 & $C^{{\sigma}XY}_{{\sigma}{\epsilon}{\sigma}}$ & 0.6553 & $C^{{\sigma}{\sigma}X}_{X{\sigma}{\sigma}}$ & -0.0625 & $C^{{\sigma}{\sigma}X}_{Y{\sigma}{\sigma}}$ & 0.0121 & $C^{{\sigma}{\sigma}X}_{Z{\sigma}{\sigma}}$ & 0.3209 & $C^{{\sigma}XZ}_{{\sigma}{\sigma}{\sigma}}$ & 0.6553 \\
\hline
$C^{{\sigma}XZ}_{{\sigma}{\sigma}X}$ & 0.1011 & $C^{{\sigma}XX}_{{\sigma}X{\sigma}}$ & -0.0884 & $C^{{\sigma}XX}_{{\sigma}XZ}$ & 0.4903 & $C^{{\sigma}X{\sigma}}_{{\sigma}ZX}$ & 0.3467 & $C^{{\sigma}YY}_{{\sigma}I{\sigma}}$ & 1.0 & $C^{{\sigma}YX}_{{\sigma}{\epsilon}Z}$ & -0.6419 \\
\hline
$C^{{\sigma}{\sigma}Y}_{X{\sigma}{\sigma}}$ & 0.1011 & $C^{{\sigma}{\sigma}Y}_{Y{\sigma}{\sigma}}$ & 0.0121 & $C^{{\sigma}{\sigma}Y}_{Z{\sigma}{\sigma}}$ & -0.3209 & $C^{{\sigma}YZ}_{{\sigma}Z{\sigma}}$ & 0.4082 & $C^{{\sigma}YZ}_{{\sigma}ZX}$ & -0.441 & $C^{{\sigma}ZZ}_{{\sigma}I{\sigma}}$ & 0.7071 \\
\hline
$C^{{\sigma}{\sigma}Z}_{X{\sigma}{\sigma}}$ & 0.1011 & $C^{{\sigma}{\sigma}Z}_{Y{\sigma}{\sigma}}$ & -0.0121 & $C^{{\sigma}ZX}_{{\sigma}{\sigma}Z}$ & 0.3209 & $C^{{\sigma}Z{\sigma}}_{{\sigma}XZ}$ & 0.3467 & $C^{{\sigma}ZZ}_{{\sigma}Y{\sigma}}$ & 0.0154 & $C^{{\sigma}ZZ}_{{\sigma}ZX}$ & -0.441 \\
\hline
$C^{{\sigma}{\sigma}{\sigma}}_{XXX}$ & -0.5624 & $C^{{\sigma}{\sigma}Z}_{XXX}$ & 0.91 & $C^{{\sigma}ZZ}_{XIX}$ & 0.7071 & $C^{{\sigma}ZX}_{X{\sigma}Z}$ & 0.3467 & $C^{{\sigma}{\sigma}{\sigma}}_{YZZ}$ & 0.3149 & $C^{{\sigma}{\sigma}X}_{YZZ}$ & -0.3149 \\
\hline
$C^{{\sigma}{\sigma}X}_{ZZZ}$ & -0.6419 & $C^{{\sigma}XX}_{ZIZ}$ & 0.7862 & $C^{XXX}_{III}$ & 0.7862 & $C^{XXY}_{I{\epsilon}{\epsilon}}$ & 0.7862 & $C^{XXZ}_{I{\sigma}{\sigma}}$ & 0.7862 & $C^{XXX}_{IXX}$ & 0.7862 \\
\hline
$C^{XYY}_{{\epsilon}I{\epsilon}}$ & 1.0 & $C^{XXY}_{X{\epsilon}{\epsilon}}$ & -0.8579 & $C^{XYZ}_{{\epsilon}Z{\sigma}}$ & 0.8165 & $C^{XZZ}_{{\sigma}I{\sigma}}$ & 0.7071 & $C^{XXZ}_{X{\sigma}{\sigma}}$ & 0.143 & $C^{XZZ}_{{\sigma}Y{\sigma}}$ & -0.0154 \\
\hline
$C^{XZZ}_{{\sigma}Z{\sigma}}$ & -0.4082 & $C^{XXX}_{XXX}$ & -0.7954 & $C^{YYY}_{III}$ & 1.0 & $C^{YYZ}_{IZZ}$ & 1.0 & $C^{YZZ}_{ZIZ}$ & 0.7071 & $C^{YZZ}_{ZYZ}$ & 0.4006 \\
\hline
$C^{YZZ}_{ZZZ}$ & -0.8165 & $C^{ZZZ}_{III}$ & 0.7071 & $C^{ZZZ}_{IYY}$ & 0.7071 & $C^{ZZZ}_{IZZ}$ & 0.7071 & $C^{ZZZ}_{YZZ}$ & -0.4006 & &\\
\hline
\end{tabular}}
\caption{Structure constants of the tetracritical Ising CFT.}\label{tab:tetra}
\end{table}

\end{document}